\documentstyle[11pt,aaspp4]{article}
\tighten 
\begin{document}
\newcommand{\lya}{Lyman~$\alpha$}
\newcommand{\lyb}{Lyman~$\beta$}
\newcommand{\za}{$z_{\rm abs}$}
\newcommand{\ze}{$z_{\rm em}$}
\newcommand{\cmtwo}{cm$^{-2}$}
\newcommand{\nhi}{$N$(H$^0$)}
\newcommand{\degpoint}{\mbox{$^\circ\mskip-7.0mu.\,$}}
\newcommand{\kms}{\,km~s$^{-1}$}      
\newcommand{\minpoint}{\mbox{$'\mskip-4.7mu.\mskip0.8mu$}}
\newcommand{\peryr}{\mbox{$\>\rm yr^{-1}$}}
\newcommand{\secpoint}{\mbox{$''\mskip-7.6mu.\,$}}
\newcommand{\sqdeg}{\mbox{${\rm deg}^2$}}
\newcommand{\squig}{\sim\!\!}
\newcommand{\subsun}{\mbox{$_{\twelvesy\odot}$}}
\newcommand{\et}{{\it et al.}~}

\def\ltsima{$\; \buildrel < \over \sim \;$}
\def\simlt{\lower.5ex\hbox{\ltsima}}
\def\gtsima{$\; \buildrel > \over \sim \;$}
\def\simgt{\lower.5ex\hbox{\gtsima}}
\def\arcs{$''~$}
\def\arcm{$'~$}
\def\erf{\mathop{\rm erf}}
\def\erfc{\mathop{\rm erfc}}
\title{THE REST-FRAME OPTICAL PROPERTIES OF ${\rm \bf z \simeq 3}$ GALAXIES 
\altaffilmark{1}}
\author{\sc Alice E. Shapley and Charles C. Steidel\altaffilmark{2}}
\affil{Palomar Observatory, California Institute of Technology, MS 105--24, Pasadena, CA 91125}
\author{\sc Kurt L. Adelberger}
\affil{Harvard-Smithsonian Center for Astrophysics, 60 Garden Street, Cambridge, MA 02138}
\author{\sc Mark Dickinson and Mauro Giavalisco}
\affil{Space Telescope Science Institute, 3700 San Martin Drive, Baltimore, MD 21218}
\author{\sc Max Pettini}
\affil{Institute of Astronomy, Madingley Road, Cambridge CB3 0HA, UK}
\altaffiltext{1}{Based on data obtained at the 
W.M. Keck Observatory, which 
is operated as a scientific partnership among the California Institute of Technology, the
University of California, and NASA, and was made possible by the generous financial
support of the W.M. Keck Foundation.
} 
\altaffiltext{2}{Packard Fellow}
\begin{abstract}
We present the results of a near-infrared imaging survey of $z\sim 3$ Lyman Break
Galaxies (LBGs). The survey covers a total of 30 arcmin$^2$ and includes 118 
photometrically selected LBGs with $K_s$ band measurements, 63 of which also
have $J$ band measurements, and 81 of which have spectroscopic redshifts. 
Using the distribution of optical ${\cal R}$ magnitudes from previous work and
${\cal R}-K_s$ colors for this sub-sample, we compute the rest-frame
optical luminosity function of LBGs. This luminosity
function is described by an analytic Schechter fit with a very steep
faint end slope of $\alpha=-1.85\pm 0.15$, and it
strikingly exceeds locally determined optical luminosity functions
at brighter magnitudes, where it is fairly well constrained.
The $V$-band luminosity density of only the observed bright end of the $z \sim 3$ LBG luminosity
function already approaches that of all stars in the local universe. 

For the 81 galaxies with measured redshifts, we investigate
the range of LBG stellar populations implied by the photometry
which generally spans the range 900--5500 \AA\ in the rest-frame.
The parameters
under consideration are the star-formation rate as a function of time,
the time since the onset of star-formation, and the degree of reddening
and extinction by dust.  
While there are only weak constraints on the parameters
for most of the individual galaxies, there are strong trends in the sample as
a whole. With a wider wavelength baseline than most previous studies at similar
redshifts,
we confirm the trend that intrinsically more luminous galaxies are dustier.
We also find that there is a strong correlation between extinction and
the age of the star-formation episode, in the sense that younger galaxies
are dustier and have much higher star-formation rates. 
The strong correlation between extinction and age, which we 
show is unlikely to be an artifact of the modeling procedure, 
has important implications for an evolutionary sequence among LBGs. 
A unified scenario which accounts for the observed trends in bright LBGs is one in which
a relatively short period of very rapid star-formation (hundreds of M$_{\sun}$ yr$^{-1}$) lasts for roughly 50--100 Myr, after
which both the extinction and star-formation rate are considerably reduced 
and stars are formed at a more quiescent, but still rapid, rate
for at least a few hundred Myr.
In our sample, a considerable fraction ($\sim 20$\%) of the LBGs have 
best-fit star-formation ages $\simgt 1$\,Gyr,
implied stellar masses of $\simgt 10^{10}$ M$_{\sun}$, and are
still forming stars at $\sim 30$ M$_{\sun}$ yr$^{-1}$.
  
\end{abstract}


\section{INTRODUCTION}

In the past few years, considerable progress has been made in our understanding of
the nature of high redshift galaxies, driven mostly by the availability of large samples
that have been photometrically selected using rest-frame UV spectral features.
The largest existing sample at present is at $z \sim 3$, where it is efficient to
use ground-based imaging through optical filters designed to isolate the Lyman
limit spectral discontinuity at 912 \AA\ in the rest-frame, and where the
spectroscopic follow-up has proved to be straightforward because many important
spectral features are well-placed at wavelengths where optical spectrographs
are most sensitive
(Steidel \et 1996a, Steidel \et 1999). Very deep Hubble Space Telescope (HST) imaging
such as that provided by the Hubble Deep Fields has also proved to be very effective,
using similar techniques for $z \simgt 2$ (Steidel \et 1996b, Madau \et 1996, Lowenthal \et 1997).
The HST data reach much deeper into the high redshift galaxy UV luminosity function, but
are confined to very small areas on the sky; as such, the ground-based and space-based
surveys have been largely complementary (cf. Steidel \et 1999). 

To date, most of the work on these high redshift galaxy samples has focused on
the large-scale clustering properties of the galaxies (e.g., Adelberger \et 1998, Giavalisco \et 1998, Steidel \et 1998, Giavalisco \& Dickinson 2001), 
on the properties of individual galaxies as deduced from their 
spectra (e.g., Pettini \et 1998, 2000, 2001), or on
inferences concerning the universal star-formation history as deduced from the global UV luminosity density (e.g., Madau \et 1996, Steidel \et 1999, Meurer, Heckman, \& Calzetti 1999, Adelberger \& Steidel 2000). 
The strong clustering of the bright Lyman Break Galaxies (LBGs) 
has generally been interpreted as indirect
evidence that the observed galaxies reside in relatively massive dark matter halos, but    
how these galaxies are linked to present-day galaxies is far from clear, and has been quite
controversial. While considerable, the directly-observed UV luminosity of LBGs is clearly very significantly
modified by extinction (see Adelberger \& Steidel 2000, hereafter AS2000, for an extensive discussion of this
topic), and at best provides information on the instantaneous formation rate of O and B stars.

While the far-UV properties of the LBGs are the most straightforward to
study observationally, it is also of interest to explore 
additional properties of these galaxies that require
longer wavelength observations, to be used in concert with the existing far-UV measurements. 
For example, one would like to determine the distribution
of rest-frame optical luminosities, and the star-formation
histories, ages, dust content, and stellar masses--- 
all necessary for understanding the range of objects
selected with the typical Lyman Break photometric
criteria, and the relationship of these objects to galaxies in
the lower redshift universe. Indeed, different models provide quite divergent
descriptions of the nature and fate of the objects which are identified as LBGs.
According to one model, the galaxies selected
as LBGs are bright in the rest-frame UV because they
are experiencing merger-induced starburst events.
The intense starburst events occur on relatively short timescales
(less than 100 Myr), and produce 10-100 times less stellar mass 
than what is seen in a typical $L^*$ galaxy. Such low-mass
bursting objects would then be the precursors of local
low-mass spheroids, unless they merge with similar objects to
form more massive systems (Lowenthal \et 1997; Sawicki \& Yee 1998;
Somerville, Primack, \& Faber 2001). Alternatively,
LBGs are the central objects in relatively massive dark matter halos, 
which form stars steadily but relatively quiescently over longer
than 1 Gyr timescales, accumulate $\geq 10^{10} M_{\odot}$
by $z\sim 3$, and evolve eventually into the ellipticals 
and spiral galaxies at the bright end of the local luminosity
function (Steidel \et 1996a; Baugh \et 1998).

It is not possible to determine the nature of
LBG stellar populations--- and thereby distinguish
between the above scenarios--- with only optical observations
probing the rest-frame UV. Both an aging stellar
population and increasing dust extinction result
in redder rest-frame UV colors, so that the
effects of age and dust are degenerate without longer wavelength data. 
Recent studies (Sawicki \& Yee 1998; Dickinson 2000;
Papovich, Dickinson, \& Ferguson 2001)
have shown that the addition of near-IR photometric measurements at
rest wavelengths longer than the location of an 
age-sensitive spectral break at $\sim 3600 \: {\rm \AA}$ 
removes some of the degeneracy between
dust and age in the modeling of LBG stellar populations. 
Here we present the results of a moderately large near-IR survey of optically selected LBGs,
designed to determine the distribution of rest-frame optical
luminosities of LBGs, and, where possible, information about their star-formation histories. 
In \S 2, we present the details
of the optical and near-IR observations and data reduction.
\S 3 summarizes the distribution of optical-IR colors
of LBGs. In \S 4, we describe the derivation of the LBG rest-frame
optical luminosity function, and compare it to local 
galaxy luminosity functions. \S 5 presents
the procedure and results of population synthesis modeling
of the LBG spectral energy distributions between 900 \AA\ and 5500 \AA\ 
in the rest-frame, using the measured optical/IR colors. 
Extensions of the model results to rest-frame UV spectroscopy
are discussed in \S 6, and 
an evolutionary sequence for LBGs is proposed in \S 7.
Our general conclusions are summarized in \S 8. 

\section{OBSERVATIONS AND DATA REDUCTIONS}
\subsection{Optical Imaging}
Optical images were obtained for all the fields included 
here as part of our extensive survey for $z \sim 3$ galaxies; 
the field centers are given in Table 1. The optical imaging data were
collected during the interval 1995 - 1998 at the William
Herschel Telescope (3C 324, B2 0902+34, and CDFa), the Palomar
200 inch Hale Telescope (CDFa, DSF2237a, DSF2237b, HDF, 
Q0201, Q0256, SSA22a, and SSA22b), and the Kitt Peak
Mayall 4 meter Telescope (Westphal). The details of
our survey have been presented elsewhere 
(Steidel \et 1996a, Giavalisco \et 1998,
Steidel \et 1999), so
here we present only a summary of its
relevant features. In deep $U_n$, $G$, ${\cal R}$ (Steidel \& Hamilton 1993)
images, with typical 1$\sigma$ surface brightness
limits of 29.1, 29.2, 28.6 $AB$ magnitudes per arcsec$^2$,
respectively, we identify $z \sim 3$ galaxy candidates
by their distinctive colors. 
To select the objects in the sample centered at $z \sim 3$,
the photometric criteria consist of the following:
$${\cal R}\le 25.5,\quad G-{\cal R}\le 1.2,\quad U_n-G\ge G-{\cal R}+1$$

These criteria effectively isolate star-forming
galaxies whose redshifts place the Lyman continuum discontinuity within the
$U_n$ band, and result in a redshift distribution that is reasonably well-described
by a Gaussian with $\langle z \rangle = 2.97$ and a standard deviation of
$\sigma(z)=0.27$. The various biases inherent in such photometric selection
have been discussed extensively by Steidel \et (1999). 
At the time of this writing, there are approximately 1000 LBGs in this redshift
range that have been spectroscopically confirmed using the Low Resolution Imaging
Spectrometer (LRIS; Oke \et 1995) at the W. M. Keck Observatory. The full optical
survey and spectroscopic catalogs will be presented elsewhere. 

\subsection{Near-IR Imaging}

To explore the range of optical/near-IR colors present in our 
sample of high redshift galaxy candidates and spectroscopically
confirmed $z>2$ galaxies, a subset of objects was selected for follow-up
near-IR imaging. Objects bright in ${\cal R}$
were preferentially selected for near-IR imaging,
as can be seen by the relative distributions
of ${\cal R}$ apparent magnitude in the near-IR sample and that in 
our $z \sim 3$ sample as a whole (Figure 1a). 
We deliberately selected objects whose rest-UV-inferred extinction properties 
(see Steidel \et 1999 and AS2000 for a discussion) spanned the entire range 
seen in the full LBG survey, from zero to several magnitudes
of extinction in the rest-frame UV. 
Within the LBG $z \sim 3$ selection function, galaxies with higher redshifts
suffer greater absorption by intergalactic H~I, resulting in systematically
redder $G-{\cal R}$ colors for a fixed intrinsic spectral energy distribution (SED). 
Figure 1b shows the distributions of $(G-{\cal R})_0$, the
$G-{\cal R}$ color statistically corrected for IGM absorption,
for both the NIRC LBG sample and the LBG $z \sim 3$ sample as a whole.
This figure shows that, in an effort to target galaxies with extreme
values of implied extinction with NIRC, we over-sampled the reddest 
$(G-{\cal R})_0$ bins by roughly a factor of 2 relative to the
rest of the distribution. 

All of the near-IR data were obtained with
the facility near-IR camera (NIRC) (Matthews \& Soifer 1994)
on the W.M. Keck I Telescope during the course of 9 separate observing
runs in the interval 1997 May - 1999 May.
NIRC has a $256 \times 256$ InSb array, 
with a pixel scale of $0.15 \arcsec/{\rm pixel}$,
resulting in a $38.4 \arcsec$ field. 
This field size represents less than 1\% of the
large optical pointings used to find LBGs (typically 9\arcm\ by 9\arcm) which 
contain 100-150 $z \sim 3$ candidates to ${\cal R}=25.5$.
To maximize efficiency, we preferentially targeted 
objects having at least one other LBG within $\sim 40$\arcsec.
A typical pointing included 2--3 LBGs, but this number varied from 1 to 7
LBGs per pointing.  
Each NIRC pointing contained at least one primary target galaxy with a spectroscopic
redshift, and, in many cases, additional LBG candidates for which spectra have not yet been obtained. 
There are spectroscopic redshifts for about 75\% of
the full ${\cal R}-K_s$ sample. 

We observed at $1.25 \mu{\rm m}$ and $2.15 \mu{\rm m}$, using the 
standard $J$ and $K_s$ filters. 
Using integration times of $4 \times 15$ seconds $(J)$ and 
$6 \times 10$ seconds $(K_s)$ per exposure, 
we dithered the telescope between exposures to
form a 9-point box pattern on the sky, 
each box position separated by a few arcseconds. 
In most cases, we aimed to complete 6 or more sets 
of the 9-point dither pattern for each NIRC pointing.
During the dither pattern the telescope was guided with an off-set CCD camera,
with the orientation of the NIRC detector chosen so that a suitable guide star would
fall on the guider chip. 
Table 1 summarizes the coverage of the NIRC
survey with respect that of the optical survey for LBGs,
listing the angular area and number of LBGs covered in $K_s$ and $J$
for each optical survey field. The optical fields listed in Table
1 do not represent the entire optical LBG survey, but only the fields
which contain NIRC pointings.
Table 2 summarizes all of the NIRC target pointings,
the objects contained in each pointing, and the total integration times
for each pointing. 

The data were reduced using standard procedures with the aid of the
DIMSUM\footnote{Deep Infrared Mosaicing Software,
a package of IRAF scripts by Eisenhardt, Dickinson, Stanford, and Ward,
available at ftp://iraf.noao.edu/contrib/dimsumV2/} package.
On clear nights, we determined the photometric zeropoints for the $J$ and $K_s$  
bandpasses with observations of faint (11th - 12th magnitude) near-infrared standard
stars from the list of Persson \et (1998). 
During 5 of our 9 NIRC runs, we experienced
stable, photometric conditions, while the remaining
4 runs contained variable cirrus.
On runs with variable conditions, we carefully calibrated observing sequences
with observations obtained when the conditions were judged to be photometric. 
As measured from the FWHM of standard
stars, the seeing during our several NIRC runs ranged from 
$0.3 \arcsec \; {\rm to} \; 0.7 \arcsec$ in both $J$ and $K_s$, with FWHM$\simeq 0.5$\arcsec\ being typical.
Figure 2 shows two different examples of ${\cal R}$ and $K_s$ images
of LBGs surveyed with NIRC. Westphal-MMD11 has the
reddest ${\cal R}-K_s$ color and brightest $K_s$ magnitude
in the NIRC LBG sample, while B20902-C6 is in the bluer and
fainter half of the NIRC LBG sample.

\subsection{Optical/Near-IR Photometry}

Prior to the measurement of optical/near-IR colors,
each NIRC frame was smoothed with 
a Gaussian to match the image quality in the corresponding optical
images, which generally had $FWHM \simeq 1.0 \arcsec$. 
The photometry was then performed in a manner analogous to 
that used for $z \sim 3$ galaxy searches, described elsewhere 
(Steidel \et 1995; Steidel \& Hamilton 1993). In brief, using a modified
version of the FOCAS (Valdes 1982) image detection and
analysis routines,  a catalog of isophotal and ``total'' 
object detections was generated in the ${\cal R}-{\rm band}$ 
(where ``total'' refers to the FOCAS definition, 
which is the flux measured within an aperture
grown to twice the area of the initial detection isophote). 
The ${\cal R}-{\rm band}$ isophotal detection apertures were applied 
to the $J$ and $K_s$ images,
to measure ${\cal R}-J$ and ${\cal R}-K_s$ colors. 
Table 3 lists the ${\cal R}$ {\it total} magnitudes, 
$G-{\cal R}$, ${\cal R}-J$, ${\cal R}-K_s$ {\it isophotal} colors,
and redshifts for all objects in the NIRC sample. 
Optical magnitudes ($G$ and ${\cal R}$)
and colors are referenced to the $AB$ system, whereas the near-IR magnitudes
($J$, and $K_s$) are on the Vega system.
\footnote{To convert from Vega magnitudes to $AB$ magnitudes,
we adopt the transformations: $K_{s}(AB) = K_{s}(Vega) + 1.82$, 
$J(AB)=J(Vega) + 0.90$. The standard relation between $AB$ magnitude 
and flux-density, $f_{\nu}$, is: $m_{AB} = -48.6 - 2.5\log f_{\nu}$ where
$f_{\nu}$ is in units of ergs s$^{-1}$ cm$^{-2}$ Hz$^{_-1}$.}

\subsection{Photometric Uncertainties}
To quantify the uncertainties in the  measured ${\cal R}-J$ and
${\cal R}-K_s$ colors, we ran Monte Carlo simulations mimicking the actual
process used to measure magnitudes and colors from the real data. 
Artificial galaxies, with a reasonable range of intrinsic sizes and with
${\cal R}$ magnitudes and ${\cal R}-K_s$ colors
drawn randomly from the observed range of both quantities, were added to
the images after convolution with the seeing disk.
We then produced detection catalogs in ${\cal R}$ and
measured the ${\cal R}-K_s$ colors with matched apertures in the
$K_s$ image. 
The process of adding fake galaxy sets (the number of galaxies added in a given
trial was kept small enough so as not to alter the systematics of the observed
data frames) and recovering the magnitudes and colors using the same procedures
used for the real data  
was repeated until
enough detections were obtained to study the scatter between
true and recovered ${\cal R}-K_s$ color as a function of recovered ${\cal R}$ and
${\cal R}-K_s$. The Monte-Carlo generated photometric uncertainties are significantly
larger than those that would result from the application of simple Poisson
counting statistics that neglect systematics.  

The detections of artificial objects  were binned in 0.5 mag steps in measured ${\cal R}$ 
magnitude and 0.2 mag steps in measured ${\cal R}-K_s$ color. 
The mean and standard deviation were then computed for the distribution of 
$\Delta ({\cal R} - K_s) = ({\cal R} - K_s)_{true} - ({\cal R} - K_s)_{measured}$ 
in each $({\cal R}, {\cal R}-K_s)$ bin. In general we did not find a significant
systematic offset between the measured and true colors, so the color
uncertainties are treated as symmetric.  Accordingly, we adopted the
standard deviation of the $\Delta ({\cal R} - K_s)$ distribution as
the $1 \sigma$ color uncertainty, $\sigma({\cal R}-K_s)$ for each 
$({\cal R}, {\cal R}-K_s)$ bin. 

To be conservative, we initially treated each NIRC pointing independently 
for the estimation of photometric uncertainties in order to explore
possible variations due to  
slightly different depths, seeing conditions, and sky background.
We subsequently determined that the trend of $\sigma({\cal R}-K_s)$ with 
$({\cal R}, {\cal R}-K_s)$ was similar enough in all $K_s$ 
NIRC pointings that we simply
combined the simulation results from all pointings
to calculate an average trend of $\sigma({\cal R}-K_s)$ with 
$({\cal R}, {\cal R}-K_s)$ for our entire ${\cal R}-K_s$ sample.
Uncertainties were then assigned to the measured 
${\cal R} - K_s$ colors in the NIRC sample, based on the
magnitude and color of each object.

An identical procedure was used to generate uncertainties in the ${\cal R}-J$
colors of all of the observed galaxies.
The $G-{\cal R}$ uncertainties were computed 
from a separate, but similar, set of Monte Carlo simulations
(Adelberger 2001). All the color uncertainties are listed along 
with the optical/IR photometry in Table 3.

\section{THE OPTICAL/IR COLORS OF LYMAN BREAK GALAXIES}

The distribution of measured optical-IR colors for
the 118 galaxies with NIRC $K_s$ data is shown in Figure 3. 107 of these
measurements are $K_s$ detections significant at the $5\sigma$ level, 
while 11 of them are only
upper-limits, whose ${\cal R}- K_s$ limits correspond to the typical $5 \sigma$ detection
limit of $K_s = 22.5$ in our images.
We find a mean ${\cal R}- K_s$ color of $< {\cal R}- K_s > = 2.85$ and a standard
deviation $\sigma({\cal R}-K_s) =0.59$.
Of these 118 galaxies, 81 have measured redshifts, with a mean of
$<z> = 2.996$. 
There are $J$ band measurements for 
63 of the galaxies in the ${\cal R} - K_s$ sample. The mean
$J-K_s$ color is $<J - K_s> = 1.52$ with $\sigma(J-K_s) = 0.77$.
At $z \sim 3$, the $J$ and $K_s$ filters correspond to 
$\lambda_{eff}(J)=3100 {\rm \AA}$, and
$\lambda_{eff}(K_s)=5400 {\rm \AA}$ in the rest-frame, respectively, 
whereas the optical passbands $U_n$, $G$, and ${\cal R}$ sample
the galaxies at rest-frame $\lambda_{eff}\simeq$ 900, 1200, and 1700 \AA, respectively. 

Clearly, there is a wide range of optical/near-IR color among the rest-UV
selected $z \sim 3$ LBGs; the bluest galaxies in ${\cal R}-K_s$ have colors that
are essentially flat in $f_{\nu}$ units from the far-UV to the optical in
the rest-frame, whereas the reddest LBGs approach the colors of ``extremely
red objects''.  The typical
${\cal R}- K_s$ color of $z \sim 3$ LBGs is more than 2.5 magnitudes redder than
that expected for an unreddened instantaneous burst of star-formation at the
mean redshift of our sample (${\cal R}- K_s$ = 0.30), but is significantly bluer
than most galaxies in the present-day universe (cf. Papovich \et 2001).  
Unfortunately, in the absence of high quality spectra covering the same
large wavelength range, the interpretation of the colors of LBGs
in the context of understanding their stellar populations and extinction
must rely heavily on models.
In \S 5, we use the full set of optical-IR colors 
($G-{\cal R}, {\cal R}- J, \: {\rm and } \: {\cal R}- K_s$) 
of LBGs in the NIRC sample in an attempt to disentangle the degenerate effects
of dust and age on LBG broadband spectral energy distributions.

\section{REST-FRAME OPTICAL LUMINOSITY FUNCTION}

The distribution of redshifts and observed ${\cal R}$ and $I$ magnitudes
have been used to construct rest-frame far-UV luminosity functions for 
LBGs at $z \sim 3$ and $z \sim 4$ 
respectively (Steidel \et 1999).
With near-IR $K_s$ magnitudes and optical-IR ${\cal R}- K_s$ colors
for a sample of 118 LBGs, we have the 
necessary information to construct the rest-frame {\it optical}
luminosity function of $z \sim 3$ LBGs,  which
is much more easily compared with galaxies in the present-day
universe. 
Optical emission is much less attenuated than far-UV light by the presence of dust, 
and expected to be less directly linked to the instantaneous star-formation
rate, as the stars giving rise to the optical
emission sample a larger swath of the main sequence and include stars with
longer lifetimes than the O/B
stars producing the far-UV light. Thus, one might hope that the optical luminosities
provide more information on the integrated  stellar populations of the LBGs than
can be obtained from far-UV measurements. 

At $z \simeq 3$, the mean redshift
of the LBG spectroscopic sample (and of the NIRC sample as well), 
the $K_s$-band central wavelength of $2.15 \mu {\rm m}$ 
is quite a good match to the
central wavelength of the standard $V$ optical filter. 
While the NIRC sample
contains galaxies with a range of redshifts, 
the $K_s$-band central wavelength falls within the rest-wavelength range
$5100 - 5700 {\rm \AA}$ for the bulk of the sample. 
Furthermore, the ``photometric depth'' of the LBG redshift selection function corresponds
to a difference in absolute magnitude of 
only $\Delta M_V \sim 0.30$ magnitudes for a given apparent magnitude;
consequently, given our typical measurement uncertainties, we view it as a reasonable
approximation for this analysis to treat all the galaxies in the NIRC
sample as if they were located at the mean redshift of the entire 
LBG spectroscopic sample. This of course allows us to use the whole NIRC sample, and
not just the objects with spectroscopic redshifts, for the luminosity function analysis. 

Two pieces of information are required 
to construct the rest-frame optical luminosity function for
LBGs at $z \sim 3$: the LBG rest-frame UV luminosity
function (i.e. the distribution of ${\cal R}$ apparent magnitudes),
and the LBG distribution of ${\cal R}- K_s$ as a 
function of ${\cal R}$ magnitude for the NIRC sample.
The best-fit ${\cal R}$ apparent magnitude luminosity function parameters 
for LBGs (uncorrected for the effects of dust extinction) are a 
faint-end slope of $\alpha = -1.57$, a characteristic apparent magnitude
of $m_{{\cal R}}^{*} = 24.54$, and an overall normalization
of $\phi^{*} = 4.4 \times 10^{-3} h^{3}{\rm Mpc ^{-3}}$ 
(AS2000).\footnote{
These parameters were fit assuming the currently favored cosmology of
$\Omega_{m} = 0.3, \Omega_{\Lambda} = 0.7$.} 
In \S 3, we presented the overall distribution of 
${\cal R}- K_s$ colors for the NIRC sample. A correlation with $98 \%$
confidence (better than $2\sigma$) is detected between
${\cal R}- K_s$ color and ${\cal R}$ magnitude,
such that fainter galaxies have redder ${\cal R}- K_s$ colors
(see Figure 4). This correlation probability 
was computed including the 11 upper limits in the sample of 118
galaxies with ${\cal R} - K_s$ measurements, so the result
should not be biased by our typical detection limit of
$K_s = 22.5$. The trend of ${\cal R}-K_s$ with ${\cal R}$ is
included in the luminosity function analysis by using the relationship
implied by the best-fit regression slope to the correlation:
$\frac{d\langle{\cal R}- K_s\rangle}{ d{\cal R} } = 0.17$. There is a lot of
scatter around this regression slope, but it provides a means
of encoding the trend between the two variables. 

We generated a large sample (15000) of 
LBG ${\cal R}$ apparent magnitudes between ${\cal R}=22.5$ and 
${\cal R} = 27$, by randomly
drawing luminosities from the LBG $z \sim 3$ rest-frame UV 
luminosity function placed at z=2.972, the mean redshift
of the current $z\sim 3$ LBG spectroscopic sample. 
An ${\cal R}- K_s$ color was then assigned
to each of the ${\cal R}$ magnitudes,
drawn randomly from the 
distribution of ${\cal R}- K_s$ colors for galaxies in 
the NIRC sample with $24 \le {\cal R} \le 24.5$, and then shifted
by the amount,
$\Delta({\cal R}- K_s)=0.17\times ({\cal R}-24.25)$,
according to the correlation between ${\cal R}$ and ${\cal R}-K_s$.
The fiducial $({\cal R}, {\cal R}- K_s)$ distribution was restricted 
to ${\cal R}=24.0-24.5$ because this $0.5$ magnitude range
contained the largest number of ${\cal R}- K_s$ measurements (36).
Also, the ${\cal R}- K_s$ measurements for ${\cal R}=24.0-24.5$
were virtually all detections,
rather than mixture of upper limits and detections.
Combining each randomly generated pair of
${\cal R}$ and ${\cal R}- K_s$ measurements, we obtained
an ensemble of $K_s$ magnitudes, which were grouped into 0.5 magnitude
bins.

To determine how the photometric uncertainties of 
both the ${\cal R}$ magnitudes and ${\cal R}- K_s$ colors
translated into uncertainties in the derived $K_s$ luminosity function,
the procedure of generating an ensemble of random 
(${\cal R}$, ${\cal R}- K_s$) pairs was repeated 
a large number (10000) of times.
In each trial, we perturbed
the sample of actual (${\cal R}$, ${\cal R}- K_s$) measurements 
by random amounts consistent with the photometric uncertainties.
In so doing, we assumed that the errors in ${\cal R}$ and
${\cal R} - K_s$ were uncorrelated, which is a valid
approximation since the error in ${\cal R} - K_s$
for galaxies in our sample is dominated by the
$K_s$ photometric error.
Due to the slight differences in the sample of ${\cal R}$ magnitudes
drawn from the LBG ${\cal R}$ apparent luminosity function,
and the more significant differences of the perturbed 
(${\cal R}$, ${\cal R}- K_s$) measurements, 
a different $K_s$ distribution was produced in each trial. 
The analytic Schechter function (1976) was fit to each perturbed
distribution of $K_s$-band apparent magnitudes. 
We restricted the fitted region to span only $19 \le K_s \le 22.5$, 
reflecting the $K_s$ range for which we have actual measurements. 
The average number from all the trials 
was adopted for each 0.5 $K_s$ magnitude bin, 
and the adopted $1\sigma$ uncertainty
was the standard deviation among the realizations (which
does not include the uncertainties
in the LBG ${\cal R}$-band luminosity function best-fit parameters (AS2000)).

The Schechter function was then fit to the average luminosity function values.
The best-fit parameters we obtain are an overall normalization of 
$\phi^{*} = 1.8 \pm 0.8 \times 10^{-3} \;
h^{3}{\rm Mpc}^{-3}$, a faint end slope of
$\alpha = -1.85 \pm 0.15$, and a characteristic apparent magnitude of
$m^*_{K_s} = 20.70 \pm 0.25$. 
The uncertainties on the fitted parameters represent the confidence
intervals generated by fitting the Schechter function 
to each perturbed $K_s$ distribution.
Since it was assumed that all the galaxies comprising the $K_s$ luminosity
function were at the mean redshift of
the LBG spectroscopic sample, it is straightforward to
convert the distribution of $K_s$ apparent magnitudes into
a $5400 {\rm \AA}$ (i.e. rest-frame $V$-band) absolute magnitude distribution.
With our adopted $\Omega_{m} = 0.3, \Omega_{\Lambda} = 0.7$ cosmology, 
$m^*_{K_s} = 20.70$ corresponds to $M^*_{V} = -22.21+ 5\log h$. 
The overall shape of the rest-frame optical luminosity function is determined
by the way in which the ${\cal R} - K_s$ distribution as
a function of ${\cal R}$ magnitude redistributes ${\cal R}$
magnitudes into $K_s$ magnitudes. Accordingly,
the faint end slope of the LBG rest-frame optical luminosity 
function is steeper than that of the UV luminosity function, 
due to the positive correlation between ${\cal R}$ and ${\cal R} - K_s$.
Figure 5 shows the apparent $K_s$
luminosity function (and the absolute optical 
magnitudes to which it corresponds),
as well as the confidence intervals on the best-fit Schechter
luminosity function parameters.
Immediately apparent from Figure 5 is the fact that the well-constrained
bright end of the LBG luminosity function 
greatly exceeds locally determined optical luminosity functions.

To calculate the rest-frame optical co-moving luminosity density for LBGs brighter
than $K_s=22.5$, we integrate over the luminosity-weighted luminosity
function in the appropriate magnitude range to obtain 
$\rho_{5400 {\rm \AA}} = 6.86 \times 10^{26}\: {\rm erg} 
\: {\rm s}^{-1}$Hz$^{-1}h \: {\rm Mpc}^{-3}$. 
We use the $V$-band absolute magnitude
of the Sun, $M_{V\odot}=4.83$ (Binney \& Merrifield 1998), 
to express this quantity in solar units:
$\rho_V=1.35 \times 10^8  \: L_{\odot} h \: {\rm Mpc}^{-3}$.
If the luminosity function were integrated to $K_s=25.0$,
the derived luminosity density would be more than 2 times larger
than the value determined down to the typical
survey detection limit of $K_s=22.5$. However, such an extrapolation
depends sensitively on the value of the faint end slope of the
luminosity function, which is poorly constrained by our observed
sample. Fukugita \et 1998 summarize recent local determinations
of the $B$-band luminosity density, most of which are in the range
$\rho_B= 1.8 - 2.2 \times 10^{8} L_{\odot} \: {\rm Mpc}^{-3}$. Thus, the
LBG $V$-band optical luminosity density down to only $K_s=22.5$ is within
a factor of 2 of local measurements of the $B$-band optical luminosity
density, which were obtained by integrating over the
entire $B$-band luminosity function.

The optical luminosity function of local galaxies has been
computed from numerous surveys over the past thirty years.
Some recent determinations from magnitude limited redshift surveys
include the Two-Degree Field Galaxy Redshift Survey (2dFGRS) $b_j$ luminosity
function (Folkes \et 1999), the Las Campanas Redshift Survey (LCRS)
$R$-band luminosity function (Lin \et 1996), 
and the Sloan Digital Sky Survey (SDSS) luminosity functions in 
five optical bands (Blanton \et 2001). 
These local optical luminosity functions have much shallower
faint end slopes $(\alpha)$, fainter characteristic luminosities $(M_{*})$, 
and higher overall normalizations $(\phi^*)$ than the rest-frame
optical luminosity function of LBGs. To demonstrate these differences,
we plot the 2dFGRS $b_j$ luminosity function in Figure 5 along with 
the LBG rest-frame optical luminosity function. 
The best-fit parameters of the 2dFGRS
luminosity function are $\phi^{*} = 1.69 \pm 0.17 \times 10^{-2} \:
h^{3}{\rm Mpc}^{-3}$, $\alpha = -1.28 \pm 0.05$,
and $M^{*} = -19.73 \pm 0.06 + 5 \log h$.  For a direct comparison
with the LBG luminosity function, which specifically probes
the rest-frame $V$ band, the 2dFGRS luminosity function has been shifted
0.5 magnitudes brighter (i.e. $M^{*} = -20.23 + 5 \log h$), 
reflecting the typical $b_j - V$ color of local galaxies. 

At the brightest absolute magnitudes, where the LBG $V$-band luminosity
function is well-constrained by our observations, 
the $z \sim 3$ LBG $V$-band luminosity function greatly exceeds
the local function, and 
has a different shape. For this reason, we caution against
over-interpreting the differences in the best-fit Schechter function
parameters relative to other luminosity functions; for example, the
very steep faint end slope, very bright value of $M^*$, and
relatively low value of $\phi^*$ may well result from 
fitting the function over a small range of luminosity that
does not actually extend as faint as the true ``knee'' in the luminosity distribution.
In practice, it will be extremely difficult to extend this distribution to much
fainter near-IR magnitudes from the ground, but it should be quite possible from
space.  
 
The obvious implication of the bright end of the $V$-band
luminosity function of LBGs is that the mass-to-light ratio $M/L_V$
of LBGs must be very different from that of galaxies in the
local universe. Since the co-moving optical luminosity density in 
the $z \sim 3$ universe is evidently as large or larger than that 
in the $z \sim 0$ universe despite the fact that most of the stars 
observed in the local universe probably formed after $z\sim 3$,
the $V$ band light at $z \sim 3$ is apparently dominated
by relatively massive stars associated with greatly enhanced current
star-formation rates, and carries little information about total stellar
mass. Unfortunately, this also means that no observations from the 
ground are likely to act as a direct proxy for stellar mass
in objects that are actively forming stars at high redshift. 
Indeed, these observations, at $2.15$ $\mu$m, are still
dominated by the light from ongoing star-formation.
In any case, {\it the $V$-band luminosity density of LBGs at $ z\sim 3$
probably exceeds that of all stars in the present-day universe}.

\section{THE STELLAR POPULATIONS OF LYMAN BREAK GALAXIES}

Given photometry that extends from $\sim 900 -5500$ \AA\ in the rest-frame
for a reasonably large sample of $z \sim 3$ LBGs, it is of interest
to use the data to constrain the star-formation histories and extinction
properties of the galaxies with reference to models. Because the galaxy luminosity
at even the longest wavelengths observed in our sample is still dominated by
stars with small mass-to-light ratios, it is impossible to constrain
the integrated stellar mass of galaxies without understanding something about
the star-formation history. As it turns out, our sensitivity to the
star-formation history using optical and near-IR data is coarse, at best---
at the corresponding rest-frame 
far-UV to optical wavelengths, active star-formation can easily mask the evidence
for previous generations of stars (see, e.g., Papovich \et 2001), 
and dust extinction has a significant effect on the emergent SED.
In order to interpret the color information
of the LBGs in terms of interesting constraints on physical parameters, 
we must rely on models of both dust extinction {\it and} 
the intrinsic spectral energy distributions of stellar populations as a function of
metallicity, initial mass function (IMF), star-formation history, and age. 
There is a large range of possible parameterizations of the extinction law and the
intrinsic stellar population, 
and in general these cannot be well-constrained independently from 
fits to the galaxy SEDs. Consequently, in what follows, we adopt simple models for
the attenuation law, and for the metallicity, IMF, and star-formation history of the 
stellar population, 
which external observations have shown to be reasonable for star-forming
galaxies in the nearby and high redshift universe. Where
relevant, we point out where different assumptions yield significantly
different results. Extensive grids of models have been fitted to (generally fainter)
LBGs observed in the HDF-North by Papovich \et 2001; we refer the reader to that
paper for a more detailed discussion of the effects of varying the modeling assumptions.

\subsection{Age-Dust Degeneracy}

With only optical ($U_n$, $G$, ${\cal R}$, and ${I}$) photometry
for $z \sim 3$ galaxies,
it is not possible to distinguish between a 
young galaxy spectral energy distribution ($\le 30$ Myr) which is 
considerably reddened by dust,
and the spectral energy distribution of a galaxy which
has been continuously forming stars over a much longer time scale ($\sim 1$ Gyr),
but with much less dust extinction (Sawicki \& Yee 1998). 
Figure 6 shows the results
of fitting only the $G-{\cal R}$ color of the galaxy 3C324-C2 with 
Bruzual \& Charlot (1996)
constant star-formation galaxy spectral templates of different
ages. As the template age increases, the amount of dust
reddening necessary to fit the observed $G-{\cal R}$ color
decreases. Both the 1 Myr model with $A_{1600} = 2.84$ and the
1 Gyr model with $A_{1600} = 1.61$ fit the observed
optical photometry equally well (where $A_{1600}$ is
magnitudes of extinction at $1600 {\rm \AA}$ rest-frame). 
Indeed, the model rest-frame UV slope can be modified to match the 
observed ($U_n$, $G$, ${\cal R}$,  ${I}$) colors, 
by simultaneously tuning its age and dust-content.
The degeneracy results from 
the fact that, at $z \sim 3$,  commonly used optical
filters probe a short-wavelength region of the rest-frame
spectrum which does not contain any age-sensitive features that can
be easily recognized from broad-band photometry. However, there is an
age-sensitive break in the spectral energy distributions 
of actively star-forming galaxies that occurs near
$3648 {\rm \AA}$, often referred to as the ``Balmer break''. 
The Balmer break feature, which is  is age sensitive 
over star-formation timescales
of $\sim 50$ Myr$-1$ Gyr, cannot be 
described equally well by any combination of stellar 
population aging and dust-content. Longer wavelength
photometry which includes the discontinuity can in principle be used to break the
degeneracy between model fits. 

The Balmer break feature in galaxy spectra
is due to the stellar bound-free absorption
coefficient, which has a sharp edge at $3648 {\rm \AA}$, (the wavelength
of the $n=2 \longrightarrow continuum$ transition in neutral hydrogen), 
causing photons with wavelengths shortward of the edge to
be absorbed much more efficiently than those with longer wavelengths.
The prominence of this feature in a stellar spectrum depends on the temperature
and gravity of the stellar atmosphere, which determines the form of hydrogen 
that is prevalent in the stellar atmosphere, and therefore which component 
of the total stellar absorption coefficient is dominant. 
In cooler late-type stars (such as F, G, and K), 
$H^{-}$ opacity dominates the total absorption coefficient,
and the Balmer decrement is not very pronounced. In the early-type massive
O and B stars the high temperatures result in the increased ionization of
hydrogen, so again, the Balmer absorption coefficient is reduced. The
Balmer break is most prominent in A-stars with 
temperatures around $10,000^{\circ}{\rm K}$ (Gray 1976).

At the onset of star-formation, a galaxy spectrum is dominated in the rest-frame
UV and blue by the most-massive (O and B) stars, most of which
have lifetimes shorter than 10 Myr. 
In a continuous star-formation model, 
after roughly 10 Myr, the rate at which O stars are being formed 
balances the rate at which they are dying, and
the number of O stars levels off to a constant value (Leitherer \et 1999). 
Meanwhile, the lifetimes of less massive A stars
range from several hundred Myr to 1 Gyr. Therefore, 
as the O star population levels off, the number of A stars
continues to increase. 
At the age where the increasing number of
A stars relative to the constant number of O stars causes the galaxy
spectrum at $\sim 3600 {\rm \AA}$ to be dominated by an A-star spectrum,
we start to see an increasing Balmer break in the galaxy spectrum. 
After 100 Myr the magnitude 
difference on either side of the Balmer break in an unreddened
galaxy model is 0.3 mag, and by 1 Gyr, the break has increased
to a 0.6 magnitudes.  

Thus, the strength of the Balmer break can be used as a rough guide to the
duration of the current episode of star-formation on timescales 
$\simlt 1 \: {\rm Gyr}$. The details will
depend on the star-formation history (e.g., whether the star-formation rate (SFR)
is an increasing, or declining, function of time), the shape of the
initial mass function, and the metallicity of the stellar population.
Because the Balmer break even at its strongest is
a relatively subtle spectral feature, and because only broad-band photometry
is available for the galaxies, a measurement of the Balmer break strength
requires removing the effects of continuum reddening. This requirement
introduces some covariance into the simultaneous estimates of extinction
and the age of the star-formation episode; we discuss this covariance 
extensively in \S 5.3.2. 

\subsection{Population Synthesis Modeling}

\subsubsection{Spectroscopic Sample}

In order to examine the range of stellar population properties
present in the NIRC sample of LBGs,
we considered only the subsample of galaxies
with measured redshifts and near-IR detections
at least in $K_s$. 
The spectroscopic sample so defined consists of 81 galaxies. 
We included the $G-{\cal R}$, ${\cal R} - J$, and ${\cal R}-K_s$
colors of these galaxies in the modeling; the $U_n$ band is strongly
affected by H~I absorption in both the galaxy and the IGM. A substantial
fraction of the spectroscopic sample also has available I band photometry,
but for these high redshift objects the I band adds little information
because of its proximity to ${\cal R}$ and because the measurements
are generally noisier than in other bands.  
Nevertheless, 
$U_n$ and $I$ data points (when they exist) are included in 
figures indicating the best-fit
spectral template with respect to the observed LBG photometry, 

The model galaxy templates used to fit
the LBG photometry include no Lyman $\alpha$ emission, and typically
contain Lyman $\alpha$ absorption with observed equivalent widths of 
$40 - 60 {\rm \AA}$ ($10-15 {\rm \AA}$ in the rest-frame).
However, the sample of LBGs for which there are rest-frame UV spectra
exhibit a wide range of Lyman $\alpha$ equivalent widths,
some of which are ten times larger than the model line widths,
and appear in either absorption, emission, or a combination of both.
The {\it observed} Lyman $\alpha$ equivalent
width was used to correct the $G-{\cal R}$ color for line emission for
each of the 81 galaxies in the spectroscopic sample. 
Figure 7 displays the distribution
of Lyman $\alpha$ observed equivalent widths for the LBGs
with both NIRC data and Keck LRIS spectra (cf. Steidel \et 2000). 

Nebular line emission redshifted into the
$K_s$-band might also bias the results of population synthesis modeling.
The [OIII] $\lambda 5007, 4959 \: {\rm \AA}$ doublet and
${\rm H} \beta$ $\lambda 4861 \: {\rm \AA}$ line all fall within the $K_s$-band
window for $z \geq 3.094$. In the NIRC LBG spectroscopic sample, 49 out of
81 galaxies ($60 \%$) have $z \geq 2.974$ such that at least
[OIII] $\lambda 5007 \: {\rm \AA}$, the strongest of the three lines,
is located within the $K_s$-band. Nebular line strengths have been
measured for only a small sample of LBGs at the current
time (see Pettini \et 2001), nine of which are in the NIRC
sample. 
The combined equivalent width from all the nebular
lines affecting the $K_s$ magnitude was calculated by comparing the
line fluxes from Pettini \et 2001 to the continuum flux, 
measured with NIRC. For each
galaxy, the total nebular equivalent width in the $K_s$ 
filter was compared to the effective width of the $K_s$ filter,
$\Delta\lambda(K_s)_{eff}=3300 {\rm \AA}$, to evaluate the necessary
correction to the $K_s$ magnitude and ${\cal R}-K_s$ color.
In every case, the indicated corrections for nebular emission
reduce the $K_s$ continuum flux, resulting in a bluer 
${\cal R}-K_s$ color. The ${\cal R}-K_s$
corrections ranged from $0.03 - 0.27$ magnitudes with a median
$\Delta({\cal R}-K_s) = 0.14 \; {\rm mag}$. 

In order to determine the effect of these ${\cal R}-K_s$ color
corrections on best-fit model parameters, 
we fit Bruzual \& Charlot (1996) constant star-formation models
to the colors of the nine galaxies with measured nebular equivalent
widths. Models were fit to both the uncorrected colors and the
colors corrected for nebular line emission.
The ${\cal R}-K_s$ correction resulted
in little or no change to the best-fit $E(B-V)$ value (since most of
the lever arm for extinction estimation comes from the rest-UV data). 
The effect on the best-fit age was more significant, in that the nebular line
correction will result in a systematically younger inferred age. The median
correction for the nebular line contamination results in stellar population
ages that are $\sim $35\% younger than if no corrections for line emission are
applied. 
It  should be emphasized that nebular line contamination is negligible for
the 40\% of the spectroscopic NIRC sample with $z < 2.974$.
With such a small sub-sample having near-IR spectroscopy at the time of
this writing, there is not enough information available to determine how
corrections for emission line contamination should be implemented for
objects without near-IR spectroscopic measurements. 
We have chosen, therefore, not to apply
these corrections, but to emphasize that in some cases the inferred
ages may be influenced by the emission line contamination of the broad-band
$K_s$ measurements.

\subsubsection{Modeling Procedure}

The two main theoretical elements which we used were (1) model galaxy
spectra generated with the Bruzual \& Charlot 1996, (hereafter BC96) 
population synthesis code, and (2) a dust effective attenuation
law empirically derived by Calzetti (1997) for local starbursts. 
\footnote{The Calzetti attenuation law
relates the observed flux, $F_{\lambda}$, in units of 
${\rm ergs}\;{\rm s}^{-1}{\rm \AA}^{-1}$, to the intrinsic flux,
$F_{0\lambda}$, with the relation:
$F_{\lambda} = F_{0\lambda}10^{-0.4E(B-V)k(\lambda)}$, where $k(\lambda)$
is a decreasing function of wavelength, reflecting the fact that 
shorter wavelengths suffer more extinction than longer wavelengths.
The attenuation curve is parameterized by an overall normalization, 
$E(B-V)$.} 
There are few constraints on the appropriate reddening
model to use for high redshift star forming galaxies, and so we have
adopted the form that works quite well for local starburst galaxies, which
are arguably the best local analogs of LBGs.   
The validity of this reddening/extinction law at high redshift has
yet to be tested adequately, due to the difficulty in measuring bolometric
star-formation luminosities at high redshift
(see Meurer \et 1999; AS2000). 
We simply use it as the most realistic representation
of the attenuation in in $z \sim 3$ LBGs; we will briefly discuss how the results
depend on the assumed attenuation law in \S 3.2.  

At the time this work began, the newest Bruzual and Charlot software
available to us was the BC96 version, so that the results
discussed in detail are based on BC96 model fits.
In order to verify that the population synthesis fits 
did not change significantly
with the most recent version of the Bruzual and Charlot package,
BC2000 (Charlot 2000, private communication; Liu, Charlot, \& Graham
2000), we compared the BC96 and BC2000 results for a subset of models.
BC96 results were also compared with those from the Starburst99
models (Leitherer \et 1999). 

Models with both solar and sub-solar metallicity 
were fit to the observed colors of LBGs, but we discuss results
for only solar metallicity models. Recent measurements of LBG metallicities
are not very well constrained, ranging from $0.1 - 0.5 Z_{\odot}$
(Pettini \et 2001), indicating that these galaxies are more metal-rich
than Damped Lyman $\alpha$ systems at the same redshift, but have
slightly lower metallicites than H~II regions in the Milky Way near the Sun. 
Even so, we prefer to use solar metallicity stellar models, 
since they are the only models which have been directly calibrated
against empirical stellar spectra of many spectral types. 
Specifically, in the grid of theoretical stellar spectra compiled by Lejeune,
Buser, and Cuisinier (1996, 1997), which
was used for the BC96 modeling, the only spectra
which have been extensively tested against empirical stellar spectra
are those with solar abundance.
Papovich \et 2001 report that the effects of using
$0.2 \: Z_{\odot}$ rather than solar metallicity models to fit
LBG colors include best-fit ages systematically younger by a 
factor of 2, best-fit UV extinction factors systematically higher
by a factor of 3, and derived stellar masses smaller by a factor of
2. These effects stem from the fact that the uncalibrated
sub-solar metallicity models have intrinsically bluer UV continuum
slopes and slightly larger Balmer breaks than solar metallicity models, 
for a given stellar population age. However, Leitherer \et 2001 
recently presented observations of O stars in the Large and Small
Magellanic clouds with an average metallicity of $Z=\frac{1}{4}Z_{\odot}$,
showing that metallicity does not have very drastic effects on the 
empirical rest-frame
UV spectra of hot stars. Leitherer \et concluded that modeling the spectra
of  $\sim \frac{1}{4}Z_{\odot}$ galaxies 
with solar metallicity models is an approximately valid approach.
Until theoretical stellar libraries are updated with empirically
calibrated sub-solar metallicity models of many spectral types,
we believe that solar metallicity models are preferable. 

For all of the BC96 models, a Salpeter IMF extending from 
$0.1 M_{\odot} \; {\rm to} \; 125 M_{\odot} \:\:$ was 
assumed. Papovich \et (2001) have investigated the
effects of varying the IMF on the estimated stellar population
parameters. Acceptable model fits are obtained for 
IMFs with various slopes and lower mass cutoffs, 
with slightly different results.
For example, use of the steeper Scalo IMF results in younger ages, 
lower attenuation values, higher star-formation rates, and
larger formed stellar masses.
The effect of changing the IMF on the formed stellar mass 
is small, however, compared with the associated uncertainties.
We choose not to include the form of the IMF as another parameter in our model, although
clearly the shape of the IMF, if allowed to vary arbitrarily, can have a very
significant effect on the stellar population parameters.\footnote{An analysis
of the gravitationally-lensed LBG MS 1512-cB58 by Pettini \et 2000 has shown that a 
Salpeter IMF is very successful in reproducing the details of the far-UV spectrum.}  

Evolutionary tracks distributed by the ``Padova School'' 
(the default for BC96) were used to describe the evolution
of all the formed stars through the H-R diagram. Several
different simple star-formation histories were considered:
a continuous star-formation rate, $\Psi(t)=\Psi_0$;
and exponentially declining star-formation rates,
$\Psi(t) = \Psi_0\exp(-t/\tau)$, with $\tau=100,50,10 \: {\rm Myr}$.
The set of time constants, $\tau$, was chosen to span the
range between the two extremes of an instantaneous burst and
continuous star-formation.
For each star-formation history, two parameters were allowed to vary:
1) the {\it extinction}, parameterized by $E(B-V)$, 
and 2) the {\it age} since the onset of the most recent episode
of star-formation, $t_{sf}$.

The model dust-attenuated, rest-frame galaxy spectra were shifted to the
measured redshift of the actual galaxy, and further attenuated 
in a manner simulating absorption by the intergalactic medium of
neutral hydrogen (Madau 1995). Each internally reddened, 
redshifted, IGM-absorbed galaxy spectrum was then integrated through
the ($G, {\cal R}$, $J$, and $K_s$) filter transmission curves 
to calculate its predicted colors.
For each $(E(B-V), t_{sf})$ combination, the
predicted ($G-{\cal R}$, ${\cal R} - J$, and ${\cal R}-K_s$)
colors were compared with the observed colors using the $\chi^2$ statistic.
The best-fit $(E(B-V), t_{sf})$ combination was chosen to minimize
$\chi^2$, and the intrinsic instantaneous star-formation rate
was determined by normalizing the best-fit model to the observed
${\cal R}$ magnitude (taking into account the best-fit extinction). 
By our definition, the ``formed stellar mass'' $m_{star}$ is simply the integral
of the star-formation rate and the age of the star-formation episode obtained
from the best-fit model.
  
Once the best-fit $E(B-V)$, $t_{sf}$, $\Psi(t_{sf})$, and 
$m_{star}$, were obtained for the galaxies in the
spectroscopic sample, we computed the $68.3 \%$  and $90 \%$
confidence intervals associated with each parameter.
For each galaxy we generated a large sample of fake ``observed'' colors, 
by perturbing the observed colors in a manner consistent
with the photometric errors. We assumed the errors were 
Gaussian, based on the results of Monte Carlo simulations (\S 2.4).
We also assumed that the color
errors were uncorrelated, a valid assumption given that the 
near-IR uncertainties tend to be much larger than the ${\cal R}$
uncertainty.
The large sample of fits to the perturbed colors indicated
the region of $(E(B-V), t_{sf}, \Psi(t_{sf}), m_{star})$
parameter space allowed by each galaxy's
observed colors and photometric uncertainties.

\subsubsection{Comparison of Models}

As described above, 
a range of BC96 star-formation histories
was considered, and the differences between BC96, BC2000, and
Starburst99 were investigated. 
For the purpose of this analysis, and given our relatively large photometric
uncertainties, the BC96 and BC2000 model fits are virtually
indistinguishable. For the solar metallicity constant 
star-formation models, the BC2000 model SEDs 
have slightly bluer rest-frame far-UV slopes and slightly
larger Balmer breaks than the BC96 SEDs for a given star-formation
history. Therefore, best-fit
ages from BC2000 are typically $70$\% as old, while extinction values are larger   
($\Delta E(B-V) \simlt 0.06$).
The BC2000 best-fit extinction-corrected star-formation rates $\Psi(t_{sf})$ are 
typically $25\%$ larger than in the BC96 models, and so the
BC2000 $m_{star}$ is typically $85\%$ of the BC96 $m_{star}$.
These differences are insignificant, given the large confidence intervals
allowed for the best-fit and derived parameters for any
given set of models.

The differences between the BC96 and the Starburst99 
constant star-formation models were
also explored.  The Starburst99 models use
stellar evolutionary tracks from the ``Geneva School'', and the 
model atmosphere grid of Lejeune \et 1997. For the sake of consistency, we 
chose solar metallicity Starburst99 templates with a Salpeter IMF,
which we re-normalized by roughly a factor of $0.5$ in luminosity,
to remove the effects of the larger minimum mass cutoff in 
the Starburst99 Salpeter IMF $(1 \: M_{\odot})$.
As in the case of BC2000, the Starburst99 best-fit 
$E(B-V)$ and $t_{sf}$ parameters tended towards
younger ages ($45$\% of the BC96 $t_{sf}$), 
and slightly larger values of $E(B-V)$, ($0 < \Delta E(B-V) \leq 0.07$).
The higher $E(B-V)$ values in the Starburst99 fits lead to 
extinction-corrected $\Psi(t_{sf})$ values which are $65 \%$ larger
than the BC96 values on average, 
and therefore the $m_{star}$ values for the Starburst99 models were
typically $70\%$ of the BC96 $m_{star}$ values. The differences
between the BC96 and Starburst99 $E(B-V)$ and $t_{sf}$
parameters are also small compared with the allowed confidence intervals.

For all three sets of models: BC96, BC2000, and Starburst99,
the relative ordering of $E(B-V)$ and $t_{sf}$
is preserved--- galaxies best described by higher $E(B-V)$ [$t_{sf}$] galaxies 
in the BC96 models also have higher $E(B-V)$ [$t_{sf}$] according 
to the Starburst99 and BC2000 models.
Finally, we note that for each galaxy in our
sample, the BC96, BC2000, and Starburst99
constant star-formation models all provide statistically indistinguishable
fits--- i.e. there are no cases where one set of models has significantly
lower $\chi^2$ values than the others.
Evidently, even for identical star-formation histories, there are
significant differences in the predictions of each of the models
that can amount to a factor of 2 in inferred age and
a factor somewhat less than 2 in formed stellar mass. Thus, 
it would be a mistake to take the results from any of the models as
anything but approximate until both the models and the data improve. 

The model results for different BC96 star-formation histories 
were also compared in detail. We did not consider a very large number
of star-formation histories, but enough to bracket a range of
star-formation decay time $(\tau)$ parameter space. 
For the majority of the galaxies successfully modeled, 
{\it all} of the star-formation histories considered 
yielded best-fit models with $\chi^2$ values
allowed at the $90 \%$ level. For 8 of the galaxies we 
were able to rule out the models with time constants smaller
than $\tau =100$ Myr---
this sub-sample is comprised of galaxies which are simultaneously
blue in $G-{\cal R}$ and red in ${\cal R}-K_s$, and which have among the
largest inferred continuous star-formation ages. 

For the rest of the galaxies, whose star-formation histories cannot be
usefully constrained, the best-fit $E(B-V)$ and, especially, $t_{sf}$, parameters 
demonstrate quite a large spread depending on which star-formation
history is chosen to fit the photometry. In contrast, the
formed stellar mass, $m_{star}$, derived from such best-fit parameters is 
a more robust quantity (cf. Papovich \et 2001). We investigated the $m_{star}$
implied by each best-fit $(E(B-V), t_{sf})$ combination as a function
of star-formation history, and find that for the 
$\tau=100 \: {\rm Myr}$ and $\tau=50 \: {\rm Myr}$ models,
the $m_{star}$ derived is typically within a factor of 2 of the
$m_{star}$ derived for the constant star-formation model. For the
$\tau=10 \: {\rm Myr}$ models (these are effectively equivalent
to an ``instantaneous burst'' model), the derived $m_{star}$ is typically
a factor of four smaller than the constant star-formation prediction--- this
relatively large difference is caused mostly by the degeneracy between
an aging starburst and one which is relatively strongly reddened by dust.
Star-formation histories with star-formation rates that are 
{\it increasing} functions of
time pose even more significant problems for the models; for example,
Papovich \et (2001) have shown that an underlying maximally old
stellar population (hidden by the most recent episode of star-formation)
can contain up to $5 \times$ the stellar mass
derived from the rest-frame UV-to-optical SED. In this sense, as emphasized
by Papovich \et, the inferred $m_{star}$ based on the modeling should be
treated as lower limits on the total stellar mass of the galaxies.   

In short, broad-band photometry is not particularly effective in constraining
star-formation histories---
the root cause of
this problem is that the far-UV flux is determined solely by
the instantaneous rate of O and B star-formation, whereas the rest-frame
optical flux is dominated by previous star-formation if the SFR is
declining, and by the instantaneous star-formation if the rate is increasing.  
There is of course no good {\it a priori} reason to favor declining star
formation rates over increasing ones. The simplest assumption is
that the observed instantaneous star-formation rate is representative
of the rate since the onset of the current episode of star-formation.
Under these simplifying assumptions, short-lived ``bursts'' of star-formation
for which the SFRs are much higher than the past average, will appear ``young'',
and protracted periods of star-formation with relatively constant
SFRs on timescales $\simgt 100$ Myr will have significant Balmer breaks that
can be recognized in the broad-band photometry.   

We summarize the best-fit and derived parameters for each galaxy assuming  
the BC96
constant star-formation models in Table 4.
In Figure 8a, the measured photometry for three NIRC LBGs 
is plotted along with the best-fit constant star-formation models
for the galaxies. These examples demonstrate the range of
stellar population parameter space which is probed with NIRC LBGs.
Figure 8b shows the $E(B-V)-t_{sf}$ confidence regions 
for these three galaxies.

\subsubsection{Anomalous Galaxies}

Most of the NIRC LBG spectroscopic sample was adequately
described by the simple models detailed above. Statistically
acceptable fits were obtained for 74 of the 81 LBGs in the NIRC
sample with redshifts. Three of the remaining seven galaxies have suspect
$J$ and $K_s$ photometry upon closer inspection,\footnote{
the ``suspect'' photometry was obtained during variable observing conditions.
an attempt was made to flux calibrate these measurements, but the optical-IR SED's
for the relevant galaxies still indicate anomalous near-IR photometric points.}
and fourth galaxy has a clear Seyfert-2-like spectrum. The remaining
galaxies have colors that defy the simple models considered above.
Two of the most anomalous galaxies are Westphal-MMD11
and DSF2237b-MD81.  MMD11 and MD81 are the two reddest galaxies
in ${\cal R} - K_s$ and ${\cal R}-J$ in the entire NIRC LBG sample, 
with ${\cal R}-K_s$ colors of 4.54 and 4.27, respectively, and
${\cal R}-J$ colors of 2.67 and 2.23. MMD11 is also in the
reddest quartile of IGM-absorption corrected
$G-{\cal R}$ measurements for the sample, 
while MD81 is in the bluest quartile (see Figure 1b).
Neither MMD11 nor MD81 has a LRIS rest-frame UV spectrum
with anomalously strong Lyman $\alpha$ emission or absorption,
or notable interstellar or stellar features (though the
signal-to-noise ratio in these spectra does not enable
a quantitative examination). 
As discussed by AS2000,  at $S_{850} = 5.5 \pm 1.4 \: {\rm  mJy}$ 
(Chapman \et 2000), the $850 \mu {\rm m}$ flux of MMD11
is the strongest detection from the sample
of optically selected LBGs that have been observed with SCUBA.
Furthermore, MMD11 has a complex [OIII] 5007
nebular line profile, extended into two components along the
spatial direction. While there is substructure present,
the [OIII] equivalent width is
one of the smallest which has been measured in LBGs (Pettini \et 2001), 
and could not have contaminated the ${\cal R}-K_s$ colors by any
significant amount. MD81 is not detected at 850$\mu$m at the $\sim 3$mJy level 
(Chapman \et, in preparation),
and has not been studied with near-IR spectroscopy.

When we attempt to fit simultaneously the $G-{\cal R}$, ${\cal R}-J$, 
and ${\cal R}-K_s$ colors for either MMD11 or MD81, we are unsuccessful,
mainly due to the extremely red optical-IR colors.
Since MMD11 and MD81 are so bright in the near-IR,
their well-determined colors have more power
to discriminate the quality of a model fit than
the more poorly determined optical-IR photometry for 
the majority of the NIRC sample. 
However, even if MMD11 and MD81 had 
${\cal R}-J$ and ${\cal R}-K_s$ uncertainties which
were more typical of the NIRC LBG sample, no satisfactory
model fits would be obtained for MD81, and only
marginally acceptable fits would be obtained for MMD11. 
Figure 9 shows various attempts to fit the colors of 
these anomalous galaxies. Estimates of dust reddening are
obtained from fitting only the $G- {\cal R}$ color,
assuming BC96 constant star-formation, and ages of
1 Myr and 1 Gyr. Clearly, the $J$ and $K_s$ points for 
both galaxies are significantly brighter than
the predictions for even the 1 Gyr model. While a model with
an age older than 1 Gyr might be able to fit the $K_s$ points, 
the Balmer break present between $J$ and $K_s$
in such an old template spectrum would not be consistent
with the well-determined, and bright, $J$ measurements.
For most of the galaxies in the NIRC LBG
sample, this is not the case--- even when fitting the $G-{\cal R}$
color alone, it is easy to find a model with the appropriate age
and dust extinction to describe all of the photometry within
the errors. 

If we neglect the $G-{\cal R} $ color, and
fit ${\cal R}-J$ and ${\cal R}-K_s$ alone, 
the best-fit models predict more than 7 magnitudes
of extinction in the rest-frame UV and would significantly under-predict
the rest-frame far-UV fluxes.
The very red
optical-IR colors of these galaxies may indicate a problem
with the Calzetti extinction law in cases where dust extinction
is very extreme. In these cases, as in the case of local
ultra-luminous infrared galaxies (ULIRGs), the details of the geometry
and distribution of the dust relative to where star-formation is
occurring become more important and the simple recipe
for extinction breaks down (Trentham, Kormendy, and Sanders 1999;
Meurer and Seibert 2001). 
For example, based on the $G-{\cal R}$ color alone, we would infer
a moderate to low amount of dust extinction in MD81 (which may still
be true). But such a
small amount of extinction, with any age or star-formation
history considered in this work, cannot account for the
very red ${\cal R}-J$ and ${\cal R} - K_s$ colors of MD81.
Alternatively, the inability to model these unusual
galaxies might be due to the over-simplified nature of the star-formation
histories considered. 

\subsection{Model Results}

Using the sample of 74 galaxies for which acceptable BC96 continuous
star-formation fits were obtained, we explore the trends present among
observed and modeled quantities. 
These galaxies, all of which have redshifts, 
are brighter by an average of $\sim 0.2$ magnitudes
in ${\cal R}$ than the NIRC LBG sample as a whole, and 
brighter by $\sim 0.5$ mag than the larger optical LBG 
photometric sample. 
It should be emphasized that the sample of HDF-N LBGs recently analyzed by
Papovich \et (2001) includes only four (of 33) galaxies brighter than 
$L^*$ in the rest-frame UV LBG luminosity
function, whereas $75$\% of the galaxies in the sample 
analyzed here are brighter than $L^*$. Differences between the two
samples might then be ascribed to UV-luminosity-dependent effects, although
we see only subtle trends with UV luminosity within our full ground-based sample
spanning a factor of $\sim 10$ in UV luminosity.  

Steidel \et 1999 and AS2000 have computed the
distribution of LBG $E(B-V)$ values based on IGM-absorption-corrected $G-{\cal R}$
colors alone, assuming a fixed intrinsic stellar population that corresponds
to a BC96 continuous star-formation model with $t_{sf}=1 \: {\rm Gyr}$. 
With the addition of near-IR data, we have been able to model the extinction and
variations in the stellar populations simultaneously. The inferred distribution
of reddening, assuming the starburst attenuation relation of Calzetti (1997) and
a suite of BC96 continuous star-formation models, has
a median $E(B-V) = 0.155$, corresponding to $A_{1600}=1.62$ magnitudes, 
similar to the median extinction presented in the 
earlier works. This reflects the fact that
the previously-assumed galaxy SED is a reasonable approximation for the bulk of
the sample in the rest-frame UV, 
and that most of the information on reddening comes from the far-UV
data points even for a relatively ``gray'' extinction law. 

The median age for our sample is $t_{sf}=320 \: {\rm Myr}$. 
More than $40\%$ of the galaxies have $t_{sf} > 500 \: {\rm Myr}$,
while $25 \%$ have $t_{sf} < 40 \: {\rm Myr}$. There were nine galaxies with
formal best-fit $t_{sf}$ values smaller than $10 \: {\rm Myr}$. Such
small ages are not physically plausible for episodes of star-formation,
given the dynamical timescales of the star-forming regions
(see \S 7). Furthermore, the conversion between extinction-corrected
UV luminosity and star-formation rate becomes highly non-linear at
ages shorter than $10 \: {\rm Myr}$. On these grounds, we
restricted the best-fit constant-star-formation
$t_{sf}$ parameter space to ages of at least $10 \: {\rm Myr}$,
which still provides acceptable fits for the nine youngest galaxies.
We also applied the constraint that best-fit $t_{sf}$
values could not be older than the age of the universe at
$z\sim 3$ (assuming an $\Omega_m=0.3$, $\Omega_{\Lambda}=0.7$, 
$h=0.7$ cosmology), which affected the fits for seven galaxies
whose formal best-fit $t_{sf}$ values were then too old.
These galaxies, however, all had colors which were 
statistically consistent with being younger than the age of the
universe at $z\sim 3$.
As discussed in \S 5.2.3, the $t_{sf}$ values for the NIRC LBG sample are affected
in a systematic way by varying the 
star-formation history or the population synthesis models,
but the sample as a whole should broadly represent the significant range in
these parameters spanned by the LBGs at $z \sim 3$.
We have noted that the distribution of $t_{sf}$ values does not reflect
corrections to the ${\cal R}- K_s$ color
for nebular line emission for the $60 \%$ of 
the galaxies in the NIRC LBG sample which are at high enough redshift
that the $K_s$ measurement could be affected. However, 
there is no apparent systematic offset towards higher 
$t_{sf}$ values for the higher-redshift portion of the sample as might
be expected if the higher redshift objects were significantly biased 
due to nebular line contamination. 
Histograms of best-fit $t_{sf}$ and $E(B-V)$ are shown in Figure 10a. 

The median derived star-formation rate for the
sample is $\Psi(t_{sf}) =45 h^{-2} M_{\odot} {\rm yr}^{-1}$, and the
distribution of values is quite broad, ranging from
$\sim 5 h^{-2} M_{\odot} {\rm yr}^{-1}$  to as high as
$\sim 940 h^{-2} M_{\odot} {\rm yr}^{-1}$.
Multiplying the best-fit $t_{sf}$ and $\Psi(t_{sf})$
yields a distribution of $m_{star}$, with a
median of $m_{star} = 1.2 \times 10^{10} h^{-2} M_{\odot}$,
smaller but comparable to the present-day $L^*$ stellar mass,
$m_{star}(L^*)\sim 4.0 \times 10^{10} h^{-2} M_{\odot}$
(Cole \et 2000; Papovich \et 2001). 
However, we find that
about $20 \%$ of the sample have $m_{star}$
smaller than $4 \times 10^9 h^{-2} M_{\odot}$ (roughly the same sample
with best-fit $t_{sf} <  40  \: {\rm Myr}$). 
Histograms of the inferred 
$m_{star}$ and $\Psi(t_{sf})$ distributions are shown in Figure 10b. 

As mentioned above, a similar modeling technique for extracting physical
parameters from observed colors of LBGs was used in recent work by
Papovich \et (2001); their sample of 33 galaxies in the HDF-N is a superset
of earlier work by Sawicki \& Yee (1998) on 17 HDF-N galaxies.
The Papovich \et HDF study, while containing all of the Sawicki \& Yee
galaxies, finds somewhat different results from Sawicki \& Yee, including lower
typical extinction factors, older ages, and larger formed stellar masses.
The authors attribute some of these differences to their higher-quality
HST NICMOS $J$ and $H$ data (compared to the Sawicki \& Yee ground-based 
$J$ and $H$), more accurate matching of optical and infrared 
photometry to compute colors, and larger galaxy sample.
Here and above we have focused primarily on the more recent HDF-N analysis of Papovich \et (2001).

The main difference between our modeling procedure and that of Papovich \et
is that the latter work allowed $\tau$, the exponential star-formation decay time constant,
to vary as a free parameter in fitting the models. As discussed above, there were few cases
in which our data could discriminate between different values of $\tau$ that were significant
compared to the inferred value of $t_{sf}$. However, a direct comparison of the model
results is possible for 19 of the 33 galaxies in the Papovich \et work because 
the best-fit $\tau$ values are either longer than or much shorter than the inferred $t_{sf}$, so
that the star-formation history is essentially indistinguishable from a constant
star-formation model.
For these 19 HDF-N galaxies, the median inferred age is $t_{sf}=453 \: {\rm Myr}$, and the
median $E(B-V)=0.095$. The median $t_{sf}$ is thus higher, while similar,
to the NIRC LBG median $t_{sf}$, and the median $E(B-V)$ is slightly smaller than that
of the NIRC LBG sample. The remaining 14 HDF-N galaxies have best-fit
$t_{sf} > \tau$, and $t_{sf}$ values which are 
systematically lower than the $t_{sf} < \tau$ sample. 
However, the $\tau$ values for these galaxies are not
very well constrained, and larger $\tau$ values also provide acceptable
fits. The older $t_{sf}$ values resulting from larger $\tau$ parameters
would be roughly consistent with the $t_{sf}$ values 
for the other 19 galaxies. Thus, given the uncertainties, it is probably fair
to compare broadly the whole Papovich \et sample to the NIRC-LBG sample.
Our results are consistent with the inferred parameters for the small subset
of the  brightest
HDF galaxies, including two galaxies in common between the two samples 
(see Table 3).  Generally speaking, the UV-brighter
NIRC LBG sample has a higher proportion of objects with large inferred extinction
and with large values of the inferred stellar mass 
(these are
generally distinct sets of objects, as discussed further below), both of
which may be due to the different range of UV luminosity spanned within the two samples.  

\subsubsection{Extinction and Luminosity}

Using observed optical and far-infrared (FIR) data,
AS2000 and Meurer \et (1999) demonstrated that at 
both low and high redshifts,
there appears to be a correlation between dust-obscuration
and bolometric luminosity 
($L_{\rm bol,dust}/L_{UV}$\lower.5ex\hbox{$\;\buildrel\propto\over\sim\;$}$L_{\rm bol,dust}$). 
The results of our modeling of the LBG SEDs support the assertion that more heavily
obscured galaxies have larger bolometric luminosities; however, because
many of our assumptions are similar--- most notably, the adopted starburst attenuation relation, 
our new results cannot be taken as independent evidence.
As in AS2000, we find that the rest-frame UV luminosity is uncorrelated
with the inferred value of $E(B-V)$, so that applying the inferred extinction
correction naturally results in a strong correlation between bolometric luminosity
and extinction. As emphasized by AS2000, inferences about extinction in the
high redshift galaxies are very difficult to test observationally, but the trends are
very similar to what is observed for star-forming galaxies in the local universe and, so far,
the starburst attenuation relation is consistent with available cross-checks at high redshift. 
We note in passing that internally consistent application of the inferred extinction corrections
to the rest-frame optical luminosities results in a similar correlation of dust obscuration
and optical luminosity, and a net correction to the $V$-band luminosity density of a factor
of $\sim 2$ compared to the uncorrected numbers presented in \S 4.
The correlations of the UV and optical luminosities with 
best-fit $E(B-V)$ are shown in Figure 11 for BC96 continuous
star-formation models. These correlations hold 
for the all of the star-formation histories used in this work
to model the galaxy colors, and would hold for any assumed extinction relation
that is correlated with rest-UV color. 

\subsubsection{Extinction and Age}

In Figure 12 we plot the best-fit $E(B-V)$ and $t_{sf}$ parameters for the
74 galaxies successfully fit with BC96 
constant-star-formation models. There is a very strong anti-correlation
between $E(B-V)$ and $t_{sf}$ in the sense that younger galaxies
are more heavily extinguished than older galaxies.
Given the strong correlation between $\Psi(t_{sf})$
and $E(B-V)$ discussed in \S 5.3.1, the relationship between extinction and age translates into
a link between age and star-formation rate, where 
the youngest galaxies also have the highest star-formation rates.
The correlation in the sample between $E(B-V)$ and $t_{sf}$ falls roughly
in the same direction as the observed covariance of the
($E(B-V), t_{sf}$) parameters for each individual galaxy.
The covariance of $E(B-V)$ and $t_{sf}$ 
is dominated in most cases by 
response of the fitted parameters to perturbations in the $G-{\cal R}$ color. 
Typically, a positive perturbation to the $G-{\cal R}$ results in a 
{\it positive} perturbation to the best-fit $E(B-V)$ and a {\it negative}
perturbation to the best-fit $t_{sf}$, the latter due to the effect of the extinction
estimate on the predicted ${\cal R}-K$ color for a given $t_{sf}$. 
The covariance of the two parameters is clearly demonstrated by the 
NIRC LBG confidence regions shown in Figure 8b.

Given the strong covariance between best-fit $E(B-V)$ and $t_{sf}$
parameters inherent in the modeling procedure, further tests
are necessary to evaluate whether the trend of increasing extinction
with decreasing $t_{sf}$ is significant. In other words, does the
apparent correlation arise solely
due to objects' covariant confidence regions in $E(B-V)-t_{sf}$ space,
because measurement errors can scatter intrinsically uncorrelated points
along the same direction as an $E(B-V)-t_{sf}$
correlation?  In order to test the strength of such an apparent 
correlation, we constructed a sample of 74 
uncorrelated $E(B-V)$ and $t_{sf}$ pairs (to match the size of the observed 
sample of 74 galaxies). This sample was generated by 
randomly selecting $E(B-V)$ and $t_{sf}$ values independently from the 
{\it intrinsic} marginal distributions of each of the parameters.
These {\it intrinsic} marginal distributions were estimated by assuming that
the {\it observed} marginal distributions (shown in Figure 10a)
represented the {\it intrinsic}
marginal distributions broadened by photometric measurement
errors. In addition to $E(B-V)$ and $t_{sf}$, ${\cal R}$ apparent magnitude
and $z$ were randomly selected from the observed
distributions of ${\cal R}$ and $z$ for NIRC LBGs, in order to compute the
colors that would be observed from a galaxy with the 
randomly selected $E(B-V)$ and $t_{sf}$ parameters.
A fake galaxy was retained only if its $E(B-V)$ and $t_{sf}$ parameters
implied a $G-{\cal R}$ color satisfying the LBG selection criterion of 
$G-{\cal R} \leq 1.2$, and a $K_s$ magnitude
brighter than the typical NIRC LBG detection
limit of $K_s = 22.5$. 
Photometric uncertainties were assigned
to each fake galaxy's predicted set of colors, based on the galaxy's
${\cal R}$ magnitude and colors,
and the previously determined functions of
$\sigma({\cal R}, G-{\cal R})$, $\sigma({\cal R}, {\cal R}-J)$, and 
$\sigma({\cal R}, {\cal R}-K_s)$ (see \S 2.4).

The sample of fake galaxies was then ``observed'' a large 
number (1000) of times.  Each time, the colors of the 
fake galaxies were perturbed in a manner
consistent with their photometric errors (a process similar to 
the one used to construct the confidence intervals for the
74 real observed galaxies). The result of observing the
fake galaxies numerous times with measurement errors
was a large distribution of best-fit $E(B-V)$ and $t_{sf}$ values
(the number of fake galaxies (74) multiplied by the number of 
trials (1000)). The marginal distributions
of $E(B-V)$ and $t_{sf}$ for this large simulated distribution matched the
observed marginal distributions of $E(B-V)$ and $t_{sf}$ in Figure 10a
--- which verifies that our estimate of the intrinsic marginal distribution was reasonable.
We then randomly selected 
a large number of groups of 74 best-fit $E(B-V)-t_{sf}$ pairs 
from the sample of 74,000 
``observed'' galaxies, and computed the correlation coefficient 
between $E(B-V)$ and $t_{sf}$ for each randomly selected sample.  
In 1000 random samples of 74 fake best-fit
$E(B-V)-t_{sf}$ values, {\it no} sample had a correlation coefficient as strong
as the one seen in the real sample.
Thus, {\it it appears that the age/extinction correlation for NIRC LBGs has
less than a $0.1\%$ chance of being the result
of correlated measurement errors alone.}

A related question is whether the detection limits
and selection criteria for the sample could have created
an apparent dearth of young, unreddened
galaxies, and old, dusty galaxies, which would mimic
a real correlation between $E(B-V)$ and $t_{sf}$.
For the redshift range of the sample,
the BC96 constant star-formation 
models with $E(B-V) \sim 0$ and $t_{sf} \leq 100 \: {\rm Myr}$
have $G - {\cal R}$ within the range for LBGs
($-0.02 < G - {\cal R} < 0.35$),
but also very blue ${\cal R}-K_s$ colors (${\cal R}-K_s \leq 1.86$). 
A detection limit in $K_s$ translates into a limit 
in ${\cal R}-K_s$ as a function of ${\cal R}$. However, since
there are ${\cal R}-K_s$ limits rather than detections for only
$3$ out of $81$ galaxies with redshifts in the NIRC sample, we conclude
that our $K_s$ detection limit has not prevented us from detecting
a significant population of young, unreddened galaxies. 
At any redshift in the NIRC sample, old and reddened galaxies,
with $E(B-V) > 0.15$ and $t_{sf} \geq 1 \: {\rm Gyr}$,
have predicted ${\cal R} - K > 3.00$,
which means that all such galaxies with ${\cal R} < 25.5$, 
should be detected in ${\cal R} - K_s$.
At the median redshift of the NIRC sample, the LBG selection
limit of $G-{\cal R} < 1.20$ implies that we should be
able to detect old galaxies with $E(B-V) \leq 0.30$, which is more
than one standard deviation to the red of the mean of the sample. 
At increasing redshift, the upper limit on $E(B-V)$ decreases
until at the highest redshift in the NIRC sample, $z=3.396$,
it is only possible to select a 1 Gyr model with $E(B-V) \leq 0.14$
as a LBG, based on $G-{\cal R}$. Such a redshift dependent effect
discriminates against old and very dusty objects at the high redshift 
end of the LBG redshift selection function. To test the significance of such
an effect, the sample of 74 galaxies was divided into
low and high redshift subsamples, using the median redshift
as a discriminator. All of the galaxies in the low redshift
sample have $z\leq3.061$, at which redshift a 
$1 \: {\rm Gyr}$ stellar population
could be detected with $E(B-V) \leq 0.300$. Despite the fact that
$E(B-V)$ is barely restricted for the low redshift sample,
a strong correlation between $E(B-V)$ and $t_{sf}$ is recovered
from this sample, as well as from the high redshift and total samples.
We conclude that the redshift-dependent bias against
detecting old and dusty galaxies cannot account for the
strong correlation found between $E(B-V)$ and $t_{sf}$ in the NIRC sample either.

The inferred strong correlation between star-formation age and extinction 
does depend on the details of the attenuation law applied
to correct observed galaxy colors and magnitudes to their intrinsic, unobscured values.  
At low redshift, relatively ``gray''  starburst galaxy attenuation relations such as 
that of Calzetti (1997) provide a much
better predictor of bolometric luminosity from far-UV observations than the 
reddening curves derived from observations of single stars, such as
the SMC, LMC, and Galactic reddening curves (Meurer \et 1999).  
There are
very few observations to constrain the nature of dust extinction
in galaxies at high redshift; however, the limited information
which exists favors the starburst attenuation relation over  
a much steeper curve such as the SMC law (Meurer \et 1999, AS2000). 
Nevertheless, if the SMC law is used with the
BC96 constant star-formation models, instead of the Calzetti relation,
the best-fit extinction values at 1600 \AA\ $(A_{1600})$ are
systematically smaller, by a median factor of 2.7,  and the $t_{sf}$
values are systematically larger. The ratio between SMC and Calzetti
$t_{sf}$ values increases as a function of increasing extinction, 
but the median ratio is 3.7.  Qualitatively, the larger SMC $t_{sf}$ 
values result from interpreting a larger fraction
of the ${\cal R}-K_s$ color as due to the Balmer break (aging stellar
population), and a smaller fraction as due to the effects of dust
extinction.\footnote{In fact, the inverse effects on extinction and age
from using an SMC law 
tend to cancel out in the calculation of formed stellar mass,
which is typically only $30 \%$ larger
when using the SMC rather than Calzetti law.} 
The result of using the SMC law is that the correlation between
inferred extinction and inferred age is very much diminished.

\subsubsection{Stellar Mass}

The formed stellar mass, $m_{star}$, for the constant star-formation 
models is simply the product of the
best-fit $t_{sf}$ and $\Psi(t_{sf})$. 
Figure 13 shows the relationship between $m_{star}$
and observed and intrinsic (extinction-corrected)
luminosity values. 
When the sample is restricted to galaxies with
best-fit $t_{sf} > 320 \: {\rm Myr}$ (the median for NIRC LBGs),
there are significant correlations between both UV and optical
extinction-corrected luminosities and $m_{star}$. However,
when the whole NIRC LBG sample is considered,
the extinction-corrected UV and optical
luminosities are only weakly correlated with $m_{star}$,
indicating that, for the wide range of extinction and
age parameters found in the NIRC LBG sample,
even rest-frame optical luminosities (the longest
accessible wavelengths for $z \sim 3$ galaxies until the
Space Infrared Telescope Facility (SIRTF) flies) 
often have more to do with current star-formation rates than with
formed stellar masses.
Thus, as emphasized by Papovich \et (2001),
the estimates of the formed stellar mass associated with the most
recent star-formation episode must rely on the IMF-dependent
population synthesis modeling until longer wavelength
observations are possible. However, Papovich \et also explored
the dependence of inferred $m_{star}$
on the modeling assumptions for a given observed galaxy SED, concluding
that, while variation of the modeling parameters yields highly variable
results for parameters such as age and extinction, the combination of
best-fit pairs of these parameters generally results in much more 
tightly constrained $m_{star}$ estimates for the 
modeled star formation episode. While exploring a smaller volume
of parameter space, we confirm this trend for the formed stellar
mass values (cf. \S 5.2.3).

Our results should be compared directly with the solar metallicity, Salpeter IMF
models of Papovich \et (2001) (their Figure 17), where the characteristic
formed stellar mass is  $\sim$ a few  $\times 10^{10}$ M$_{\odot}$ for objects
with UV luminosities of $L^{\ast}$ in the far-UV luminosity function 
of Steidel \et (1999). Our NIRC sample, which has a median luminosity of
somewhat brighter than $L^{\ast}$, has a median inferred stellar mass 
of $2.5 \times10^{10}$ M$_{\odot}$ using the $h=0.7$ cosmology adopted by
Papovich \et al. Thus, in the small region of overlap, the ground-based
sample yields results similar to the brightest HDF-N galaxies. 
In contrast to Papovich \et, who found that the UV luminosity
was well correlated with the inferred stellar mass, we find that the 
UV luminosity is uncorrelated
with the inferred $m_{star}$. 
This may reflect a real difference between the brighter and fainter samples.
One possible explanation could be that we see a wider range of extinction among the
NIRC LBGs than is found at fainter UV luminosities (e.g., objects of a given UV 
luminosity can be either heavily extinguished very luminous objects, or modestly 
extinguished, much less luminous objects) such that the relationship
between UV luminosities uncorrected and corrected for extinction
is much less tight (see AS2000). 
Because of the very small overlap in UV luminosity between the two samples, the
differences are at present statistically insignificant.  

There are 9 objects\footnote{One is Westphal MMD11, which we were unable to
model successfully, and so it has not been included in this discussion.} from the NIRC sample
which also have nebular line width measurements from near-IR
spectroscopy (Pettini \et 2001). We use the
nebular line widths and the measured near-IR half-light radii
to compute the dynamical mass enclosed within the half-light
radius. The mass inferred from the nebular line widths
is probably not indicative of the total mass in the dark-matter
halo containing the LBG, but is likely to represent a lower
limit (see Pettini \et 2001). Similarly, since the inferred stellar mass estimates are only
sensitive to the most recent episode of star-formation,
they should also represent lower limits on the total stellar mass present. 
However, the near-infrared spectroscopic and photometric measurements
generally span the same physical
region, which contains not only stellar mass, but also
gas and dark matter. The spectroscopic line widths should be
sensitive to all this matter, while the photometry only probes the
luminous stellar matter.
Therefore, the  mass inferred from the nebular line widths should
represent a rough upper limit to the $m_{star}$ values 
derived from the BC96 model fits.
We would suspect a problem in the population synthesis modeling technique
if the derived $m_{star}$ values were much larger and statistically
discrepant with the inferred dynamical masses.
As shown in Figure 14, for the majority
of the galaxies, the $m_{star}$ value is consistent with 
the inferred dynamical mass; in two cases where it is not, the best-fit
stellar mass is significantly smaller than the inferred dynamical mass. 
We refrain from drawing conclusions from Figure 14, given the very large
uncertainties in both mass estimates, but there is no evidence that the two
mass scales are wildly inconsistent with one another.

\section{``Young'' and ``Old'' LBGs: Spectral Differences}

One of the more striking results from the modeling of 
LBG stellar populations is the 
strong correlation between extinction and age, such that
the galaxies best fit by younger stellar populations are also best
described by larger amounts of extinction and reddening. As discussed in 
\S 5.3.2, we believe
that this correlation is significant, despite the tendency for the
individual ($E(B-V), t_{sf}$) covariance intervals 
to lie in a similar direction to the observed sample correlation,
and despite a $G-{\cal R}$ selection criterion which prevents a 
galaxy older than $1 \: {\rm Gyr}$
with significant amounts of dust extinction from being classified as
a LBG at the highest redshifts in the sample.
Assuming that the correlation is real, 
there are implications
for interpreting the range of LBG best-fit stellar populations and extinction
parameters.
It appears that the distributions of LBG star-formation ages, extinction, and
inferred stellar masses
fall along a continuum in the parameter space, the
extremes of which might be viewed
as separate ``populations''. 
To illustrate this point, we isolate 16 galaxies with 
best-fit $t_{sf} \geq 1 \: {\rm Gyr}$, and 16 galaxies 
with $t_{sf} \leq 35 \: {\rm Myr}$. 
In the discussion which follows, 
the $\sim 1 $ Gyr subsample is referred to as the ``old'' sample, 
and the $\leq 35 \: {\rm Myr}$ subsample is labeled ``young.''
The ``young'' sample has 
$\langle E(B-V) \rangle = 0.260$, 
$\langle \Psi(t_{sf}) \rangle = 210 h^{-2} M_{\odot} {\rm yr}^{-1}$, and 
$\langle m_{star} \rangle = 2.9 \times 10^{9} h^{-2} M_{\odot}$. 
In contrast, the ``old'' sample has $\langle E(B-V) \rangle = 0.100 $, 
$\langle \Psi(t_{sf}) \rangle = 25 h^{-2} M_{\odot} {\rm yr}^{-1}$, and 
$\langle m_{star} \rangle = 4.0 \times 10^{10} h^{-2}M_{\odot}$. 
The young sample is characterized by
2 magnitudes more dust extinction in the rest-frame UV than the old sample, 
star-formation rates at least an order of magnitude higher (even
when restricting the best-fit $t_{sf}$ values to be above a reasonable minimum
age), and formed stellar masses more than an order of magnitude smaller.

To investigate further the distinctions between galaxies in the
young, dusty subsample and those in the older, less reddened
group, we drew from our database of LBG rest-frame far-UV spectra,
collected by our group between October 1995 and November 1999
(Steidel \et 1996a,b, 1998, 1999).
By combining individual rest-frame UV spectra of galaxies in the 
``young'' and ``old'' subsamples described above,
we constructed ``young'' and ``old'' composite spectra following a method
similar to that described in Steidel, Pettini, \& Adelberger (2001).  

The young and old composite spectra are plotted in Figure 15,
and Figure 16 shows an expanded view of four specific
regions of the spectra for more detailed comparison. There
are several clear differences to note, the most dramatic of which
is in the relative Lyman $\alpha$ line profiles. 
The old spectrum has 
strong Lyman $\alpha$ emission with rest-frame $W_{\lambda}=20 \: {\rm \AA}$, 
whereas the young spectrum has a broad absorption trough plus
much weaker emission which combine to give roughly $W_{\lambda}=0$.
The radiative transfer of 
Lyman $\alpha$ photons in galaxies is a complex process which depends
not only on the amount of dust present in the ISM but also,
and perhaps more importantly, on the geometry
and kinematics of the neutral interstellar hydrogen gas
(Charlot \& Fall, 1993; Chen \& Neufeld 1994; Kunth \et 1999;
Tenorio-Tagle \et 1999). However, due to resonant scattering, Lyman $\alpha$ photons
will on average traverse a much longer path before escaping the galaxy than nearby
continuum photons, so that they are more prone to absorption by dust in the ISM.
Therefore, the amount of
dust present in the ISM should have a non-negligible effect on the 
emergent Lyman $\alpha$ profile. 
The fact that the composite ``young'' spectrum exhibits a Lyman $\alpha$ profile
with a combination of absorption and
only weak emission, while the composite ``old'' spectrum 
has very strong Lyman $\alpha$ emission, offers independent support for
a scenario in which the youngest galaxies have significantly 
dustier interstellar media than more mature LBGs. 

Additionally, the young spectrum has much stronger Lyman $\beta$ absorption,
stronger low-ionization
interstellar metal absorption lines of Si II 
$\lambda 1192, 1260, \: {\rm and } \: 1526$, 
C II $\lambda 1334$, Fe II $\lambda 1608$,  and a stronger P-Cygni
C IV $\lambda 1549$ profile. The only interstellar metal absorption
line which is stronger in the old spectrum is Al II $\lambda 1670$.
While the S/N ratios of the composite spectra represent 
large improvements over that of the individual spectra,
the significance of the differences in line strengths between
the old and young spectra is still difficult to quantify, given the
noise in the composite spectra and the relatively small sample of objects
whose spectra were combined.
We therefore choose to make more qualitative
observations. The low-ionization interstellar absorption lines are
probably optically thick, and therefore on the flat part of
the curve of growth, where equivalent width is determined
mainly by velocity width, and not column density (Steidel \et 1996a, 
Gonzalez Delgado \et 1998, Heckman \& Leitherer 1997).
Accordingly, the stronger interstellar
absorption lines in the young spectrum may indicate 
interstellar medium velocity widths larger than those in the old
spectrum. Larger velocity widths are consistent with a
scenario in which a higher supernova rate (based
on the higher $\Psi(t_{sf})$) deposits
larger amounts of mechanical energy into the ISM, 
accelerating the interstellar gas to higher velocities.

\section{A Proposed Evolutionary Sequence for LBGs}

A picture begins to emerge from the range of
stellar populations observed in the NIRC LBG sample. 
At one extreme of the continuum, there is a group of 
galaxies with best-fit ages younger than or 
equal to the dynamical timescale associated with
the luminous portions of LBGs. 
Based on NIRC and HST-WFPC2 observations, the typical LBG angular
half-light radius is $0\secpoint25$ (Giavalisco \et 1996; Pettini \et 1998).
This angular size translates into 
$r_{hl}= 1.3 \: h^{-1}$ kpc at $z=3$. Furthermore,
from near-IR spectroscopic measurements of a sample of 15 LBGs,
the typical nebular line widths are $\sigma \simeq 80$ \kms
(Pettini \et 2001). The associated dynamical
timescale is $t_{dyn} \simeq \frac {2 r_{hl}}{\sigma}$. With 
typical values for $r_{hl}$ and $\sigma$, this calculation yields
$t_{dyn} \simeq 30 \: h^{-1}$ Myr. A starburst cannot occur on timescales
shorter than the dynamical timescale of the region experiencing
star-formation, for reasons of causality.
Since a constant star-formation model with an age less than
or equal to the dynamical timescale is indistinguishable from
a bursting mode with a decay time, $\tau$, limited
to be greater than or equal to the dynamical timescale,
the young, dusty galaxies in the NIRC LBG sample might well
be described as galaxies undergoing bursts of star-formation.
Whatever the cause of the burst of star-formation,
we see intense star-formation and enhanced dust extinction
in the youngest galaxies. 
At the other extreme are the galaxies whose best-fit
constant star-formation $t_{sf}$ values are close to a Hubble time
at $z\sim 3$, whose dust extinction is only moderate, and
whose star-formation rates are more quiescent. 
The NIRC LBG sample also contains
galaxies with ages, dust-extinction, and star-formation
rates which are intermediate between these two extremes.

Since the range of NIRC LBG stellar populations is controlled by two strong 
correlations--- one between dust and age, and another between
dust and intrinsic star-formation rate---
we use these correlations to construct a unified
evolutionary model for LBGs at high redshift. According to this model,
the ``young'' and ``old'' samples discussed in this
section constitute different evolutionary stages of the
same population, which are both selected as LBGs. 
The ``young'' galaxies represent
objects in the first stage, where the star-formation
rate is very active ($\geq 200 h^{-2} M_{\odot} \: {\rm yr}^{-1}$),
large amounts of dust obscure the sites of star-formation,
and a stellar mass on the order of $10^9 M_{\odot}$ is formed in 
a dynamical timescale.
As time passes, dust is either destroyed or blown out, (or both),
resulting in lower extinction, and the star-formation rate
decays on a $\sim 50 - 100$ Myr timescale to reach the more
quiescent rates seen in the ``old'' sample
($\sim 25 h^{-2} M_{\odot} \: {\rm yr}^{-1}$).
The ``old'' sample then represents the most advanced stage of the evolution
which we observe in $z\sim 3$ LBGs, by which stage
a stellar mass of $\simgt 10^{10} M_{\odot}$ has formed. 
As discussed earlier, for most ``old'' LBGs,
it is not possible to distinguish from a strong Balmer break
whether constant star-formation has continued for at least
1 Gyr, or whether exponentially declining star-formation has continued
well past the age of the time constant, $\tau$.
If the relatively quiescent star-formation rates of the old galaxies
are extrapolated back to the ages of the young/dusty sample
assuming an exponentially declining star-formation rate with a
time constant on the order of the dynamical timescale for
LBGs, the star-formation rates obtained are
on the order of the active rates in the young-dusty sample.

Such a simple evolutionary picture for LBGs 
is qualitatively consistent with the strong correlations we see in the 
NIRC LBG stellar population parameters.
This evolutionary picture predicts a relatively
flat distribution of $t_{sf}$ values, which is broadly consistent
with the observed $t_{sf}$ histogram (Figure 10a) except for the smallest
$t_{sf}$ bin, which is ``over-sampled'' in the NIRC LBG data set by about
a factor of two (\S 2.2). 
While correcting the youngest bin
by this amount makes the $t_{sf}$ distribution flatter, 
the youngest bin still contains more galaxies than the other bins, 
implying that some fraction of the young, dusty, actively star-forming
galaxies may fade to star-formation rates which are too low to be
detected with the (ground-based) LBG criteria.

There are several interesting issues which accompany the above scenario. First, 
there is the question of what process produces the initial, 
rapid star-formation phase. Some obvious
possibilities are major or minor merger events
(e.g., Somerville \et 2001; Mihos \& Hernquist 1996), 
or even the initial collapse of the baryonic component of the galaxy.
Second, there is the question of how the large amount of dust in the 
significantly reddened ``young'' galaxies forms, 
given the relatively short timescales implied by the ages
of these galaxies--- how much of 
the dust is formed in the current star-formation episode, 
and how much originated in previous generations of stars? 

Another important issue is the process by which dust is 
depleted as ``young'' LBGs evolve into ``old'' galaxies. 
There are a few proposed mechanisms for the removal of
dust from galaxies. The first is radiation pressure,
by which dust grains (especially larger ones $\sim 0.1 \: 
\mu{\rm  m}$ in size) can be expelled by the general radiation
field of a galaxy. Davies \et (1998) have numerically modeled
the removal of dust from a present-day $L^*$ spiral galaxy
by radiation pressure, and have shown
that up to $90 \%$ of the dust formed in such a galaxy
may be ejected from the disk. While the relevant physical
conditions might not be identical in local spiral galaxies and
the high redshift LBGs, radiation pressure provides one viable
mechanism by which dust can be removed from galaxies.
Perhaps more relevant to the conditions in LBGs is the
observational evidence for galactic-scale outflows,
so-called ``superwinds,'' in actively star-forming local galaxies.
These outflows are due to the combined mechanical energy input from
supernovae and winds from massive stars, which sweeps up interstellar
material and accelerates it to speeds of several 
hundred ${\rm km} \: {\rm s}^{-1}$ or greater (Heckman \et 1990,
Heckman \et 2000). There is strong kinematic evidence for such outflows
in LBGs at high-redshift as well (Franx \et 1997; Steidel \et 1998; Pettini \et 1998, 2000, 2001).
Relative to the nebular emission lines representing the
systemic velocity of the galaxy,
low-ionization interstellar absorption lines in LBG
rest-frame UV spectra are blue-shifted
by $\sim 300 {\rm km} \: {\rm s}^{-1}$, 
and Lyman-$\alpha$ emission is redshifted
by $200 - 1100 \: {\rm km} \: {\rm s}^{-1}$ (Pettini \et 2001). 
The implied typical outflow velocities are $\sim 400-500$ \kms,
and in the best studied case (Pettini \et 2000) the implied 
mass outflow rate equals or exceeds the current star-formation
rate (similar to what is seen in local starburst galaxies--- e.g. Martin 1999). 

Recently, there has been direct observational evidence that 
dust, as well as gas, has been swept up by starburst superwinds and is 
contained in the material outflowing from local starburst galaxies 
(Heckman \et 2000; Alton, Davies, \& Bianchi 1999).
Heckman \et (2000) determine a typical outflowing dust mass
of $10^6-10^7 M_{\odot}$ with a typical dust outflow rate
of $0.1 - 1 M_{\odot} \: {\rm yr}^{-1}$ for the starbursts
in their sample. While there are no direct observations
of dust in LBG outflows, the presence of dust in the ISM of
LBGs is required by their broad-band colors. 
It seems very reasonable that the 
dust in the ISM of LBGs becomes entrained in the outflows implied
by the kinematics of UV interstellar and nebular spectral features.
The process of dust being expelled 
in superwinds from starburst galaxies 
at high redshift is qualitatively consistent with the observed 
trend between extinction and star-formation age in LBGs. Indeed,
the superwind phenomenon has important implications not only
for the enrichment of the intergalactic medium, but
also for the evolution of dust opacity in galaxies.

\section{SUMMARY AND CONCLUSIONS}

We have presented the results of a near-IR survey of optically selected
$z\sim 3$ LBGs.  The survey includes 118 galaxies
with $K_s$ measurements, 63 of which also have $J$ measurements. 
We combined the new near-IR data with previously
obtained optical data to compute the distributions of optical-IR
colors of LBGs.  We have used the photometric data, spanning typical
rest-frame wavelengths of 900--5500 \AA\, to obtain the following results:  

1. The rest-frame optical luminosity function of UV-selected LBGs 
greatly exceeds locally determined luminosity functions at the
bright end. A Schechter function fit formally yields a
very steep ($\alpha=-1.85 \pm 0.15$) faint-end slope, but part of
this could be due to the fact that at our detection limit of $K_s=22.5$ 
($M_V=-20.41+5 {\rm log} h$ at $z \sim 3$) 
the data do not extend significantly beyond the knee in the luminosity function 
and the overall shape of the luminosity distribution is quite different from
the local functions.
Independent of the fitted parameters, the $z\sim 3$ optical luminosity function 
has a large excess of luminous galaxies relative to locally 
determined luminosity functions.
Down to our detection limit of $K_s=22.50$ (or a luminosity equivalent
to $\sim 1.2 L^{*}$ in the present-day $V$-band luminosity function),
the rest-frame optical co-moving luminosity density of LBGs
(uncorrected for dust extinction)
is $\rho_V=1.35 \times 10^8 L_{\odot} h \: {\rm Mpc}^{-3}$.
This value is already within a 
factor of 2 of values determined by integrating local 
luminosity functions down to arbitrarily small luminosities.
Thus, the optical light from LBGs at $z \sim 3$ likely exceeds
that produced by all stars in the local universe.

2. We have modeled the stellar populations and extinction of
the LBGs with both spectroscopic redshifts and near-IR measurements (81 galaxies) 
using relatively simple prescriptions.  
Most of the galaxies have star-formation histories, star-formation rates,
and star-formation ages  
that are only weakly constrained by the photometric observations
without limiting parameter space using
external constraints on the assumed extinction law and relatively simple
forms for the star-formation history.  We have chosen to report the results
assuming the solar-metallicity Bruzual and Charlot (1996) models, 
a Salpeter IMF, the Calzetti (1997)
starburst attenuation relation, and a constant star-formation history. The systematic
effects of changing the any of these assumptions have been summarized.  

3. When the interpretation of LBG stellar populations is restricted
to BC96 constant star-formation models, we find a large spread of inferred 
ages, ranging from several Myr to more than 1 Gyr. The
median best-fit constant star-formation age is $t_{sf}=320 \: {\rm Myr}$.
In model-independent terms, the distribution of best-fit
ages indicates that a significant fraction of LBGs have
been forming stars on timescales long enough that a
detectable Balmer break exists in their spectral energy distributions.
Although the specific age assigned to the Balmer break depends on the 
star-formation history used to fit the colors,
we note that very recent work
indicates that winds from LBGs similar
to those modeled above have shock-heated the intergalactic medium on
physical scales of $100-200h^{-1}$ kpc (Adelberger \et 2001, in preparation).
Such a large sphere of influence, with typical outflow velocities of
a few hundred \kms, suggests star-formation timescales of at least
a few hundred million years for typical LBGs. This constitutes completely
independent support for the timescales resulting from the stellar population 
modeling presented in this work.
The BC96 constant star-formation distribution
of best-fit $E(B-V)$ ranges from $\sim 0.0 - 0.45$ with a 
median of 0.155. This distribution depends less on star-formation
history than the distribution of best-fit ages. The median
implied extinction for this sample corresponds 
to a factor of $\sim 4.5$ attenuation in the 
rest-frame UV (1600 \AA) and a factor of $\sim $2 in the rest-frame optical.
Derived from the distributions of $E(B-V)$ and $t_{sf}$, 
the best-fit star-formation rate, $\Psi(t_{sf})$, ranges from 
$5 - 940 h^{-2} M_{\odot} \: {\rm yr}^{-1}$, with a median of 
$45 h^{-2} M_{\odot} \: {\rm yr}^{-1}$. The distribution
of formed stellar masses, $m_{star}$, ranges over more than 2 orders of magnitude
with a median of $\sim 10^{10} h^{-2} M_{\odot}$.
We note that the above parameter distributions and median values 
may not be completely general to LBGs at $z\sim3$, since the
way in which the NIRC LBGs were selected tended to enhance
the number of young, dusty, intensely star-forming, low $m_{star}$
objects, relative to the full LBG sample.

4. Regardless of the star-formation history used in this work to model
the colors of the NIRC LBGs, we find a strong
correlation between dust extinction and intrinsic luminosity,
such that dustier galaxies are intrinsically more luminous.
This correlation has been reported by other authors
(AS2000; Meurer \et 1999), but here the extinction estimates are
supported by observations over a wider wavelength baseline, 
spanning
on average from $900 - 5500 \: {\rm \AA}$ in the rest-frame. This
dependence of extinction on luminosity is consistent with a similar
relation seen for star-forming galaxies in the local universe, and has the 
implication that, for the range of galaxies which we select as LBGs,
those which are more heavily obscured by dust
are as easy to detect in the rest-frame UV and optical
as galaxies which are less dusty,
because they are also intrinsically more luminous.

5. For most galaxies, we find that the formed stellar mass derived from the
integral of the best-fit star-formation rate over time
depends relatively weakly on the star-formation history applied to 
fit the observed colors (cf. Papovich \et 2001). 
Regardless of star-formation history,
the inferred formed stellar mass is only weakly correlated with
extinction corrected UV and optical luminosities for our total ground-based sample. 
This fact, together with the results on the rest-frame optical luminosity function
of LBGs, shows that even at observed wavelengths of 2.15 $\mu$m, the luminosities
are poor proxies for stellar mass in general. 
Therefore, inferences on stellar mass from optical/near-IR photometry
are highly IMF-dependent and will apply only to the stars that have formed
in the last several hundred Myr prior to observation. 
Longer-wavelength observations 
with SIRTF should improve the situation considerably.

6. We find a strong correlation between inferred star-formation age and extinction 
in LBGs, such that younger galaxies are dustier than older galaxies, 
and have higher star-formation rates.
Establishing the significance of this correlation is difficult
because of the covariance of $E(B-V)$ and $t_{sf}$ in our fits, but
we have been unable to reproduce a similarly strong
observed correlation from any intrinsically uncorrelated
distributions of $E(B-V)$ and $t_{sf}$.  The correlation might be made
to vanish with a suitable choice of star-formation histories
or dust attenuation curve (i.e. SMC), but its existence appears
secure for the most plausible choices of both.
We also present evidence
that this correlation is not an artifact of either the LBG
$G-{\cal R}$ selection criteria, or of the typical detection
limit of $K_s=22.5$ for the NIRC observations.  

7. Composite rest-frame UV spectra of ``young'' and ``old'' subsamples
of NIRC LBGs exhibit differences, most dramatically
in their respective Lyman-$\alpha$ profiles, offering independent
evidence that the youngest galaxies have more dust extinction
and/or more active interstellar media. We interpret the dust-age correlation
in the context of a unified picture of LBGs in which the younger, dustier, more
actively star-forming galaxies evolve into the older, less
reddened, and more quiescent galaxies. The evolution of the
dust extinction towards lower values is 
probably governed by the outflow of
dust which is entrained in a ``superwind''
powered by the energy from supernovae explosions and
the winds from massive stars.
Objects at various points along this evolutionary sequence 
are identified by the LBG optical selection criteria.
The timescale for the evolution of the young-dusty galaxies
into the older-less-reddened galaxies is on the
order of $50-100 \: {\rm Myr}$, while the more quiescent phase can
last a Gyr or more.

\bigskip
\bigskip
We would like to thank the staffs of the Palomar, 
La Palma, Kitt Peak, and Keck Observatories for
their assistance in both optical and near-infrared observations.
We would also like to thank
Matthew Hunt, for assistance with the near-IR data reduction, and 
Stephane Charlot for helpful discussions and assistance in implementing
the most up-to-date Bruzual and Charlot spectral synthesis
codes.  
Finally, we wish to extend special thanks to those of Hawaiian ancestry on
whose sacred mountain we are privileged to be guests. Without their generous
hospitality, most of the observations presented herein would not
have been possible.
CCS, KLA, and AES have been supported by grants
AST95-96229 and AST-0070773 from the U.S. National Science Foundation
and by the David and Lucile Packard Foundation.

\bigskip
\newpage

\begin{deluxetable}{lcccccc}
\tablewidth{0pc}
\scriptsize
\tablecaption{NIRC Lyman-Break Galaxy Fields}
\tablehead{
\colhead{Field Name} & \colhead{Field Center}  &
\colhead{Area (${\cal R}$)} &
\colhead{Area ($K_s$)} &
\colhead{\# Galaxies ($K_s$)} & \colhead{Area ($J$)} & \colhead{\# Galaxies ($J$)} \nl
\colhead{} & \colhead{(J2000)} &
\colhead{(arc min)$^{2}$} & \colhead{(arc min)$^{2}$} & \colhead{} &
\colhead{(arc min)$^{2}$} & \colhead{}
 }
\startdata
CDFa       & 00 53 23.7  +12 34 00 & ~78.42 & ~1.72 & ~~7 & ~1.18 & ~5 \nl
Q0201 &  02 03 51.6 +11 34 09 & ~75.77 &~1.35 &~~3 & ~1.21 & ~3 \nl
Q0256 &  02 59 08.6 +00 11 41 & ~72.28 & ~1.17& ~~6& \nodata & \nodata \nl
B2 0902+34 & 09 05 30.2  +34 07 55  & ~41.93 & ~2.63 & ~10 & ~1.77 & ~6 \nl
HDF & 12 36 52.3  +62 12 59 & ~75.32 & ~2.96 & ~14 & ~2.29 & 10 \nl
WESTPHAL & 14 17 14.5 +52 24 36 & 226.93 & ~9.03 &~30 & ~4.12 & 15 \nl
3C324 & 15 49 49.6  +21 29 07 & ~44.29 & ~1.85 & ~~6  & ~0.56 & ~2 \nl
SSA22a & 22 17 34.2  +00 15 01 & ~77.75 & ~5.75 & ~22 & ~1.73 & ~6 \nl
SSA22b & 22 17 34.2  +00 06 18 & ~77.60 & ~0.59 & ~~2 & \nodata & \nodata \nl
DSF2237a & 22 40 08.5  +11 52 34 & ~83.38 & ~0.64 & ~~1 & ~0.61 & ~1 \nl
DSF2237b & 22 39 34.3  +11 51 44 & ~81.62 & ~2.43 & ~17 & ~1.71 & 15 \nl\nl
TOTAL & &  935.29 & 30.12 & 118 & 15.18 & 63\nl
\enddata
\end{deluxetable}
\newpage
\begin{deluxetable}{llcc}
\tablewidth{0pc}
\scriptsize
\tablecaption{NIRC Observations and Integration Times}
\tablehead{
\colhead{Pointing}  & \colhead{Objects}   & \colhead{Exposure ($K_s$)}  &
\colhead{Exposure ($J$)}  \nl
\colhead{}  & \colhead{}   & \colhead{(sec)}  &
\colhead{(sec)}
 }
\startdata

CDFa-C8  & C8,MD10 & 2400 & \nodata \nl
CDFa-C1  & C1,MD2 & 3780 & 1620\nl
CDFa-C22 & C22,M19,C19 & 3780 & 4620\nl
Q0201-C6  & C6,MMD21 & 3480 & 3780\nl
Q0201-B13  & B13 & 2700 & 2400\nl
Q0256-M13  & M13,C15,MD22 & 5220 & \nodata \nl
Q0256-M17  & M17,MD32,MD34 & 4260 & \nodata \nl
B20902-C6  & C6,M11,MD31 & 3780 & 3240\nl
B20902-MD21  & MD21 & 3240 & 3240\nl
B20902-D11  & D11,D12,MD32 & 1080 & 1620\nl
B20902-MD11  & MD11,MD14,MD18 & 2160 & \nodata \nl
HDF-MM23  & MM23,oC37,oMD49 & 6000  & 4680\nl
HDF-DD15  & DD15,MM28 & 3780 & 1620\nl
HDF-CC24  & CC24,oC38,oM5,MM25 & 2160 & \nodata \nl
HDF-DD3  & DD3,MM9,MM7 & 3120 & 3240\nl
HDF-MM18  & MM18,MM17 & 3240 & 3240\nl
WESTPHAL-CC70  & CC70,MM69,DD49 & 3240 & 3240\nl
WESTPHAL-DD8  & DD8,MMD14,MMD16,MM8 & 3240 & 3240\nl
WESTPHAL-C11  & C11 & 3240 & \nodata \nl
WESTPHAL-CC13  & CC13 & 2700 & \nodata \nl
WESTPHAL-CC32  & CC32,MMD53,MMD54 & 3240 & \nodata \nl
WESTPHAL-CC79  & CC79,MMD115 & 3000 & \nodata \nl
WESTPHAL-MMD17  & MMD17,MMD20,MM13 & 6420 & 2520\nl
WESTPHAL-CC63  & CC63,MMD91 & 3240  & 3180\nl
WESTPHAL-MMD11  & MMD11 & 1080 & 1620\nl
WESTPHAL-CC1  & CC1 & 3240 & \nodata \nl
WESTPHAL-CC43  & CC43,CC45,DD29 & 3180 & 3240\nl
WESTPHAL-MMD109  & MMD109,MMD112 & 1620 & \nodata \nl
WESTPHAL-MMD113  & MMD113,MMD11 & 1080 & \nodata \nl
WESTPHAL-MMD23  & MMD23,MM11 & 1620 & \nodata \nl
3C324-C1  & C1,C2,MD5 & 4200 & \nodata \nl
3C324-D7  & D7,C12 & 1080 & 1380\nl
3C324-C3  & C3 & 1620 & \nodata \nl
SSA22a-D14  & D14,MD42,MD40,M31& 1860  & \nodata \nl
SSA22a-D3  & D3,MD3 & 7800 & 2700 \nl
SSA22a-MD46  & MD46,MD50 & 3360 & \nodata \nl
SSA22a-aug96D1  & aug96D1,M8  & 3780 & \nodata \nl
SSA22a-C16  & C16,M13,M11 & 3240  & \nodata \nl
SSA22a-C36  & C36& 3240 & 3240\nl
SSA22a-C6  & C6,M4,MD4 & 6960 & 4860\nl
SSA22a-blob1  & C11,C15,M10 & 2160 & \nodata \nl
SSA22a-blob2  & M14,MD14 & 2160 & \nodata \nl
SSA22b-oct96D8  & oct96D8,MD11 & 2880 & \nodata \nl
DSF2237a-C2  & C2 & 1860 & 2700\nl
DSF2237b-M20  & M20,C26,M19,MD56,MD57,M17,M18 & 7380 & 7920\nl
DSF2237b-MD2  & MD2,MD10  & 3240 & \nodata \nl
DSF2237b-D28  & D28,MD81,MD80,C43 & 3240 & 3240\nl
DSF2237b-D3  & D3,D4,MD9,C1 & 6480 & 4860\nl
\enddata
\end{deluxetable}

\newpage
\begin{deluxetable}{lccccccc}
\tablewidth{0pc}
\scriptsize
\tablecaption{Lyman-Break Galaxy Optical/Near-IR Photometry}
\tablehead{
\colhead{Object Name} & \colhead {RA (J2000)} & \colhead{Dec (J2000)} &
\colhead{${\cal R}_{\rm AB}$} & \colhead{$(G-{\cal R})_{
\rm AB}$} &\colhead{$({\cal R}_{\rm AB}-J_{\rm Vega})$} &
\colhead{$({\cal R}_{\rm AB}-K_{s \rm Vega})$} &
 \colhead{$z$}
 }
\startdata
CDFa-C22 & 00 53 09.42 & +12 36 00.5 & 23.97 & 0.73 $\pm$ 0.08 & 0.90 $\pm$ 0.50 & 3.18 $\pm$ 0.23 & 3.046 \nl
CDFa-M19 & 00 53 11.04 & +12 36 12.3 & 25.29 & 0.84 $\pm$ 0.15 & 0.93 $\pm$ 0.75 & $<$ 2.79 & \nodata \nl
CDFa-C19 & 00 53 11.27 & +12 35 39.9 & 24.63 & 0.81 $\pm$ 0.13 & 1.96 $\pm$ 0.30 & 3.79 $\pm$ 0.24 & 2.667 \nl
CDFa-C8 & 00 53 32.85 & +12 32 11.4 & 23.72 & 0.96 $\pm$ 0.08 & \nodata & 2.16 $\pm$ 0.40 & 3.071 \nl
CDFa-MD10 & 00 53 32.22 & +12 31 56.3 & 25.42 & 0.48 $\pm$ 0.16 & \nodata & 2.10 $\pm$ 0.60 & \nodata \nl
CDFa-C1 & 00 53 34.74 & +12 30 30.6 & 23.53 & 0.69 $\pm$ 0.06 & 1.29 $\pm$ 0.18 & 2.69 $\pm$ 0.21 & 3.110 \nl
CDFa-MD2 & 00 53 36.31 & +12 30 31.2 & 23.88 & 0.75 $\pm$ 0.06 & 1.52 $\pm$ 0.23 & 2.96 $\pm$ 0.24 & 2.871 \nl
Q0201-C6 & 02 03 41.81 & +11 34 41.5 & 23.92 & 0.62 $\pm$ 0.07 & 0.40 $\pm$ 0.55 & 2.39 $\pm$ 0.42 & 3.052 \nl
Q0201-MMD21 & 02 03 41.66 & +11 34 44.8 & 24.67 & 0.51 $\pm$ 0.12 & 1.46 $\pm$ 0.47 & 2.72 $\pm$ 0.28 & \nodata \nl
Q0201-B13 & 02 03 49.23 & +11 36 10.8 & 23.34 & 0.03 $\pm$ 0.08 & 1.72 $\pm$ 0.13 & 2.81 $\pm$ 0.15 & 2.167 \nl
Q0256-C15 & 02 59 00.02 & +00 11 38.6 & 24.23 & 0.69 $\pm$ 0.09 & \nodata & 2.46 $\pm$ 0.46 & 3.385 \nl
Q0256-M13 & 02 58 58.83 & +00 11 25.4 & 24.48 & 0.96 $\pm$ 0.12 & \nodata & 2.89 $\pm$ 0.38 & 3.227 \nl
Q0256-M17 & 02 59 19.86 & +00 12 32.4 & 24.04 & 1.10 $\pm$ 0.11 & \nodata & $<$ 1.54 & \nodata \nl
Q0256-MD22 & 02 59 00.79 & +00 11 40.3 & 24.54 & 0.74 $\pm$ 0.11 & \nodata & 2.30 $\pm$ 0.51 & \nodata \nl
Q0256-MD32 & 02 59 18.11 & +00 12 41.5 & 23.61 & 0.89 $\pm$ 0.07 & \nodata & 2.61 $\pm$ 0.25 & \nodata \nl
Q0256-MD34 & 02 59 20.21 & +00 13 02.9 & 24.02 & 0.72 $\pm$ 0.08 & \nodata & 3.88 $\pm$ 0.14 & \nodata \nl
B20902-D11 & 09 05 23.01 & +34 09 40.1 & 22.97 & 0.29 $\pm$ 0.06 & 0.89 $\pm$ 0.27 & 2.97 $\pm$ 0.13 & 2.837 \nl
B20902-D12 & 09 05 23.45 & +34 09 45.0 & 25.46 & -0.12 $\pm$ 0.10 & $<$ -0.23 & $<$ 2.96 & \nodata \nl
B20902-MD32 & 09 05 26.24 & +34 09 28.8 & 24.03 & 0.73 $\pm$ 0.10 & 1.25 $\pm$ 0.37 & 2.52 $\pm$ 0.41 & 2.860 \nl
B20902-C6 & 09 05 20.58 & +34 09 07.7 & 24.13 & 0.45 $\pm$ 0.08 & 0.73 $\pm$ 0.54 & 2.39 $\pm$ 0.45 & 3.098 \nl
B20902-M11 & 09 05 19.58 & +34 09 04.0 & 24.19 & 1.18 $\pm$ 0.16 & 1.36 $\pm$ 0.43 & 2.82 $\pm$ 0.32 & 3.300 \nl
B20902-MD31 & 09 05 20.27 & +34 09 29.6 & 24.74 & 0.55 $\pm$ 0.12 & \nodata & 3.77 $\pm$ 0.24 & \nodata \nl
B20902-MD21 & 09 05 20.08 & +34 07 19.7 & 24.18 & 1.06 $\pm$ 0.16 & 1.03 $\pm$ 0.49 & 2.51 $\pm$ 0.32 & 3.017 \nl
B20902-MD11 & 09 05 35.20 & +34 06 25.2 & 24.27 & 1.04 $\pm$ 0.16 & \nodata & 2.22 $\pm$ 0.48 & 3.392 \nl
B20902-MD14 & 09 05 36.58 & +34 06 37.0 & 24.68 & 0.97 $\pm$ 0.14 & \nodata & 3.44 $\pm$ 0.28 & \nodata \nl
B20902-MD18 & 09 05 35.23 & +34 06 53.2 & 24.51 & 0.41 $\pm$ 0.12 & \nodata & $<$ 2.01 & 2.869 \nl
HDF-CC24 & 12 36 51.81 & +62 15 16.3 & 24.33 & 0.76 $\pm$ 0.12 & \nodata & 2.72 $\pm$ 0.40 & 3.333 \nl
HDF-DD3 & 12 36 47.60 & +62 10 54.1 & 24.25 & 0.58 $\pm$ 0.09 & 1.45 $\pm$ 0.30 & 2.82 $\pm$ 0.32 & 2.942 \nl
HDF-DD15 & 12 36 48.97 & +62 15 43.4 & 23.61 & 0.62 $\pm$ 0.07 & 0.36 $\pm$ 0.49 & 2.39 $\pm$ 0.37 & 3.135 \nl
HDF-MM7 & 12 36 47.75 & +62 10 32.9 & 24.79 & 0.67 $\pm$ 0.17 & 1.02 $\pm$ 0.68 & $<$ 2.29 & 2.985 \nl
HDF-MM9 & 12 36 51.45 & +62 10 42.7 & 24.73 & 0.78 $\pm$ 0.16 & \nodata & 4.04 $\pm$ 0.19 & 2.972 \nl
HDF-MM17\tablenotemark{a} & 12 36 47.67 & +62 12 56.8 & 24.46 & 1.00 $\pm$ 0.15 & 1.52 $\pm$ 0.45 & 3.54 $\pm$ 0.25 & 2.931 \nl
HDF-MM18\tablenotemark{b} & 12 36 44.01 & +62 13 11.9 & 24.10 & 1.00 $\pm$ 0.11 & 1.63 $\pm$ 0.21 & 3.46 $\pm$ 0.16 & 2.929 \nl
HDF-MM23 & 12 37 02.64 & +62 14 27.0 & 24.61 & 1.09 $\pm$ 0.16 & 1.58 $\pm$ 0.47 & 3.65 $\pm$ 0.24 & 3.214 \nl
HDF-MM25 & 12 36 50.72 & +62 14 45.6 & 24.82 & 0.74 $\pm$ 0.17 & \nodata & 2.45 $\pm$ 0.57 & 3.105 \nl
HDF-MM28 & 12 36 46.64 & +62 15 18.1 & 25.04 & 0.70 $\pm$ 0.18 & 1.01 $\pm$ 0.70 & 2.89 $\pm$ 0.53 & 3.371 \nl
HDF-oC37 & 12 37 03.18 & +62 14 52.1 & 25.25 & 0.40 $\pm$ 0.16 & 0.99 $\pm$ 0.75 & 2.14 $\pm$ 0.40 & 2.925 \nl
HDF-oC38 & 12 36 48.81 & +62 15 03.6 & 24.97 & 0.67 $\pm$ 0.18 & -0.81 $\pm$ 0.61 & 3.47 $\pm$ 0.32 & 3.114 \nl
HDF-oMD49 & 12 37 04.25 & +62 14 47.2 & 24.78 & -0.09 $\pm$ 0.11 & 2.20 $\pm$ 0.24 & 4.21 $\pm$ 0.19 & 2.212 \nl
HDF-oM5 & 12 36 50.46 & +62 14 45.5 & 24.98 & 0.93 $\pm$ 0.16 & \nodata & 2.93 $\pm$ 0.53 & 3.097 \nl
WESTPHAL-CC79 & 14 17 13.70 & +52 36 16.5 & 24.28 & 0.87 $\pm$ 0.14 & \nodata & 3.17 $\pm$ 0.40 & 3.061 \nl
WESTPHAL-MMD115 & 14 17 15.48 & +52 36 12.6 & 23.97 & 0.65 $\pm$ 0.11 & \nodata & 1.62 $\pm$ 0.41 & 3.203 \nl
WESTPHAL-MMD109 & 14 17 27.43 & +52 35 49.0 & 23.94 & 0.84 $\pm$ 0.11 & \nodata & 3.05 $\pm$ 0.20 & 2.715 \nl
WESTPHAL-MMD112 & 14 17 26.57 & +52 35 59.2 & 25.34 & 0.63 $\pm$ 0.20 & \nodata & $<$ 2.84 & \nodata \nl
WESTPHAL-MMD113 & 14 18 24.85 & +52 36 10.0 & 23.29 & 0.45 $\pm$ 0.08 & \nodata & 2.71 $\pm$ 0.24 & 2.730 \nl
WESTPHAL-MMD111 & 14 18 23.71 & +52 36 09.2 & 25.03 & 0.08 $\pm$ 0.15 & \nodata & 3.22 $\pm$ 0.49 & \nodata \nl
WESTPHAL-CC70 & 14 17 31.43 & +52 34 25.6 & 23.75 & 0.55 $\pm$ 0.08 & 1.28 $\pm$ 0.24 & 2.41 $\pm$ 0.31 & 2.992 \nl
WESTPHAL-MM69 & 14 17 30.62 & +52 34 17.2 & 25.24 & 0.83 $\pm$ 0.21 & 1.49 $\pm$ 0.57 & 3.24 $\pm$ 0.42 & \nodata \nl
WESTPHAL-DD49 & 14 17 29.23 & +52 34 31.5 & 23.50 & 0.17 $\pm$ 0.06 & 1.33 $\pm$ 0.18 & 3.35 $\pm$ 0.13 & 2.806 \nl
WESTPHAL-CC63 & 14 18 23.10 & +52 32 46.9 & 23.50 & 1.14 $\pm$ 0.12 & 1.83 $\pm$ 0.14 & 2.60 $\pm$ 0.21 & 3.133 \nl
WESTPHAL-MMD91 & 14 18 22.93 & +52 32 35.9 & 23.84 & 0.49 $\pm$ 0.09 & 1.11 $\pm$ 0.35 & 2.73 $\pm$ 0.32 & 2.738 \nl
WESTPHAL-CC43 & 14 17 25.47 & +52 29 37.9 & 23.87 & 1.04 $\pm$ 0.15 & 1.24 $\pm$ 0.33 & 3.30 $\pm$ 0.16 & 3.081 \nl
WESTPHAL-CC45 & 14 17 27.68 & +52 29 50.4 & 24.83 & 0.67 $\pm$ 0.18 & 1.59 $\pm$ 0.53 & 3.03 $\pm$ 0.41 & 2.758 \nl
WESTPHAL-DD29 & 14 17 25.92 & +52 29 32.1 & 24.82 & 0.32 $\pm$ 0.13 & 1.60 $\pm$ 0.42 & 3.86 $\pm$ 0.24 & 3.240 \nl
WESTPHAL-CC32 & 14 18 14.45 & +52 28 04.7 & 24.17 & 0.55 $\pm$ 0.11 & \nodata & 2.40 $\pm$ 0.46 & 3.197 \nl
WESTPHAL-MMD53 & 14 18 12.63 & +52 28 06.6 & 25.22 & 0.36 $\pm$ 0.16 & \nodata & 3.98 $\pm$ 0.25 & \nodata \nl
WESTPHAL-MMD54 & 14 18 14.46 & +52 28 16.0 & 24.64 & 0.45 $\pm$ 0.13 & \nodata & 2.98 $\pm$ 0.45 & 3.017 \nl
WESTPHAL-MMD23 & 14 17 21.95 & +52 23 39.1 & 24.22 & 0.82 $\pm$ 0.14 & \nodata & 2.81 $\pm$ 0.32 & 2.857 \nl
WESTPHAL-MM11 & 14 17 22.02 & +52 23 27.0 & 25.24 & 0.61 $\pm$ 0.20 & \nodata & 3.59 $\pm$ 0.32 & \nodata \nl
WESTPHAL-CC13 & 14 18 02.47 & +52 24 36.5 & 23.64 & 1.06 $\pm$ 0.14 & \nodata & 2.42 $\pm$ 0.37 & 3.396 \nl
WESTPHAL-MMD17 & 14 18 09.56 & +52 23 32.1 & 24.56 & 0.91 $\pm$ 0.17 & 2.04 $\pm$ 0.24 & 2.49 $\pm$ 0.55 & 2.869 \nl
WESTPHAL-MMD20 & 14 18 09.63 & +52 23 37.2 & 24.59 & 0.63 $\pm$ 0.16 & 1.87 $\pm$ 0.30 & 2.15 $\pm$ 0.50 & 2.799 \nl
WESTPHAL-MM13 & 14 18 12.15 & +52 23 30.4 & 25.40 & 0.69 $\pm$ 0.21 & \nodata & 3.96 $\pm$ 0.27 & 2.856 \nl
WESTPHAL-west3-C11 & 14 18 17.63 & +52 23 46.2 & 24.25 & 1.22 $\pm$ 0.16 & \nodata & 2.69 $\pm$ 0.40 & 3.137 \nl
WESTPHAL-DD8 & 14 18 25.44 & +52 23 22.4 & 24.47 & 0.68 $\pm$ 0.14 & 1.09 $\pm$ 0.56 & 2.77 $\pm$ 0.46 & 2.841 \nl
WESTPHAL-MMD14 & 14 18 24.72 & +52 23 19.8 & 24.92 & 0.82 $\pm$ 0.20 & 1.62 $\pm$ 0.42 & 3.08 $\pm$ 0.41 & \nodata \nl
WESTPHAL-MMD16 & 14 18 23.61 & +52 23 28.8 & 24.73 & 0.92 $\pm$ 0.17 & 1.88 $\pm$ 0.30 & 4.06 $\pm$ 0.19 & \nodata \nl
WESTPHAL-MM8 & 14 18 23.92 & +52 23 07.7 & 24.13 & 1.04 $\pm$ 0.16 & 1.74 $\pm$ 0.21 & 3.25 $\pm$ 0.19 & 2.829 \nl
WESTPHAL-MMD11 & 14 18 09.73 & +52 22 01.3 & 24.05 & 1.04 $\pm$ 0.16 & 2.67 $\pm$ 0.10 & 4.54 $\pm$ 0.11 & 2.979 \nl
WESTPHAL-CC1 & 14 18 21.98 & +52 21 22.0 & 23.83 & 1.02 $\pm$ 0.15 & \nodata & 2.78 $\pm$ 0.32 & 2.984 \nl
3C324-D7 & 15 49 52.96 & +21 30 59.1 & 23.88 & 0.52 $\pm$ 0.08 & 0.78 $\pm$ 0.49 & 3.01 $\pm$ 0.20 & \nodata \nl
3C324-C12 & 15 49 53.95 & +21 31 06.7 & 25.04 & 0.54 $\pm$ 0.15 & 0.78 $\pm$ 0.71 & 3.34 $\pm$ 0.45 & \nodata \nl
3C324-C3 & 15 49 47.10 & +21 27 05.5 & 24.14 & 0.85 $\pm$ 0.13 & \nodata & 2.57 $\pm$ 0.41 & 3.283 \nl
3C324-C1 & 15 49 54.27 & +21 26 33.2 & 24.33 & 0.62 $\pm$ 0.14 & \nodata & 3.12 $\pm$ 0.26 & 2.873 \nl
3C324-C2 & 15 49 53.98 & +21 26 35.6 & 24.26 & 0.54 $\pm$ 0.10 & \nodata & 3.19 $\pm$ 0.26 & 2.880 \nl
3C324-MD5 & 15 49 54.46 & +21 26 34.6 & 25.48 & -0.08 $\pm$ 0.24 & \nodata & 3.93 $\pm$ 0.27 & \nodata \nl
SSA22a-MD46 & 22 17 27.28 & +00 18 09.9 & 23.30 & 0.42 $\pm$ 0.09 & \nodata & 2.34 $\pm$ 0.29 & 3.090 \nl
SSA22a-MD50 & 22 17 26.83 & +00 18 30.2 & 25.18 & 0.20 $\pm$ 0.17 & \nodata & 3.18 $\pm$ 0.47 & \nodata \nl
SSA22a-D14 & 22 17 35.29 & +00 17 24.1 & 24.32 & 0.19 $\pm$ 0.14 & \nodata & 2.07 $\pm$ 0.46 & 3.018 \nl
SSA22a-MD42 & 22 17 35.83 & +00 17 19.8 & 25.33 & 0.06 $\pm$ 0.17 & \nodata & 4.04 $\pm$ 0.25 & \nodata \nl
SSA22a-MD40 & 22 17 35.96 & +00 17 08.3 & 24.89 & 0.70 $\pm$ 0.19 & \nodata & 2.21 $\pm$ 0.57 & 3.015 \nl
SSA22a-M31 & 22 17 36.87 & +00 17 12.4 & 25.41 & 0.42 $\pm$ 0.18 & \nodata & 3.22 $\pm$ 0.44 & \nodata \nl
SSA22a-C36 & 22 17 46.10 & +00 16 43.1 & 24.06 & 0.78 $\pm$ 0.13 & 1.23 $\pm$ 0.37 & 3.07 $\pm$ 0.23 & 3.065 \nl
SSA22a-C16 & 22 17 31.96 & +00 13 16.1 & 23.64 & 0.98 $\pm$ 0.10 & \nodata & 2.86 $\pm$ 0.20 & 3.061 \nl
SSA22a-M13 & 22 17 31.46 & +00 12 55.2 & 25.46 & 0.74 $\pm$ 0.24 & \nodata & $<$ 2.96 & \nodata \nl
SSA22a-M11 & 22 17 31.77 & +00 12 51.3 & 25.35 & 0.50 $\pm$ 0.18 & \nodata & $<$ 2.85 & \nodata \nl
SSA22a-M14 & 22 17 39.05 & +00 13 30.1 & 25.47 & 0.75 $\pm$ 0.24 & \nodata & 2.99 $\pm$ 0.62 & 3.091 \nl
SSA22a-MD14 & 22 17 37.93 & +00 13 44.2 & 24.14 & 0.86 $\pm$ 0.14 & \nodata & 1.96 $\pm$ 0.43 & \nodata \nl
SSA22a-C11 & 22 17 25.68 & +00 12 35.3 & 24.20 & 0.47 $\pm$ 0.12 & \nodata & 2.82 $\pm$ 0.32 & 3.108 \nl
SSA22a-C15 & 22 17 26.12 & +00 12 55.3 & 25.19 & 0.55 $\pm$ 0.19 & \nodata & 3.50 $\pm$ 0.32 & 3.092 \nl
SSA22a-M10 & 22 17 26.79 & +00 12 21.1 & 24.45 & 1.03 $\pm$ 0.18 & \nodata & 2.32 $\pm$ 0.51 & 3.098 \nl
SSA22a-aug96D1 & 22 17 24.00 & +00 12 02.8 & 23.60 & 0.32 $\pm$ 0.09 & \nodata & 2.79 $\pm$ 0.12 & 2.202 \nl
SSA22a-M8 & 22 17 25.11 & +00 11 56.8 & 24.72 & 0.89 $\pm$ 0.19 & \nodata & 3.64 $\pm$ 0.24 & \nodata \nl
SSA22a-D3 & 22 17 32.42 & +00 11 33.0 & 23.37 & 0.97 $\pm$ 0.10 & -0.60 $\pm$ 0.40 & 2.24 $\pm$ 0.32 & 3.086 \nl
SSA22a-MD3 & 22 17 31.89 & +00 11 38.3 & 24.60 & 0.15 $\pm$ 0.13 & 1.40 $\pm$ 0.47 & 2.17 $\pm$ 0.50 & 2.483 \nl
SSA22a-C6 & 22 17 40.93 & +00 11 26.0 & 23.44 & 0.79 $\pm$ 0.09 & 0.36 $\pm$ 0.44 & 2.01 $\pm$ 0.37 & 3.099 \nl
SSA22a-M4 & 22 17 40.92 & +00 11 27.9 & 24.83 & 0.76 $\pm$ 0.19 & 0.93 $\pm$ 0.68 & 2.68 $\pm$ 0.56 & 3.091 \nl
SSA22a-MD4 & 22 17 39.95 & +00 11 39.6 & 24.25 & 0.24 $\pm$ 0.12 & 1.03 $\pm$ 0.49 & 2.92 $\pm$ 0.32 & 2.611 \nl
SSA22b-MD11 & 22 17 23.07 & +00 03 42.2 & 25.23 & 0.37 $\pm$ 0.15 & \nodata & 3.14 $\pm$ 0.47 & \nodata \nl
SSA22b-oct96D8 & 22 17 23.52 & +00 03 57.3 & 23.53 & 0.77 $\pm$ 0.08 & \nodata & 2.87 $\pm$ 0.18 & 3.323 \nl
DSF2237a-C2 & 22 40 08.30 & +11 49 04.9 & 23.55 & 1.13 $\pm$ 0.10 & 1.47 $\pm$ 0.20 & 3.02 $\pm$ 0.17 & 3.318 \nl
DSF2237b-D28 & 22 39 20.25 & +11 55 11.4 & 24.46 & 0.32 $\pm$ 0.08 & 1.34 $\pm$ 0.49 & 2.42 $\pm$ 0.52 & 2.934 \nl
DSF2237b-MD81 & 22 39 21.72 & +11 55 10.5 & 24.16 & 0.31 $\pm$ 0.07 & 2.23 $\pm$ 0.17 & 4.27 $\pm$ 0.14 & 2.819 \nl
DSF2237b-MD80 & 22 39 22.23 & +11 55 08.3 & 25.37 & 0.94 $\pm$ 0.20 & 1.16 $\pm$ 0.77 & 3.08 $\pm$ 0.37 & \nodata \nl
DSF2237b-C43 & 22 39 21.57 & +11 54 44.9 & 24.35 & 0.70 $\pm$ 0.11 & 1.73 $\pm$ 0.34 & $<$ 1.85 & 2.885 \nl
DSF2237b-M20 & 22 39 38.88 & +11 52 22.1 & 24.20 & 1.12 $\pm$ 0.13 & 1.33 $\pm$ 0.43 & 2.96 $\pm$ 0.20 & 3.156 \nl
DSF2237b-C26 & 22 39 38.02 & +11 52 11.4 & 24.73 & 0.57 $\pm$ 0.12 & 1.03 $\pm$ 0.60 & 2.64 $\pm$ 0.50 & 3.251 \nl
DSF2237b-M19 & 22 39 38.04 & +11 52 20.6 & 24.80 & 0.96 $\pm$ 0.18 & 1.47 $\pm$ 0.53 & 2.79 $\pm$ 0.56 & 3.259 \nl
DSF2237b-MD56 & 22 39 39.00 & +11 52 19.2 & 25.43 & 0.65 $\pm$ 0.17 & 1.81 $\pm$ 0.41 & 3.10 $\pm$ 0.39 & \nodata \nl
DSF2237b-MD57 & 22 39 39.92 & +11 52 27.2 & 24.56 & 0.73 $\pm$ 0.13 & 0.95 $\pm$ 0.66 & 2.51 $\pm$ 0.55 & 3.331 \nl
DSF2237b-M17 & 22 39 39.87 & +11 52 06.5 & 24.67 & 1.03 $\pm$ 0.18 & 1.06 $\pm$ 0.60 & 2.53 $\pm$ 0.55 & 3.020 \nl
DSF2237b-M18 & 22 39 40.16 & +11 52 16.0 & 25.33 & 0.76 $\pm$ 0.16 & 0.78 $\pm$ 0.72 & $<$ 2.83 & \nodata \nl
DSF2237b-MD2 & 22 39 29.89 & +11 47 12.5 & 24.17 & 0.90 $\pm$ 0.12 & \nodata & 2.28 $\pm$ 0.36 & 2.505 \nl
DSF2237b-MD10 & 22 39 30.58 & +11 47 34.3 & 25.04 & 1.01 $\pm$ 0.19 & \nodata & 2.87 $\pm$ 0.53 & 3.115 \nl
DSF2237b-D3 & 22 39 18.91 & +11 47 40.4 & 24.19 & 0.51 $\pm$ 0.09 & 1.07 $\pm$ 0.49 & 2.02 $\pm$ 0.37 & 2.931 \nl
DSF2237b-D4 & 22 39 18.75 & +11 47 40.5 & 24.35 & 0.17 $\pm$ 0.10 & 1.13 $\pm$ 0.56 & 2.20 $\pm$ 0.29 & 2.935 \nl
DSF2237b-MD9 & 22 39 17.68 & +11 47 39.2 & 24.97 & 0.45 $\pm$ 0.07 & 0.31 $\pm$ 0.56 & $<$ 2.47 & \nodata \nl
DSF2237b-C1 & 22 39 16.90 & +11 47 47.5 & 24.99 & 0.37 $\pm$ 0.12 & 1.60 $\pm$ 0.51 & 2.95 $\pm$ 0.32 & 3.064 \nl
\tablenotetext{a}{This galaxy is in the central Hubble Deep Field (HDF), and
is referred to as ``4-52.0'' in Williams \et (1996) and ``HDF/NIC 813/814''
in Papovich \et (2001).}
\tablenotetext{b}{ This galaxy is in the central HDF, and is referred to
as ``1-54.0'' in Williams \et (1996) and ``HDF/NIC 522'' in
Papovich \et (2001).}

\enddata
\end{deluxetable}

\newpage
\begin{deluxetable}{lcccc}
\tablewidth{0pc}
\scriptsize
\tablecaption{Lyman-Break Galaxy Constant SFR Best-Fit Parameters}
\tablehead{
\colhead{Object Name} & \colhead{$E(B-V)$}& \colhead{$t_{sf}$ (Myr)} &
\colhead{SFR ($h^{-2}M_{\odot}{\rm yr}^{-1}$)} &
\colhead{$ \log(m_{star} (h^{-2}M_{\odot}))$   }    }
\startdata
CDFa-C1 & 0.08 & ~~905 & ~~~35 & 10.5 \nl
CDFa-C19 & 0.38 & ~~203 & ~~156 & 10.5 \nl
CDFa-C22 & 0.13 & ~1278 & ~~~36 & 10.7 \nl
CDFa-C8\tablenotemark{b} & 0.26 & ~~~10 & ~~275 & ~9.4 \nl
CDFa-MD2 & 0.20 & ~~321 & ~~~65 & 10.3 \nl
Q0201-C6 & 0.08 & ~~321 & ~~~23 & ~9.9 \nl
Q0201-B13 & 0.04 & ~2500 & ~~~14 & 10.5 \nl
Q0256-C15 & 0.02 & ~~905 & ~~~12 & 10.0 \nl
Q0256-M13 & 0.15 & ~~404 & ~~~32 & 10.1 \nl
B20902-C6 & 0.05 & ~~509 & ~~~15 & ~9.9 \nl
B20902-D11 & 0.09 & ~1278 & ~~~53 & 10.8 \nl
B20902-M11 & 0.21 & ~~143 & ~~~80 & 10.1 \nl
B20902-MD11 & 0.17 & ~~~30 & ~~~69 & ~9.3 \nl
B20902-MD21 & 0.23 & ~~~35 & ~~~99 & ~9.5 \nl
B20902-MD32 & 0.24 & ~~~35 & ~~110 & ~9.6 \nl
HDF-CC24 & 0.08 & ~~806 & ~~~19 & 10.2 \nl
HDF-DD15 & 0.10 & ~~227 & ~~~42 & 10.0 \nl
HDF-DD3 & 0.12 & ~~719 & ~~~23 & 10.2 \nl
HDF-MM17 & 0.30 & ~~321 & ~~104 & 10.5 \nl
HDF-MM18 & 0.32 & ~~203 & ~~179 & 10.5 \nl
HDF-MM23\tablenotemark{a} & 0.18 & ~2000 & ~~~36 & 10.8 \nl
HDF-MM25 & 0.09 & ~~360 & ~~~11 & ~9.6 \nl
HDF-MM28\tablenotemark{a} & 0.05 & ~1800 & ~~~~8 & 10.1 \nl
HDF-MM9\tablenotemark{a} & 0.24 & ~2100 & ~~~49 & 11.0 \nl
HDF-oC37 & 0.11 & ~~114 & ~~~~9 & ~9.0 \nl
HDF-oC38\tablenotemark{a} & 0.15 & ~2000 & ~~~18 & 10.5 \nl
HDF-oM5 & 0.29 & ~~~55 & ~~~81 & ~9.6 \nl
WESTPHAL-CC1\tablenotemark{b} & 0.34 & ~~~10 & ~~516 & ~9.7 \nl
WESTPHAL-CC13 & 0.11 & ~~227 & ~~~51 & 10.1 \nl
WESTPHAL-CC32 & 0.06 & ~~454 & ~~~16 & ~9.9 \nl
WESTPHAL-CC43 & 0.23 & ~~454 & ~~104 & 10.6 \nl
WESTPHAL-CC45 & 0.23 & ~~255 & ~~~33 & ~9.9 \nl
WESTPHAL-CC70 & 0.10 & ~~321 & ~~~32 & 10.0 \nl
WESTPHAL-CC79 & 0.18 & ~~641 & ~~~45 & 10.4 \nl
WESTPHAL-DD49 & 0.17 & ~1139 & ~~~68 & 10.9 \nl
WESTPHAL-CC63\tablenotemark{b} & 0.36 & ~~~10 & ~~942 & 10.0 \nl
WESTPHAL-DD8 & 0.32 & ~~~15 & ~~175 & ~9.4 \nl
WESTPHAL-MM8 & 0.37 & ~~~40 & ~~324 & 10.1 \nl
WESTPHAL-MMD109\tablenotemark{b} & 0.41 & ~~~10 & ~~662 & ~9.8 \nl
WESTPHAL-MMD113 & 0.23 & ~~~81 & ~~156 & 10.1 \nl
WESTPHAL-MMD115\tablenotemark{b} & 0.13 & ~~~10 & ~~~71 & ~8.9 \nl
WESTPHAL-MMD23 & 0.32 & ~~~20 & ~~219 & ~9.6 \nl
WESTPHAL-MMD54 & 0.10 & ~1278 & ~~~14 & 10.3 \nl
WESTPHAL-MMD91 & 0.16 & ~~286 & ~~~42 & 10.1 \nl
WESTPHAL-west3-C11 & 0.29 & ~~~15 & ~~223 & ~9.5 \nl
3C324-C1 & 0.12 & ~1434 & ~~~20 & 10.4 \nl
3C324-C2 & 0.12 & ~1700 & ~~~22 & 10.6 \nl
3C324-C3 & 0.07 & ~~641 & ~~~19 & 10.1 \nl
SSA22a-aug96D1 & 0.17 & ~~321 & ~~~37 & 10.1 \nl
SSA22a-C11 & 0.08 & ~1139 & ~~~18 & 10.3 \nl
SSA22a-C15\tablenotemark{a} & 0.12 & ~2000 & ~~~11 & 10.3 \nl
SSA22a-C16 & 0.21 & ~~181 & ~~109 & 10.3 \nl
SSA22a-C36 & 0.17 & ~~571 & ~~~48 & 10.4 \nl
SSA22a-C6\tablenotemark{b} & 0.20 & ~~~10 & ~~204 & ~9.3 \nl
SSA22a-D14 & -0.01 & ~~571 & ~~~~6 & ~9.6 \nl
SSA22a-M10\tablenotemark{b} & 0.30 & ~~~10 & ~~212 & ~9.3 \nl
SSA22a-M14 & 0.15 & ~~571 & ~~~12 & ~9.8 \nl
SSA22a-M4 & 0.13 & ~~321 & ~~~17 & ~9.7 \nl
SSA22a-D3\tablenotemark{b} & 0.29 & ~~~10 & ~~540 & ~9.7 \nl
SSA22a-MD3 & 0.24 & ~~~15 & ~~~54 & ~8.9 \nl
SSA22a-MD4 & 0.14 & ~~641 & ~~~21 & 10.1 \nl
SSA22a-MD40 & 0.21 & ~~~20 & ~~~45 & ~9.0 \nl
SSA22a-MD46 & 0.01 & ~~806 & ~~~22 & 10.2 \nl
SSA22b-oct96D8\tablenotemark{a} & 0.06 & ~1900 & ~~~32 & 10.8 \nl
DSF2237a-C2 & 0.15 & ~~719 & ~~~76 & 10.7 \nl
DSF2237b-C1\tablenotemark{a} & 0.05 & ~2100 & ~~~~6 & 10.1 \nl
DSF2237b-C26 & 0.05 & ~1139 & ~~~~9 & 10.0 \nl
DSF2237b-D28 & 0.06 & ~~719 & ~~~10 & ~9.9 \nl
DSF2237b-D3 & 0.16 & ~~~30 & ~~~46 & ~9.1 \nl
DSF2237b-D4 & -0.04 & ~1278 & ~~~~5 & ~9.8 \nl
DSF2237b-M19 & 0.20 & ~~161 & ~~~40 & ~9.8 \nl
DSF2237b-M20 & 0.21 & ~~227 & ~~~69 & 10.2 \nl
DSF2237b-MD10 & 0.29 & ~~~35 & ~~~88 & ~9.4 \nl
DSF2237b-MD2\tablenotemark{b} & 0.39 & ~~~10 & ~~377 & ~9.6 \nl
\tablenotetext{a}{The best-fit ages for marked galaxies, when unconstrained,
were older than the age of the universe at the galaxy's redshift. The
best-fit quantities listed for these galaxies reflect the constraint 
that the age of a galaxy should not be older than the age of the universe.
Such a constraint still provides statistically acceptable fits for marked
galaxies.}
\tablenotetext{b}{The best-fit ages for the marked galaxies, when
unconstrained, were younger than $10 \; {\rm Myr}$. Such young ages
are physically implausible, and therefore we restricted the best-fit
ages to be at least $10  \; {\rm Myr}$. The best-fit quantities
listed for marked galaxies reflect this constraint, and provide
statistically acceptable fits for marked galaxies.}
\enddata
\end{deluxetable}
\newpage
\clearpage


\begin{figure}
\figurenum{1a}
\plotone{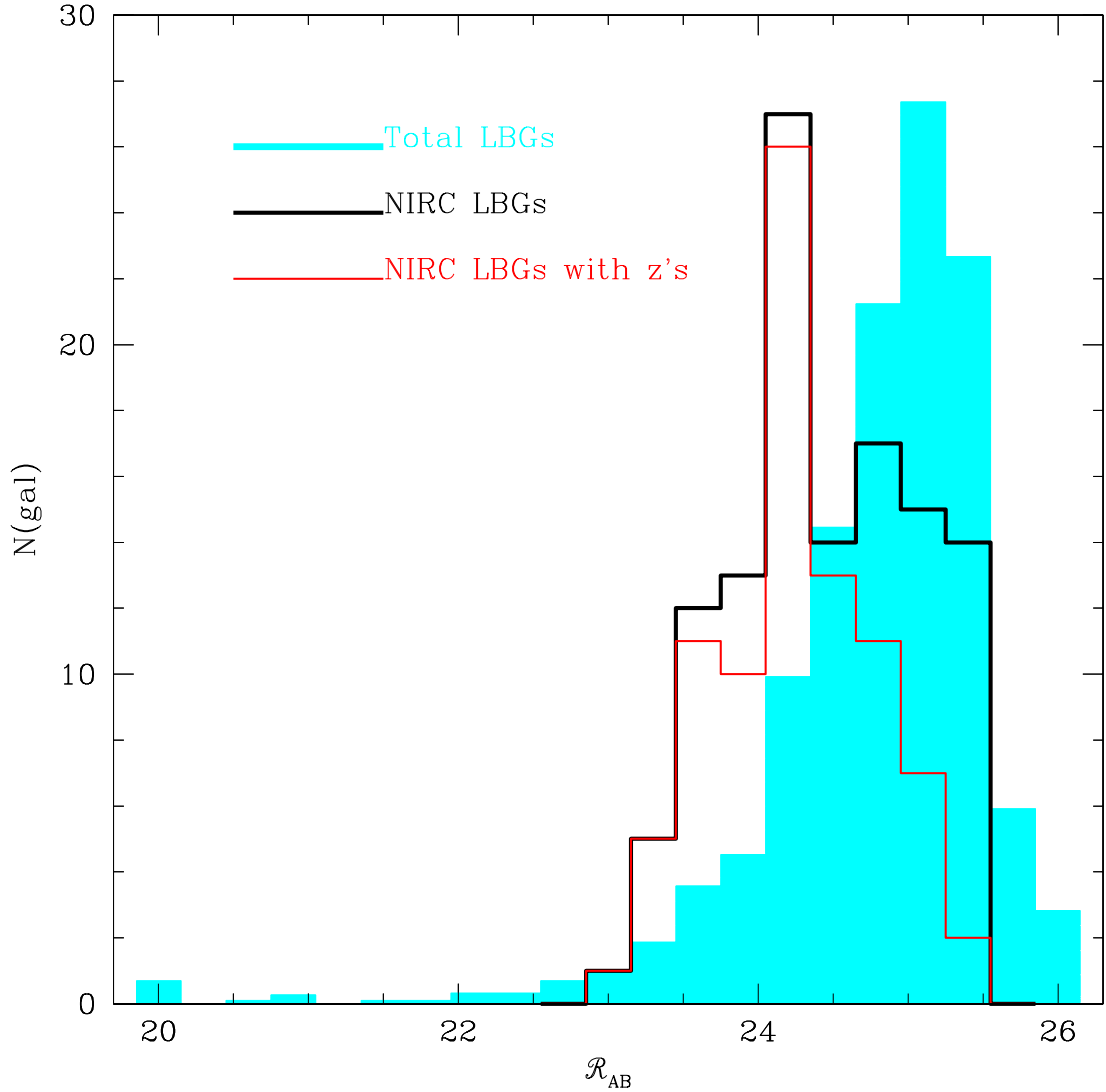}
\caption{
The distribution of ${\cal R}$ magnitudes for the NIRC LBG sample,
relative to that of the LBG survey as a whole. 
The LBG total sample histogram has been normalized to the number of galaxies
contained in the NIRC LBG subsample, for the purpose of comparison.
Also indicated is the ${\cal R}$ distribution for the NIRC LBGs
with measured redshifts, which comprise a slightly brighter
sample than the NIRC LBG sample as a whole.
 }
\end{figure}
\newpage
\begin{figure}
\figurenum{1b}
\plotone{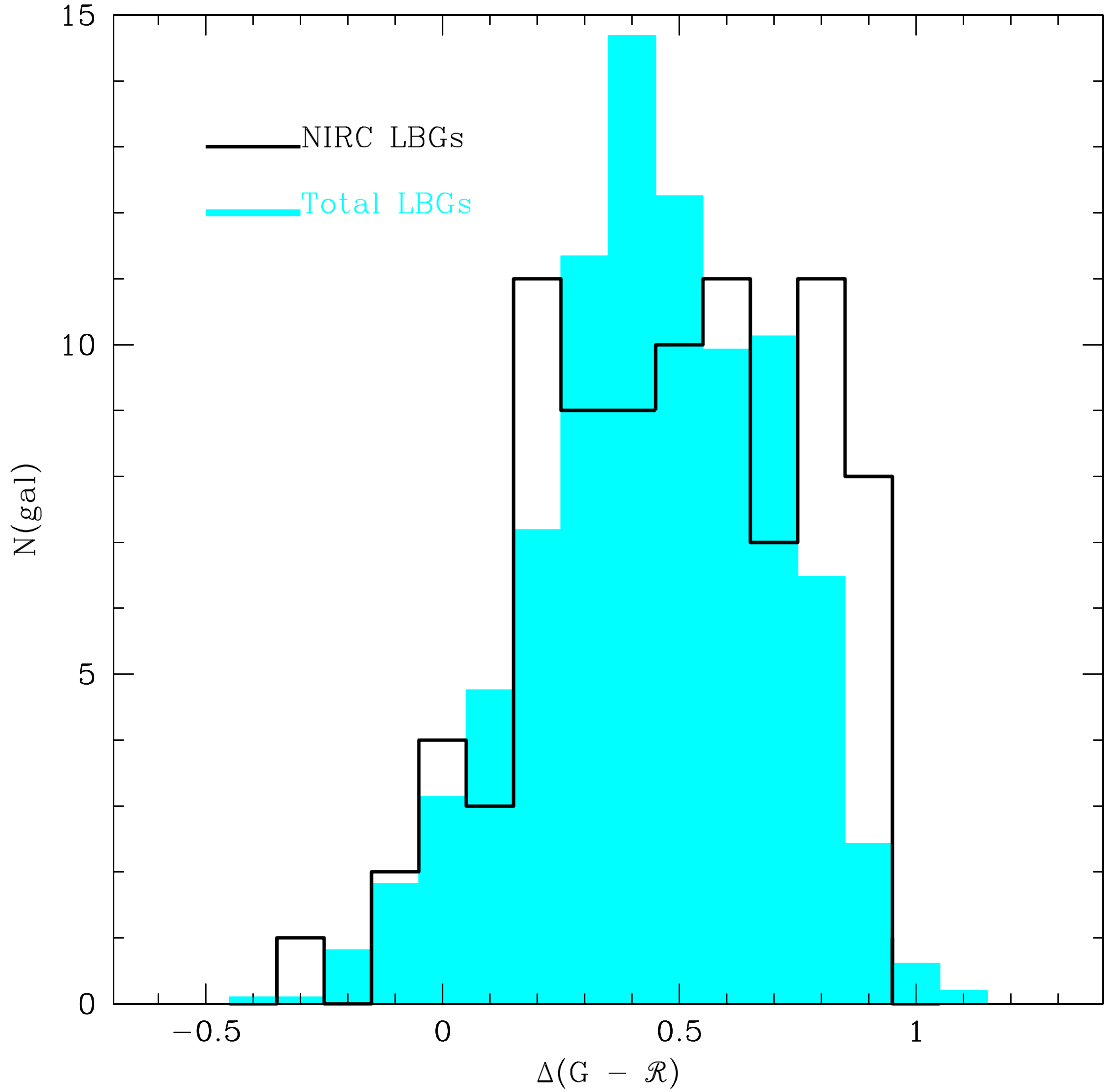}
\caption{
The distribution of $(G-{\cal R})_0$, the 
IGM-absorption-corrected $G-{\cal R}$ color, for  
galaxies in the NIRC LBG sample with redshifts, relative to that of
the LBG spectroscopic sample as a whole. 
The total LBG histogram has been normalized to the 
number of objects in the NIRC LBG histogram. 
This figure shows that the NIRC LBG spectroscopic sample contains an 
excess of highly reddened galaxies, 
relative to the total LBG spectroscopic sample.
 }
\end{figure}
\newpage
\begin{figure}
\figurenum{2}
\plotone{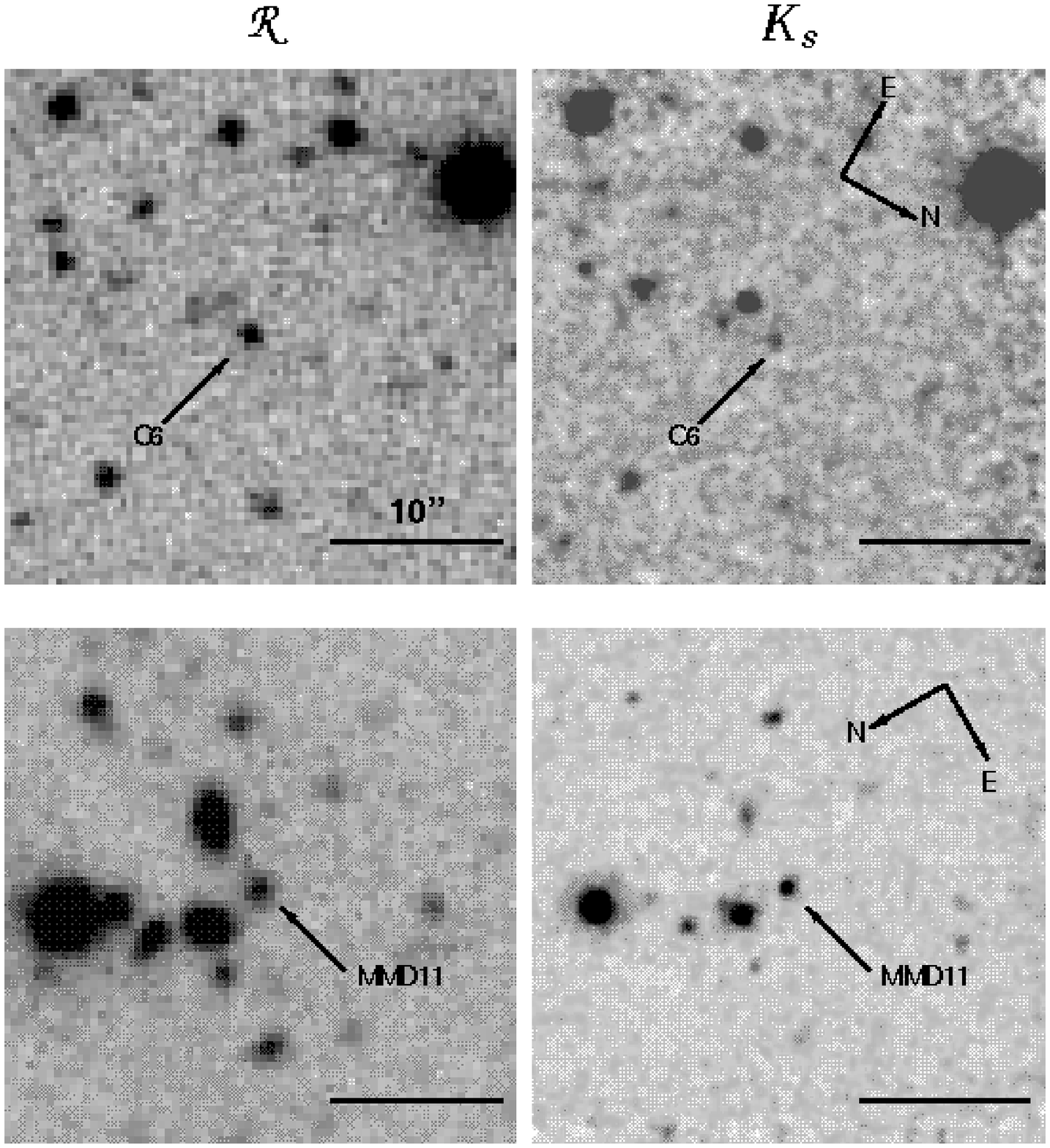}
\caption{
${\cal R}$ and $K_s$ images for two objects in the NIRC LBG
sample.  A bar indicating 10'' is in the lower right-hand corner
of each image. The N-E orientation of the ${\cal R}$ and $K_s$
images for each object is indicated in the $K_s$ images. NIRC
orientations were chosen to locate a guide star on the 
guide camera, as well as to maximize the number of galaxies per
pointing. 
Top: B20902-C6 has ${\cal R}=24.13$ and  ${\cal R}-K_s=2.39$
which puts it in the bluer half of the sample of 
${\cal R}-K_s$ measurements, with a fairly typical signal-to-noise
ratio. Bottom: Westphal-MMD11 has ${\cal R}=24.05$ 
and  ${\cal R}-K_s=4.54$, which gives it the reddest and highest
signal-to-noise ${\cal R}-K_s$ measurement in the
NIRC LBG sample.
 }

\end{figure}
\newpage
\begin{figure}
\figurenum{3}
\plotone{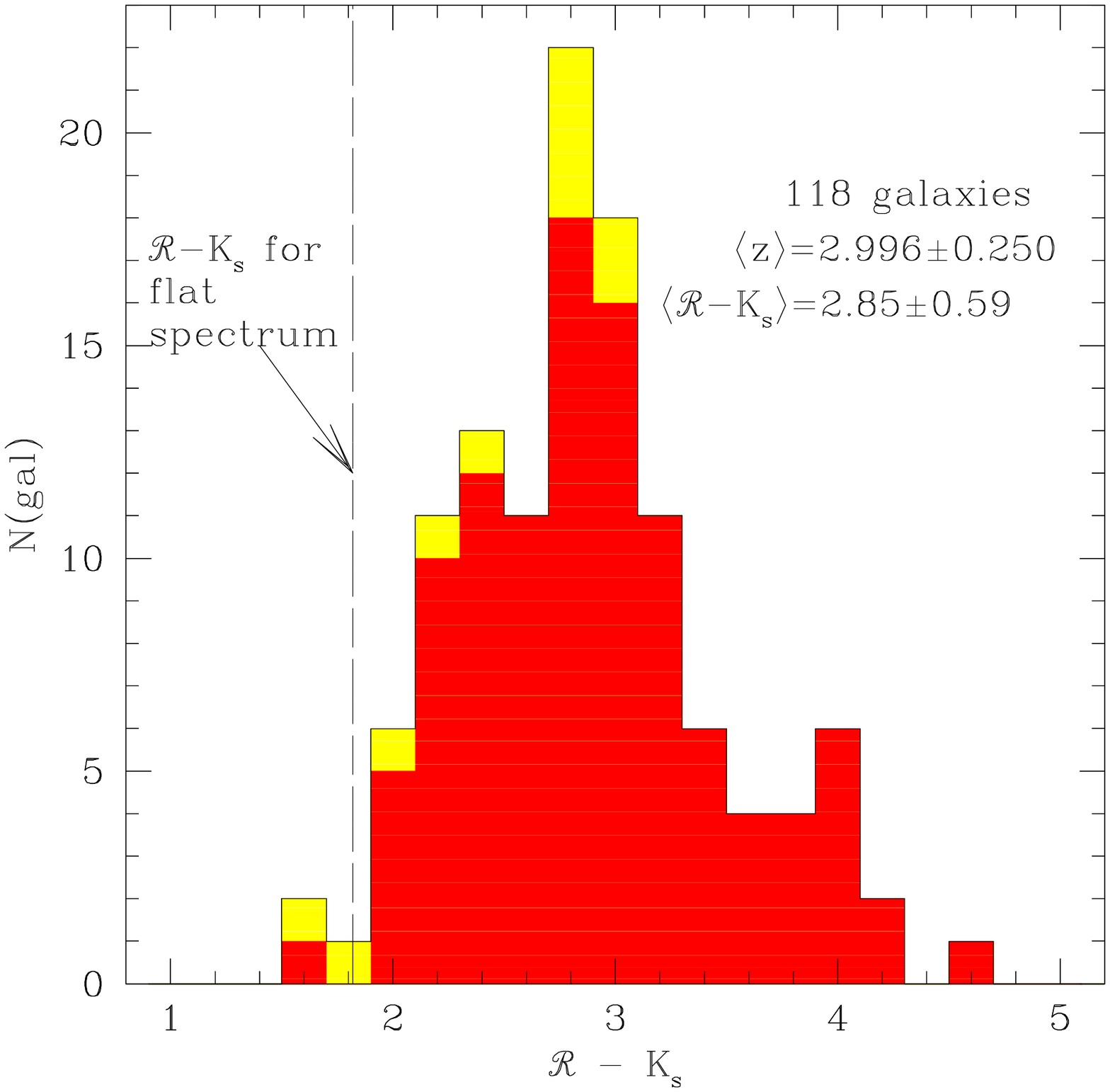}
\caption{
The distribution of observed ${\cal R}-K_s$ colors for the NIRC LBG sample.
Of the 118 galaxies, 107 have ${\cal R}-K_s$ detections, and 11 have 
upper limits, corresponding to the typical $K_s$ detection limit of
$K_s=22.5$. Detections are indicated in the red (dark) shaded histogram, while
upper limits are in the yellow (light) shaded bins. A dashed line at 
${\cal R}-K_s=1.82$ marks the color for a flat spectrum in $F_{\nu}$.
The average color in the sample is $<{\cal R}-K_s> = 2.85$. 81 of 118 galaxies
in the NIRC LBG sample have measured redshifts, with $<z>=2.996$.
 }
\end{figure}
\newpage
\begin{figure}
\figurenum{4}
\plotone{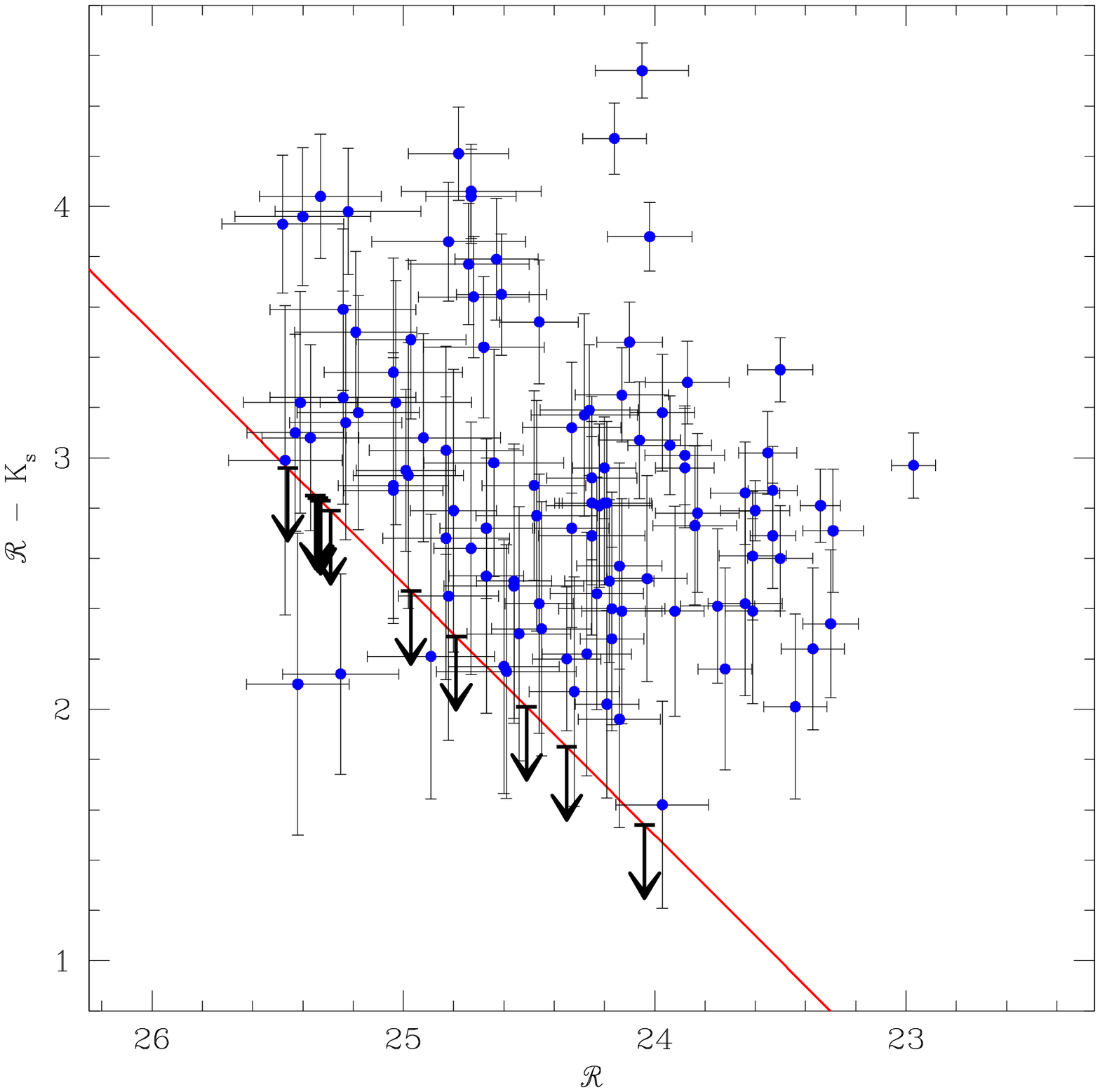}
\caption{
${\cal R} - K_s$ vs. ${\cal R}$. Dots indicate ${\cal R} - K_s$
detections, while down-arrows indicate upper limits. The diagonal line
traces the typical sample detection limit of $K_s=22.5$. There are
three points below this detection line, representing galaxies
in $K_s$ images with more sensitive detection limits than the typical one.
 }
\end{figure}

\newpage
\begin{figure}
\figurenum{5}
\plotone{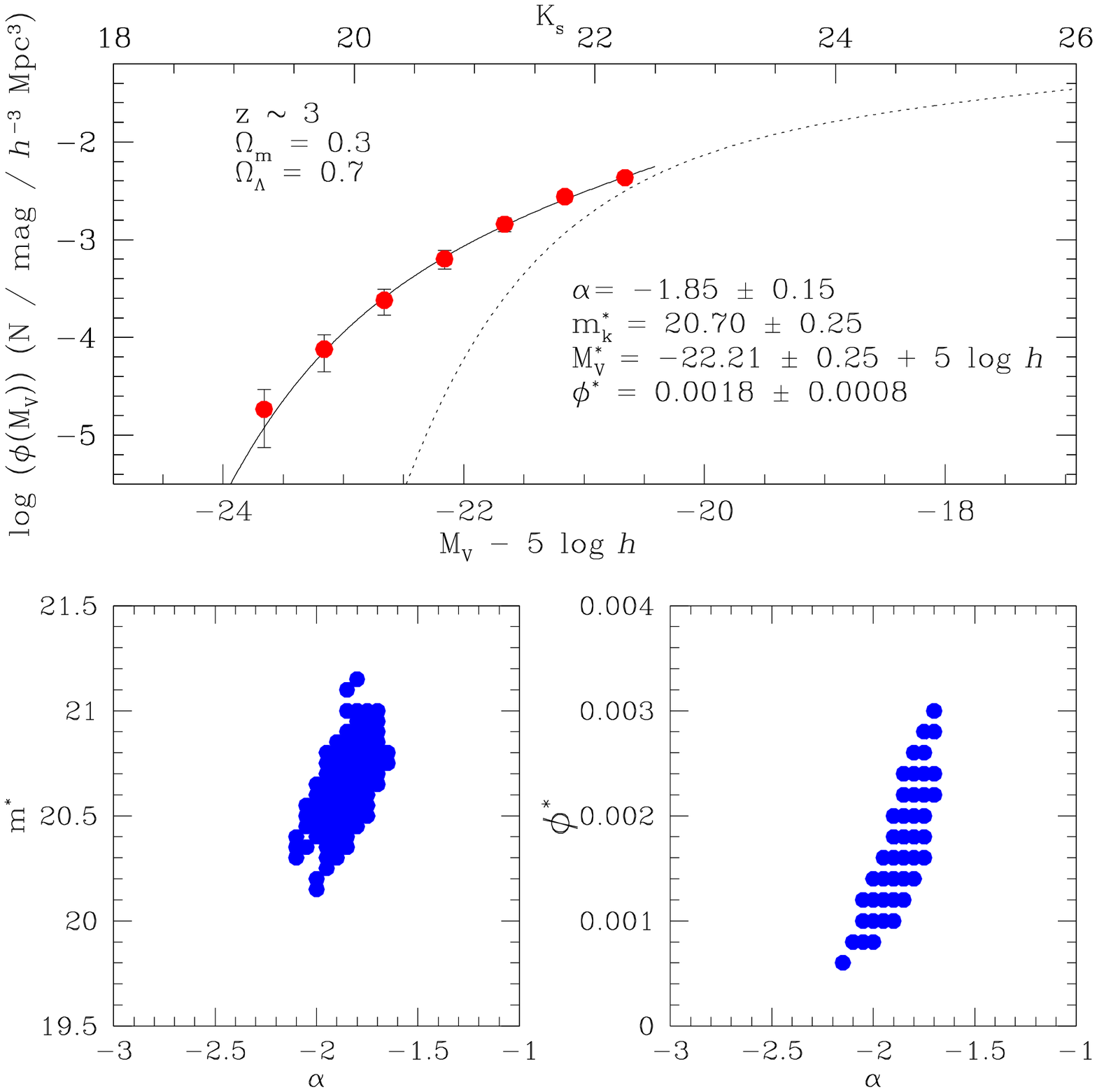}
\caption{
Top: 
The rest-frame optical luminosity function of LBGs. 
The points and solid line indicate the
magnitude range used for the Schechter function fit.
The error bars on the points include the photometric uncertainties on the
${\cal R}$ and ${\cal R}-K_s$ measurements. 
The lower x-axis indicates
the range in optical absolute magnitude, $M_V$, spanned by the
$K_s$ luminosity function (assuming $h=1$). The dotted line indicates
the locally determined 2dFGRS $b_j$ luminosity function, offset
by a color of $b_j-V=0.5$ magnitudes (for comparison at the same rest wavelength
as the LBG optical luminosity function). 
Bottom: The $68.3 \%$ confidence intervals for the best-fit Schechter 
function parameters. The covariance
between the best-fit parameters is indicated by the confidence 
regions. As the faint-end slope ($\alpha$) becomes steeper, 
the characteristic luminosity ($m^*$) increases, and the 
overall normalization ($\phi^*$) decreases.
 }
\end{figure}

\newpage
\begin{figure}
\figurenum{6}
\plotone{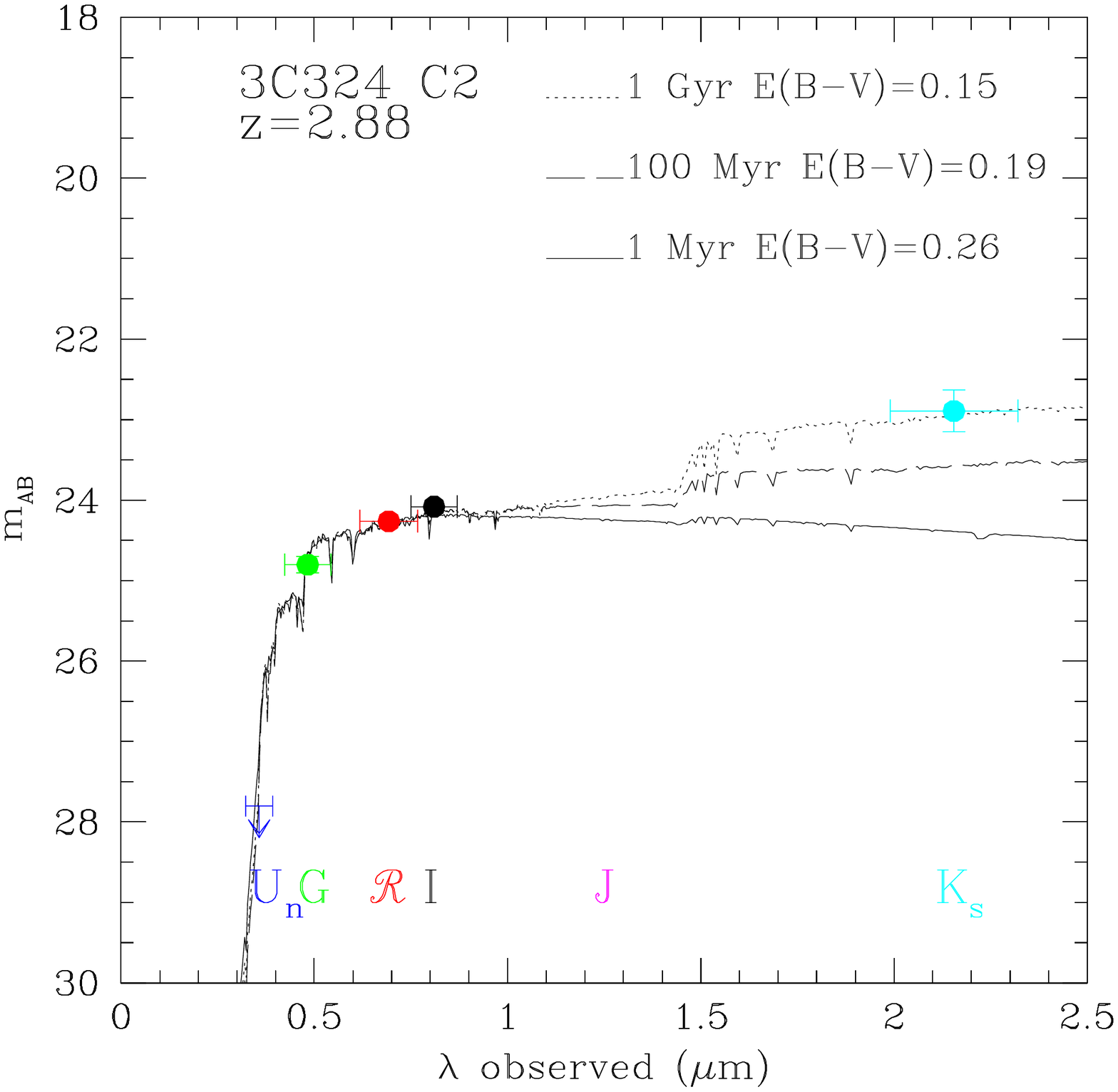}
\caption{
The Age-Dust Degeneracy. The points indicate the observed spectral
energy distribution of 3C324-C2, a LBG at $z=2.880$. Shown with the
points are BC96 constant star-formation models of different ages,
modified by the amount of dust extinction required to reproduce the
observed $G-{\cal R}$ color. The dotted line is a 1 Gyr model 
with $E(B-V) = 0.149$; the dashed line is a 100 Myr model with 
$E(B-V)=0.186$; and the solid line is a 1 Myr model with $E(B-V)=0.263$. 
All of these models describe the observed optical photometry equally
well. However, only the 1 Gyr model successfully describes the
the observed ${\cal R} - K_s$ color.
 }
\end{figure}

\newpage
\begin{figure}
\figurenum{7}
\plotone{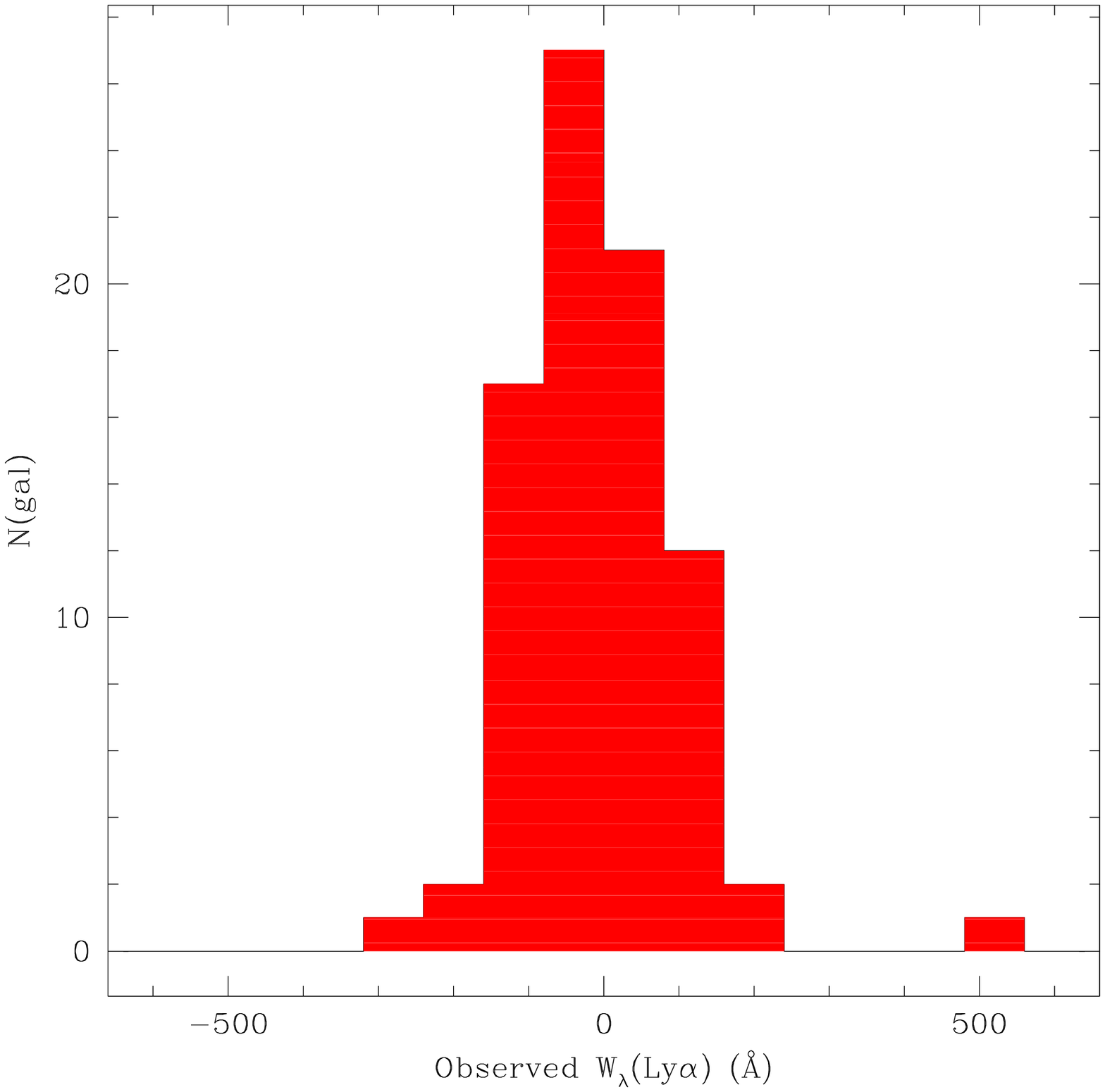}
\caption{
The distribution of observed Lyman $\alpha$ equivalent widths. 
The median equivalent width is 0. However, there are
extreme cases of observed equivalent widths greater than $300 \: {\rm \AA}$
in either emission or absorption. If unaccounted for,
such large equivalent widths would bias the modeling of 
the stellar population--- especially the estimate
of the dust extinction, but, to a certain extent, the age as well. 
 }
\end{figure}

\newpage
\begin{figure}
\figurenum{8a}
\plotone{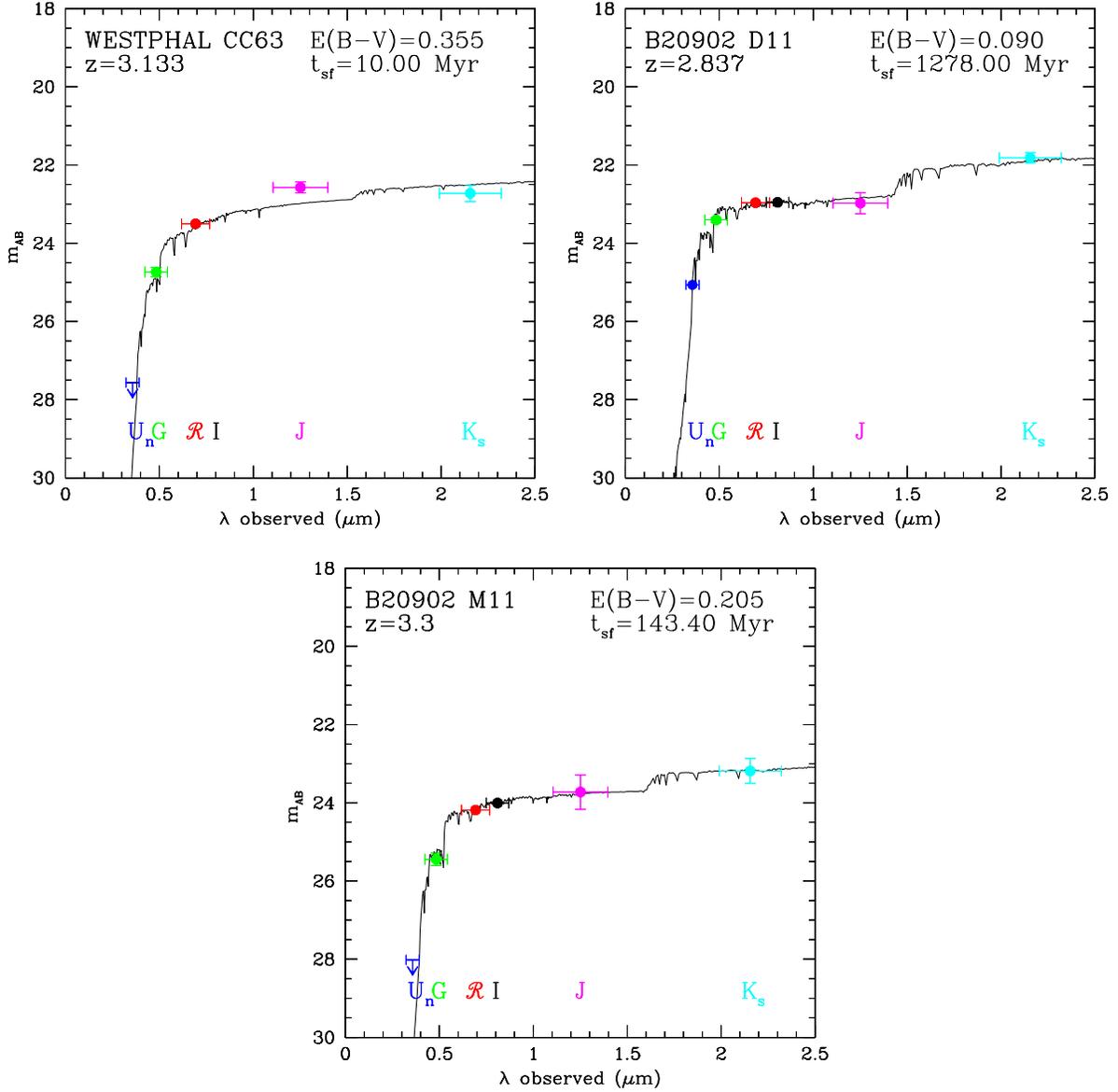}
\caption{ 
The best-fit BC96 constant star-formation models 
for three galaxies in 
the NIRC LBG sample. These three examples span the range of properties
in the sample. A young and dusty galaxy with very little 
evidence for a Balmer break, Westphal-CC63
is fit with a 10 Myr model with $E(B-V)=0.355$ (while the formal
best-fit age for CC63 is 4 Myr, we have restricted the best-fit 
age parameter space to values which are physically plausible
($\geq 10 \: {\rm Myr})$ ). At the other extreme,
B20902-D11 has a best-fit age of 1.3 Gyr and $E(B-V)=0.09$--- much older
with much less dust extinction. Intermediate between these two extremes,
B20902-M11 has a best-fit age of 140 Myr and $E(B-V)=0.205$. 
Each plot shows the galaxy's redshift and best-fit parameters.
The photometric measurements are plotted as magnitudes, yet the error
bars on the $G$, $J$, and $K_s$ points refer to the uncertainties 
in the $G-{\cal R}$, ${\cal R}-J$ and ${\cal R}-K_s$ colors,
respectively. 
The $U$ and $I$ measurements are plotted, even
though they were not used to determine the best-fit models. There
is no $I$ data for the Westphal field, hence the lack of an $I$
point for CC63.
 }
\end{figure}

\newpage
\clearpage
\begin{figure}
\figurenum{8b}
\plotone{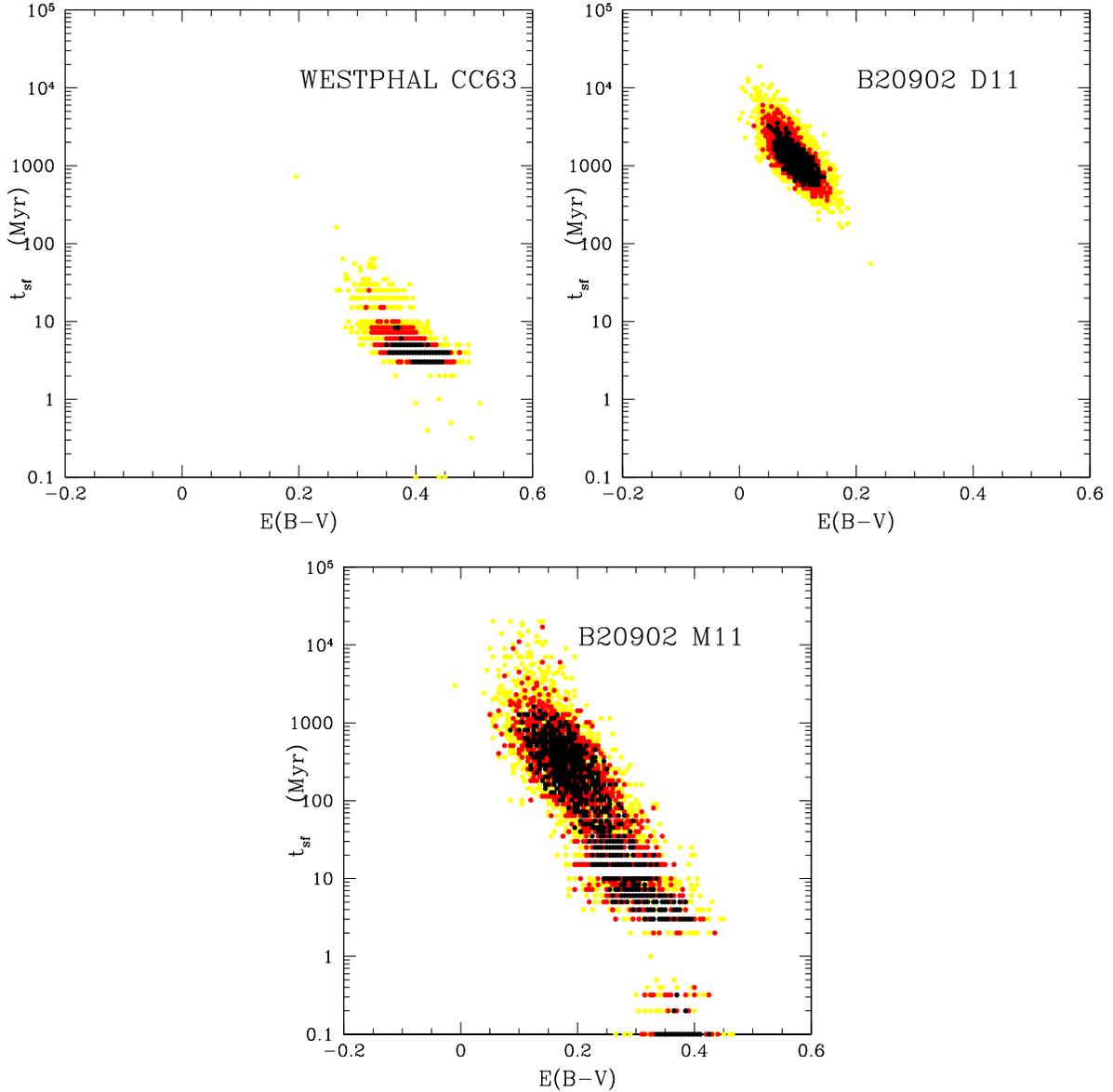}
\caption{
Confidence intervals
in the $E(B-V)-t_{sf}$ parameter space for
the three galaxies featured in Figure 8a. Each confidence
region is determined by generating a large sample of artificial
colors for a galaxy (based on its measured colors and photometric
uncertainties), and finding the best-fit $E(B-V)$ and $t_{sf}$ for
each set of artificial colors. The black region indicates the $68.3 \%$
confidence region; the red (dark grey) region indicates the $90\%$ region, and the
yellow (light grey) region indicates the remaining $10\%$ of the realizations. 
Colors were assigned by probability density--- i.e. the black points
contain the highest density of realizations, while the red (dark grey)
points contain intermediate densities, and the yellow (light grey)
points have the lowest densities. 
Westphal-CC63 and B20902-D11 both have better determined optical-IR
colors than B20902-M11, as demonstrated by the smaller confidence
regions for those galaxies.
 }
\end{figure}

\newpage
\begin{figure}
\figurenum{9}
\plotone{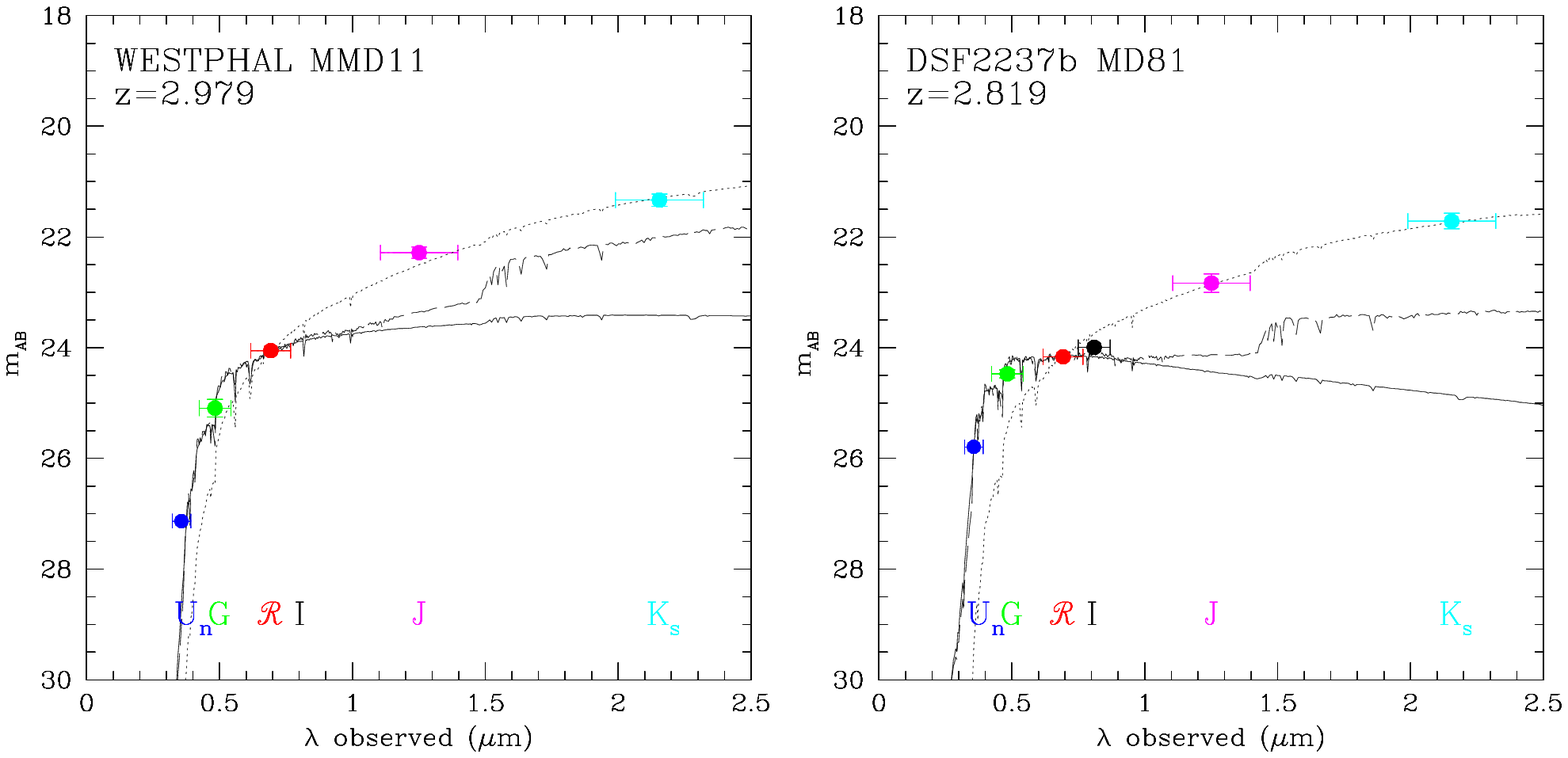}
\caption{
Anomalous Galaxies. Westphal-MMD11 and DSF2237b-MD81 are not
described by any of the simple models which successfully fit the
majority of galaxies in the NIRC LBG sample.
For both galaxies, there is no combination of dust and
age which simultaneously fits the $G-{\cal R}$, 
${\cal R}-J$ and ${\cal R}-K_s$ colors. The ${\cal R}-J$
and ${\cal R}-K_s$ colors are simply too red to be fit by any of the
simple models, including models with different star-formation histories.
1 Myr and 1 Gyr BC96 constant star-formation models were fit
to the observed $G-{\cal R}$ color alone, excluding other colors.
Solid lines represent 1 Myr BC96 constant star-formation models,
and dashed lines indicate 1 Gyr models.
Dotted lines indicate fits to the ${\cal R}-J$ and ${\cal R}-K_s$ colors,
excluding the $G-{\cal R}$ color. For both galaxies,
the ${\cal R}-J$ and ${\cal R}-K_s$ colors
alone are fit by large amounts of extinction $(E(B-V) \geq 0.6)$
and very young ages $(t_{sf} \leq 10 \: {\rm Myr})$.
The extrapolation of the dotted line
to shorter wavelengths significantly underpredicts the flux
observed in the $U_n$ and $G$ filters.
 }
\end{figure}

\newpage
\begin{figure}
\figurenum{10a}
\plotone{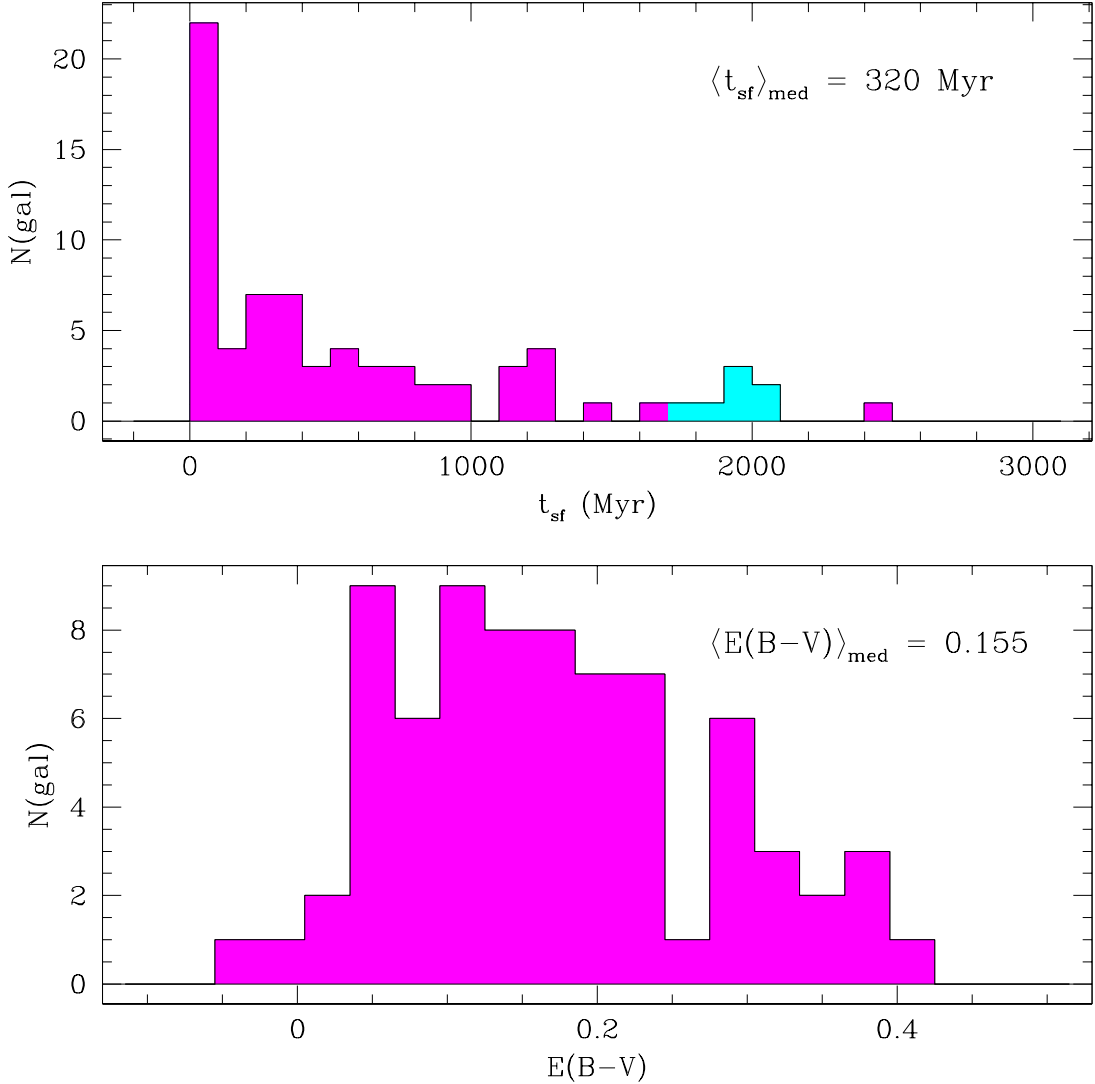}
\caption{
Histograms of $t_{sf}$ and $E(B-V)$ values derived from BC96 constant
star-formation models. Top: The $t_{sf}$ histogram. 
The cyan (light grey) bin
indicates seven galaxies whose unconstrained best-fit $t_{sf}$ values were
older than the age of the universe at $z \sim 3$. 
Constraining each of these galaxies to have a best-fit $t_{sf}$ younger than the
age of the universe at its redshift
(assuming an $\Omega_m=0.3$, $\Omega_{\Lambda}=0.7$, $h=0.7$ cosmology)
resulted in best-fit $t_{sf}$ values of roughly 2 Gyr.
The galaxy Q0201-B13 is at $z=2.167$, so its
best-fit $t_{sf}$ of 2.5 Gyr (the oldest bin) does not pose a problem.
Bottom: The $E(B-V)$ distribution, derived from both
optical and near-IR photometry. 
This distribution probably over-represents the dustiest galaxies,
relative to the LBG population as a whole, as shown by Figure 1b.
 }
\end{figure}

\newpage
\begin{figure}
\figurenum{10b}
\plotone{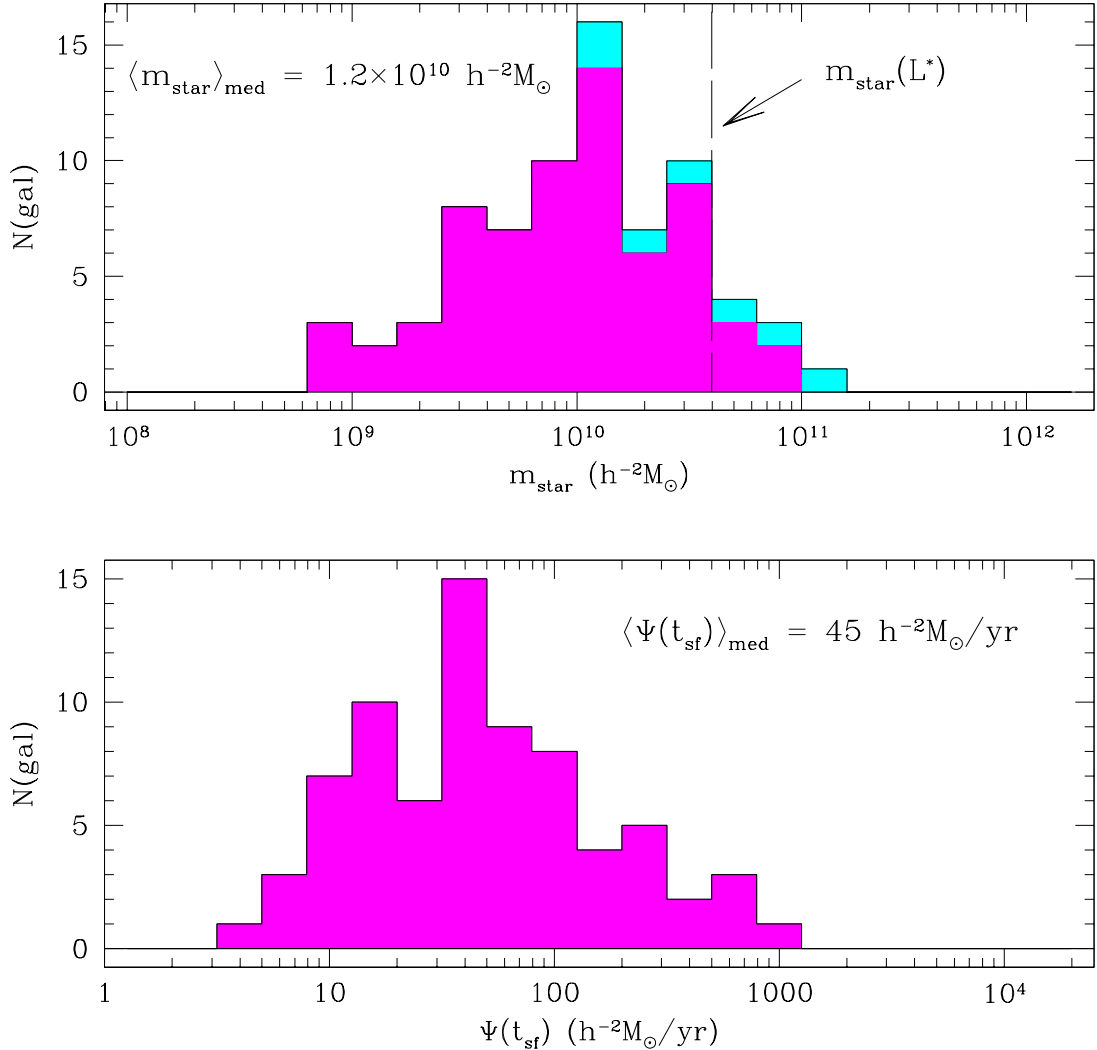}
\caption{
Histograms of $m_{star}$ and instantaneous star-formation rate,
$\Psi(t_{sf})$,
derived from the best-fit BC96 constant star-formation models. 
Top: The formed $m_{star}$ distribution. 
The cyan (light grey) bins contain the
seven galaxies with unconstrained best-fit $t_{sf}$ values older
than the age of the universe at $z\sim 3$. A significant fraction of
the NIRC LBG sample have $m_{star}$ values 
approaching the formed stellar mass in a current 
$L^*$ galaxy $(4 \times 10^{10}h^{-2}M_{\odot}$), while a significant fraction have
$m_{star}$ values that are an order of magnitude smaller. 
Bottom: The distribution of instantaneous star-formation rates,
determined by applying the inferred extinction corrections
to the distribution of rest-frame UV luminosities.
 }
\end{figure}

\newpage
\begin{figure}
\figurenum{11}
\plotone{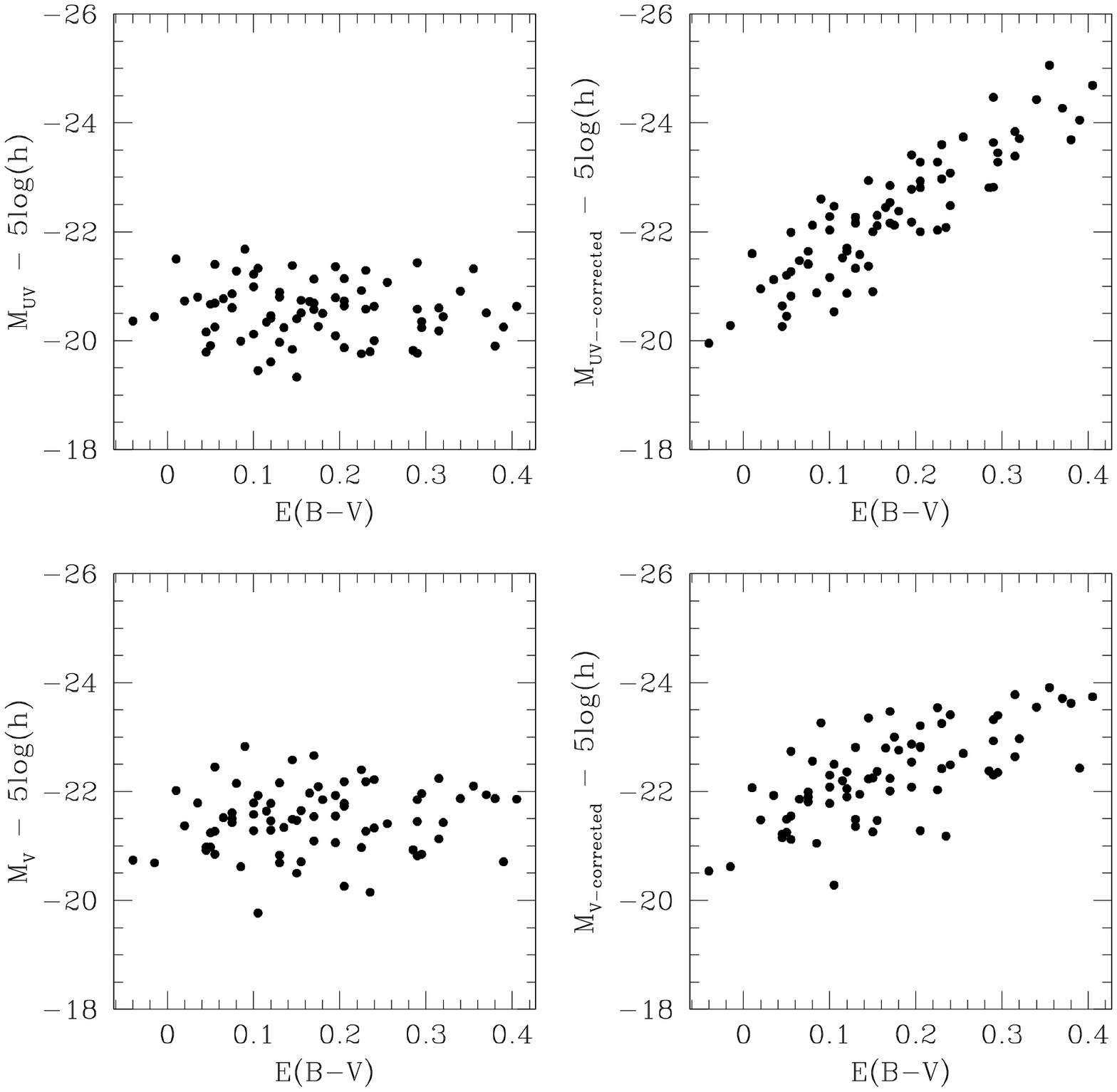}
\caption{
$M_{UV}$ and $M_V$  vs. $E(B-V)$. $M_{UV}$ and $M_V$
refer to the rest-frame UV and optical $(V)$ absolute magnitudes
which are probed by the ${\cal R}$ and $K_s$
apparent magnitudes, respectively,  at $z\sim 3$. The two left-hand panels
show the relationship of rest-frame 
UV and optical luminosities with best-fit $E(B-V)$. Both
UV and optical luminosities are uncorrelated with $E(B-V)$. 
When the luminosities are corrected for dust-extinction (shown in
the right-hand panels), strong
correlations result between intrinsic luminosity and dust extinction.
The correlation
holds not only in the rest-frame UV but also in the rest-frame optical,
where the effects of dust-extinction are less extreme
(but clearly still important, as shown by the lower right-hand
panel). The correlation between intrinsic luminosity and dust extinction
is independent of the star-formation history used to fit the observed
spectral energy distributions.
 }
\end{figure}

\newpage
\begin{figure}
\figurenum{12}
\plotone{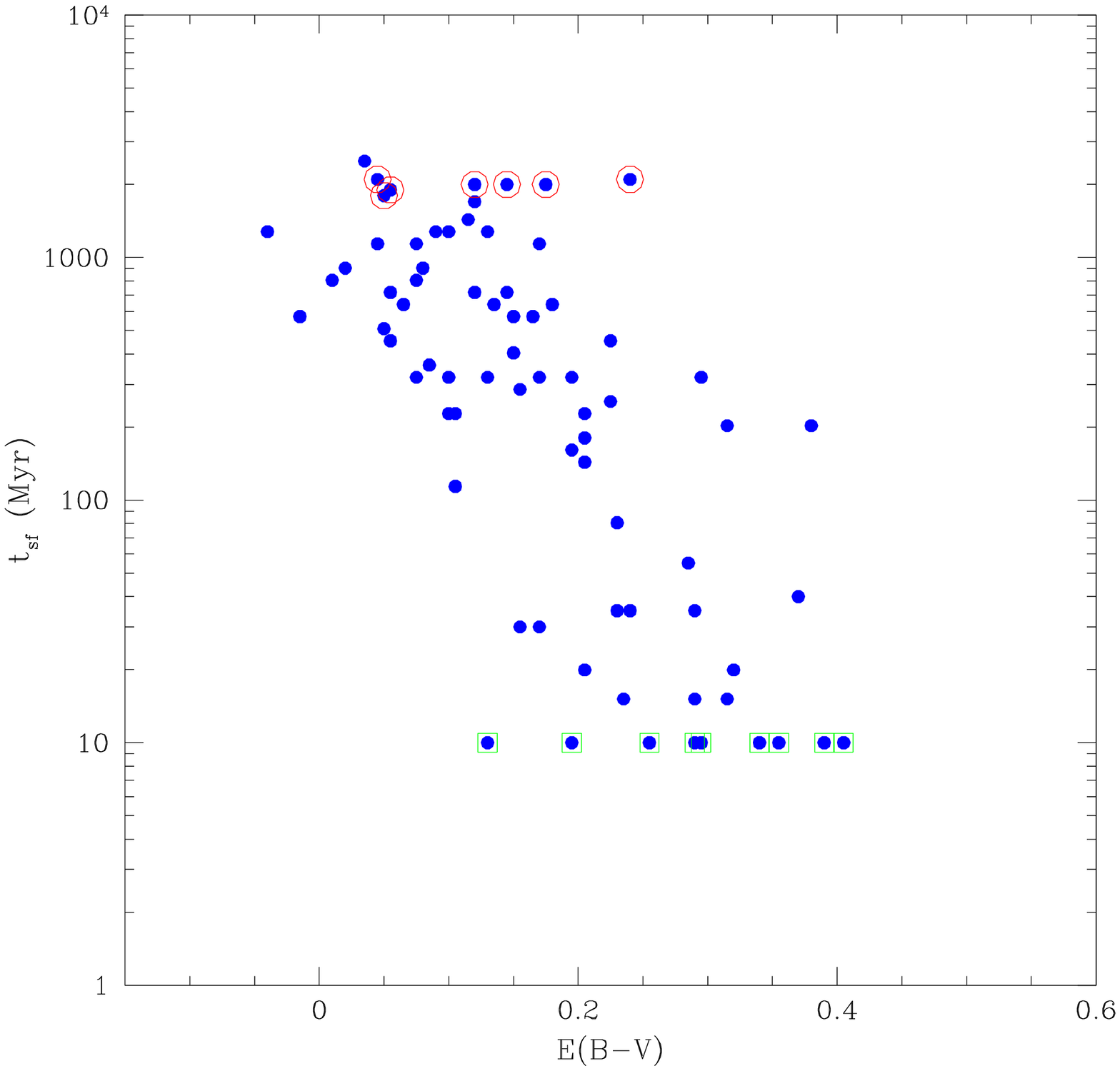}
\caption{
The Extinction-Age Correlation. This is the joint distribution of best-fit
BC96 constant star-formation $E(B-V)$ and $t_{sf}$ parameters. There
is a significant correlation between the two best-fit parameters,
which holds for all of the star-formation histories which we used
to fit the observed spectral energy distributions. However, the strength
of the correlation does depend on the assumed dust attenuation law. 
This plot assumes a Calzetti dust law, but if an SMC curve is used 
instead, the correlation is greatly reduced. Open circles 
indicate galaxies whose unconstrained best-fit $t_{sf}$ values were 
older than the age of the universe at $z\sim 3$. Open squares indicate
galaxies whose unconstrained best-fit ages are $< 10 \: {\rm Myr}$.  
}
\end{figure}
\newpage
\begin{figure}
\figurenum{13}
\plotone{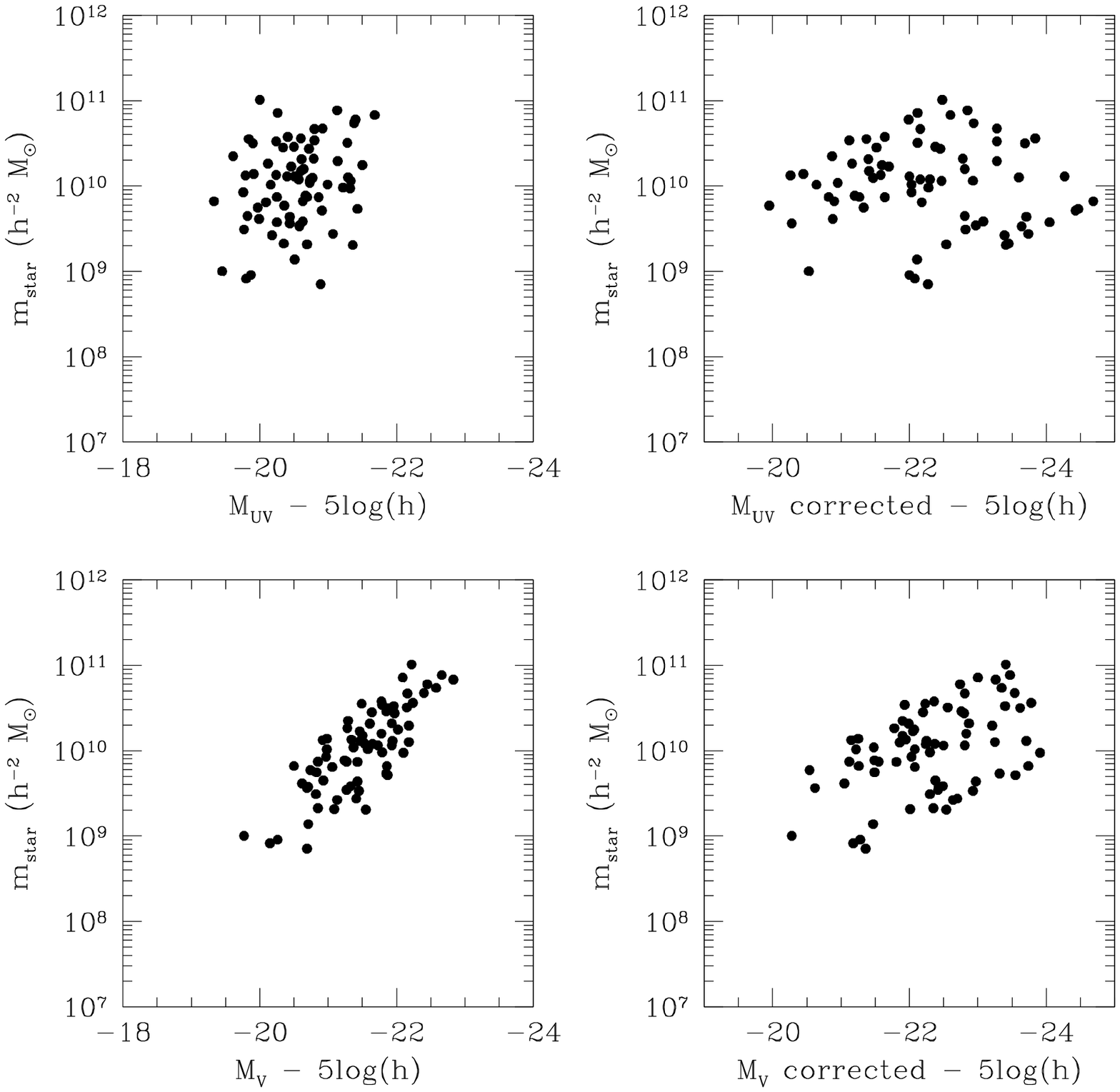}
\caption{
$m_{star}$ vs. $M_{UV}$ and $M_{V}$.
Relationships are shown between $m_{star}$ derived from the
best-fit BC96 constant star-formation models, and the UV and
optical luminosities. $M_{UV}$ and $M_V$
refer to the rest-frame UV and optical $(V)$ absolute magnitudes. 
The left-hand panels show the relationships for
luminosity which is uncorrected for extinction, while the right-hand
panels show the relationships for
extinction-corrected luminosity.
 }
\end{figure}
\newpage
\begin{figure}
\figurenum{14}
\plotone{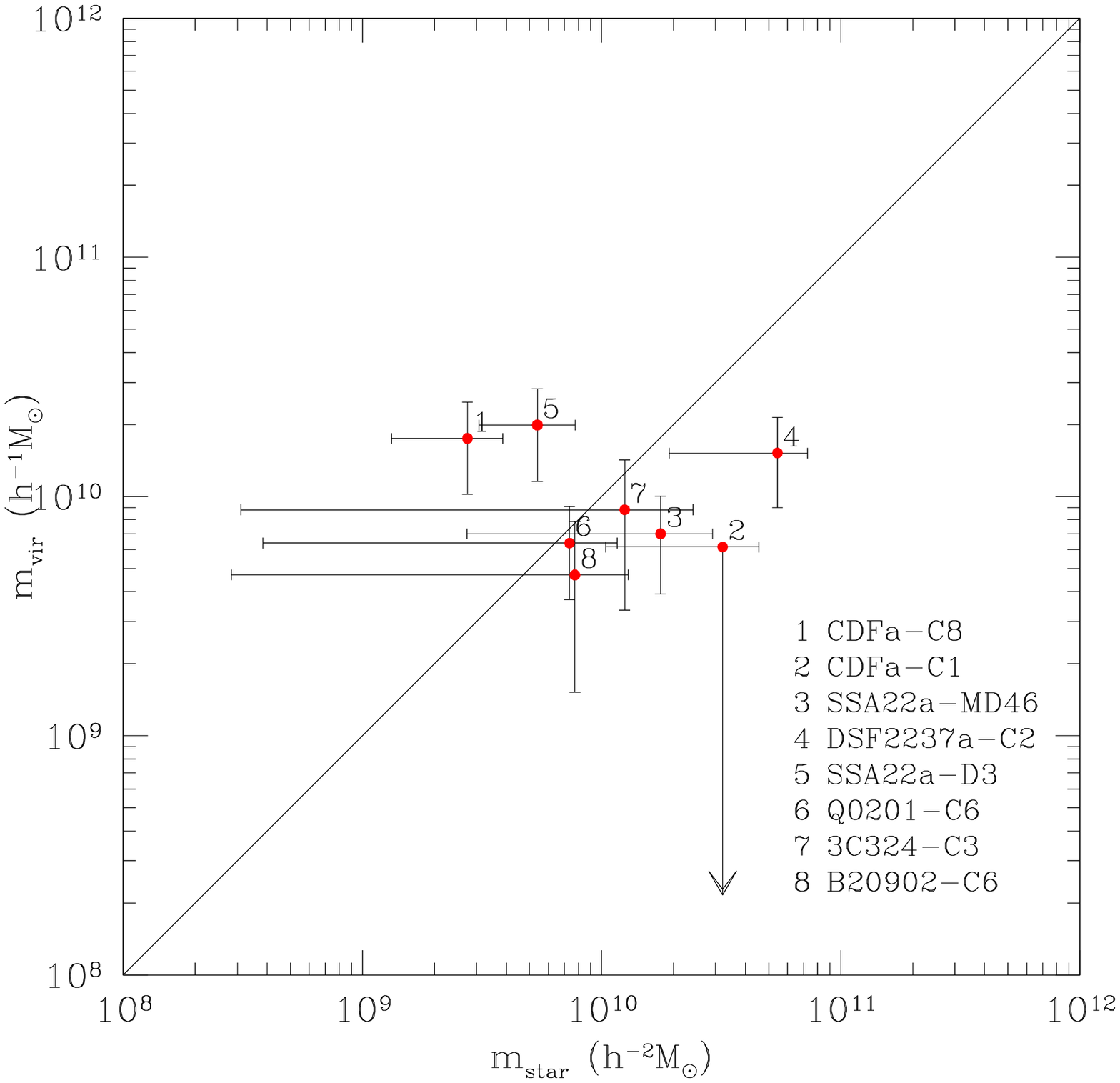}
\caption{
$m_{vir}$ vs. $m_{star}$. This plot shows the relationship
between the dynamical mass implied by nebular line widths,
assuming virial equilibrium ($m_{vir}$),
and the best-fit constant star-formation
formed stellar mass ($m_{star}$). The solid line describes 
$m_{vir} = m_{star}$.  Horizontal error bars 
reflect the $1\sigma$ model confidence region of $m_{star}$,
while vertical error bars reflect the uncertainties in the 
measured nebular line width and angular half-light radius.
The inferred dynamical masses should reflect approximate
upper limits to the formed stellar mass in the physical region
probed by the NIRC observations. We find broad consistency
between the mass scales inferred from near-infrared
spectroscopic and photometric measurements. In the two
cases where there are significant differences,
the best-fit stellar mass is significantly smaller than 
the inferred dynamical mass.
 }
\end{figure}
\newpage
\begin{figure}
\figurenum{15}
\plotone{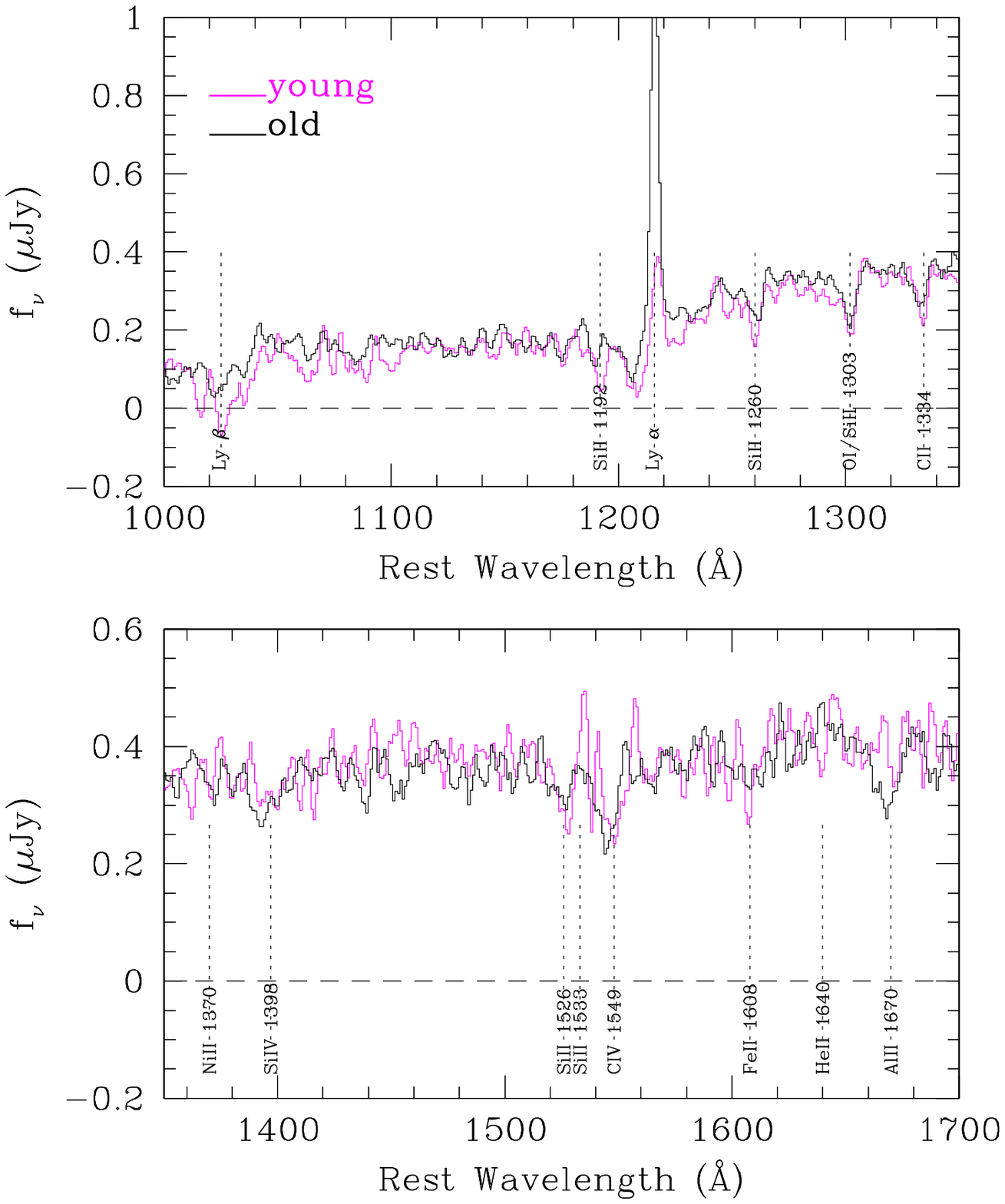}
\caption{
Comparison of ``young'' and ``old'' LRIS spectra. The composite spectrum
of a subsample of galaxies with best-fit constant star-formation 
$t_{sf} \leq 35 \: {\rm Myr}$ is shown in magenta (grey), while the composite
spectrum of a subsample of galaxies with $t_{sf} \geq 1 \: {\rm Gyr}$
is shown in black. The most striking differences between the
two composite spectra are the relative strengths of Lyman-$\alpha$
emission; the relative strengths of the interstellar features
Si II $1192, 1260, 1526 \: {\rm \AA}$,
CII $1334 \: {\rm \AA}$, and Al II $1670 \: {\rm \AA}$;
and the relative CIV $1549 \: {\rm \AA}$ P-Cygni profiles.
 }
\end{figure}
\newpage
\begin{figure}
\figurenum{16}
\plotone{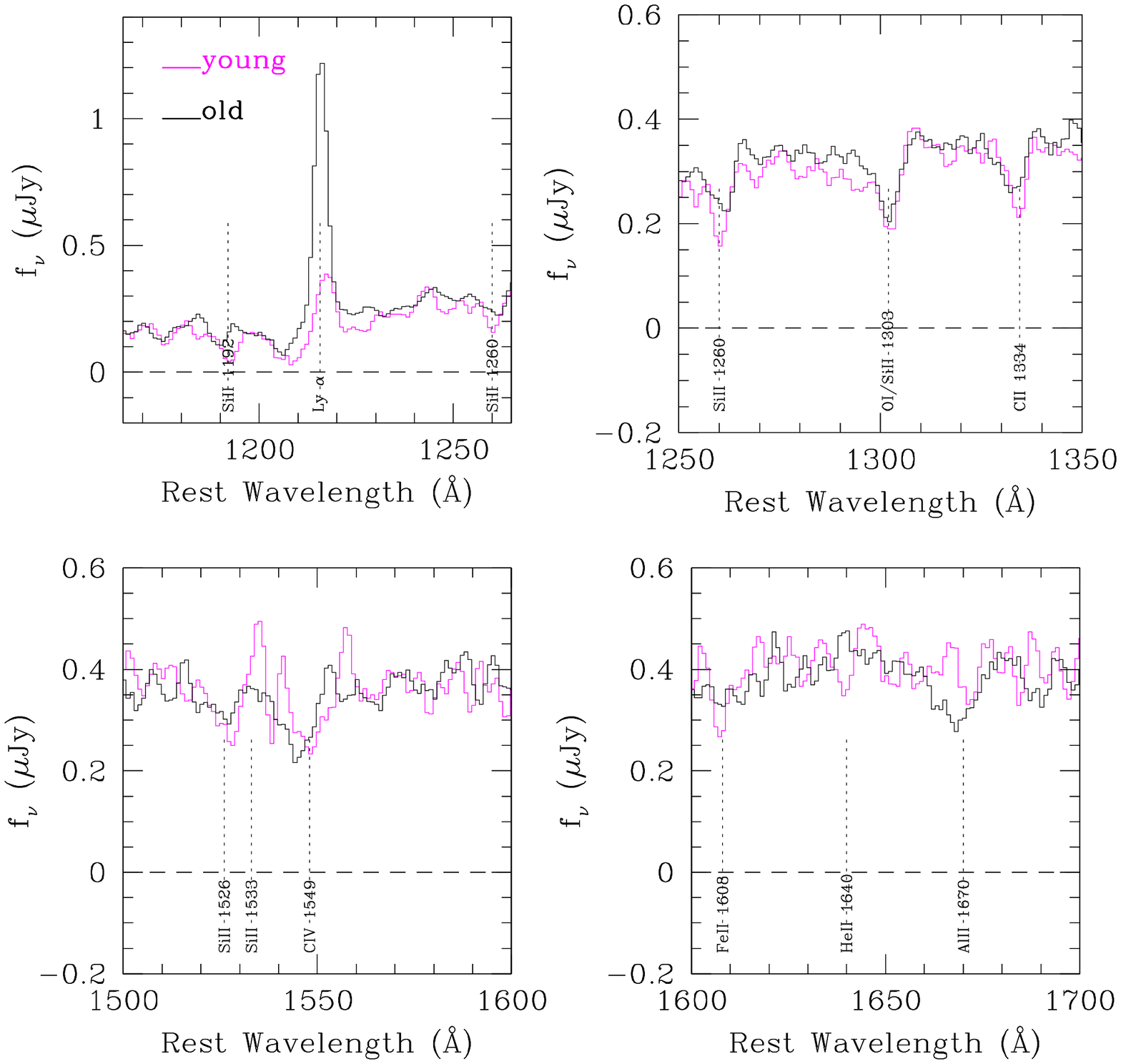}
\caption{
Zoomed in comparison of ``young'' and ``old'' LRIS spectra. Four
regions of the composite young and old spectra are expanded for
a more detailed view. The vertical axis of the detailed plot
of Lyman-$\alpha$ spans a larger range in intensity than the 
other plots, for the purpose of showing the full extent of the
strong Lyman-$\alpha$ emission in the old spectrum. 
 }
\end{figure}

\end{document}